\documentclass[11pt,a4paper]{article}

\usepackage{jheppub}

\usepackage{multirow, graphicx,amssymb,url,mathrsfs,amsmath}
\usepackage{wrapfig,boxedminipage,setspace,subfigure,epsfig}
\usepackage{amsxtra,amstext,latexsym,dsfont,amsfonts}
\usepackage{color,eucal}
\usepackage[dvipsnames]{xcolor}
\usepackage{float}
\usepackage{slashed,comment}
\usepackage{kotex}
\usepackage{tensor}

\usepackage{indentfirst}

\newcommand{\eg}{{\it e.g.}}
\newcommand{\ie}{{\it i.e.}}

\title{More on the upper bound of holographic n-partite information}

\author[a]{Xin-Xiang Ju,}
\author[b]{Wen-Bin Pan,}
\author[a,c]{Ya-Wen Sun}
\author[d,e]{Yuan-Tai Wang}
\author[a]{and Yang Zhao}

\emailAdd{juxinxiang21@mails.ucas.ac.cn}
\emailAdd{panwb@ihep.ac.cn}
\emailAdd{yawen.sun@ucas.ac.cn}
\emailAdd{wangyuantai@ustc.edu.cn}
\emailAdd{zhaoyang20a@mails.ucas.ac.cn}

\affiliation[a]{School of Physical Sciences, University of Chinese Academy of Sciences, Zhongguancun east road 80, Beijing 100190, China}
\affiliation[b]{Institute of High Energy Physics, Chinese Academy of Sciences,\\19B Yuquan Road, Shijingshan District, Beijing 100049, China}
\affiliation[c]{Kavli Institute for Theoretical Sciences, University of Chinese Academy of Sciences, Beijing 100049, China}
\affiliation[d]{Interdisciplinary Center for Theoretical Study, University of Science and Technology of China, Hefei, Anhui 230026, China}
\affiliation[e]{Peng Huanwu Center for Fundamental Theory, Hefei, Anhui 230026, China}

\abstract{
{We show that there exists a huge amount of multipartite entanglement in holography by studying the upper bound for holographic $n$-partite information $I_n$ that $n-1$ fixed boundary subregions participate. We develop methods to find the $n$-th region $E$ that makes $I_n$ reach the upper bound. Through the explicit evaluation, it is shown that $I_n$, an IR term without UV divergence, could diverge when the number of intervals or strips in region $E$ approaches infinity. At this upper bound configuration, we could argue that $I_n$ fully comes from the $n-$partite global quantum entanglement. Our results indicate: fewer-partite entanglement in holography emerges from more-partite entanglement; $n-1$ distant local subregions are highly $n$-partite entangling. Moreover, the relationship between the convexity of a boundary subregion and the multipartite entanglement it participates, and the difference between multipartite entanglement structure in different dimensions are revealed as well.

}

}

\begin{document}
\maketitle

%%%%%%%%%%%%%%%%%%%%%%%%%%%%%%%%
%    Section: Introduction
%%%%%%%%%%%%%%%%%%%%%%%%%%%%%%%%
\section{Introduction} \label{sec1}
\noindent In a holographic system, due to the strong coupling nature, the dual quantum system is expected to be strongly entangled \cite{Maldacena:1997re,Ryu_2006,Maldacena_2013,VanRaamsdonk:2010pw}. To reveal the entanglement structure of the dual quantum system, various entanglement measures could be employed. The basic measure is the entanglement entropy, which could be given by the area of the Ryu-Takayanagi (RT) surface in holography \cite{Ryu_2006}. The entanglement entropy measures the amount of entanglement that one subregion participates. For two subregions $A$ and $B$ of the boundary quantum system, mutual information $I(A:B)$ quantifies both the classical and quantum entanglement between $A$ and $B$. However, mutual information alone does not give the full quantum entanglement structure that  $A$ and $B$ participate: it cannot tell the multipartite entanglement that $A$ and $B$ are involved in simultaneously, i.e. how $\rho_{AB}$ is embedded in the whole boundary system. This could be reflected via other measures \cite{Akers:2019gcv,Ju:2024hba,Ju:2023dzo,Basak:2024uwc,Bao:2017nhh}, \eg, the conditional mutual information (CMI) between $A$ and $B$ under the condition of a third subregion $E$, \begin{equation}\label{CMIformula}
    I(A:B|E)=S_{AE}+S_{BE}-S_E-S_{ABE}.
\end{equation}

For two small subregions at a distance, the mutual information between them is always zero (at the order of $\frac{1}{G_N}$), while CMI could be nonzero, revealing possible multipartite entanglement \cite{Ju:2024xcn} that both regions participate together. CMI serves as a basic building block in the definition of differential entropy \cite{Hubeny_2014,Czech_2015,Czech:2014wka,Myers:2014jia,Headrick:2014eia,Balasubramanian:2018uus,Balasubramanian:2013lsa,Czech:2014ppa,Balasubramanian:2013rqa,Ju:2023bjl} and is a candidate for partial entanglement entropy \cite{Vidal:2014aal,Wen:2018whg,Wen:2019iyq}, which could be utilized to characterize the entanglement density \cite{Nozaki:2013wia, Bhattacharya:2014vja,Shimaji:2018czt} and plays an important role in studying the entanglement spreading in a nonequilibrium system. In \cite{Ju:2024xcn}, it was also shown that CMI of two small subregions at different distances could be related to bulk geometry at the corresponding radial positions for certain geometries.

In all these circumstances above, the conditional region $E$ has been chosen to be the interval situated between the two subregions $A$ and $B$ in the $1+1$-dimensional boundary. However, apparently, varying the region $E$ would result in different behavior of CMI and reveal different entanglement structures among different subregions. Therefore, observing how the CMI changes with different selections of $E$ can provide crucial insights into the underlying entanglement structures, particularly when $E$ is chosen such that it maximizes the CMI. %Finding the region $E$ which give the largest CMI with regions $A$ and $B$ fixed would provide us important information about the entanglement structure of the dual quantum many-body system. 
As $I(A:B|E)=-I_3(A:B:E)+I(A:B)$, where $I_n(A:B:...)$ denotes the n-partite information \cite{Alishahiha:2014jxa}, for fixed regions $A$ and $B$ this would be the same as finding the maximum of $-I_3(A:B|E)$ with $I(A:B)$ fixed. 

For the region $E$ chosen to be the interval in-between $A$ and $B$, it can be verified that this is not the optimal choice for maximizing the CMI. Instead, choosing $E$ to be a set of disconnected intervals could potentially result in a higher CMI. In this paper, we develop two different methods to find the region $E$ that maximizes the CMI or equivalently maximizes $-I_3$ . We start from considering E to be a subregion with a given number $m$ of disconnected intervals and find the maximum value of CMI for each number $m$ of disconnected intervals. We find that as the number $m$ of disconnected intervals increases, the maximum value of CMI or equivalently $-I_3$ grows and diverges when $m$ goes to infinity. To analyze this problem more systematically, we develop a more general and universal method which imposes significant constraints on the configuration at the maximum $-I_3$, thereby simplifying the task considerably. The result shows that $-I_3$ could reach its quantum information theoretical upper bound. We then generalize these calculations to the upper bound of $I_4$, $-I_5$ and to higher dimensions. We find that when $n-1$ subregions are fixed, tuning the $n$-th region $E$, the upper bound of $(-1)^n I_n$ with $n=4,5$ are both finite in AdS$_3$/CFT$_2$ and could reach the quantum informational upper bound in higher dimensions, which diverges as the number $m$ of disconnected strips in $E$ goes to infinity.

The maximum configuration of CMI indeed reveals important features of the underlying entanglement structure. $-I_3$ is a measure for tri-partite entanglement, which however, is not a perfect measure in general as it may get contributions due to classical correlations, making its sign indefinite. Purely classical contributions of the tripartite information may result in a negative $-I_3$ \cite{LoMonaco:2023xws}. In holographic systems, $-I_3$ has been shown to be always non-negative from the monogamy of mutual information \cite{Hayden_2013}.

In our holographic maximum configuration of $-I_3$, $-I_3(A:B:E)$ fully captures the tripartite global entanglement that $A$ and $B$ participate together with $E$. This could be seen from the fact that at the maximum configuration, $I(A:E)=I(A:B)=0$ while $I(A:BE)=-I_3(A:B:E)=2 S_A$. This indicates that $A$ does not entangle with $B$ or $E$, but fully entangles with the combination of $BE$, which is a symbol of tripartite global entanglement \cite{bengtsson2016brief} fully captured by $-I_3(A:B:E)$. This fact, together with other stronger and more general evidences that we will present in this work, imply that any two distant and small subregions are highly tripartite entangled and all bipartite entanglement in holographic systems could emerge from tripartite global entanglement. The maximum configurations for general $(-1)^nI_n $ could generalize this conclusion to general $n$-partite global entanglement.

It is worth noting that the multi-partite entanglement we investigate is only one type of the  multipartite entanglement structure that could be revealed by the entanglement entropy of the subsystems. However, even if two multipartite quantum systems have all subsystems with the same entanglement entropy, those two systems could still not be local unitary equivalent, or could not transform into each other by local operations and classical communication (LOCC) \cite{bengtsson2016brief}. To distinguish such delicate difference between those two quantum systems, measures which are beyond the combination of entanglement entropy should be introduced, such as reflected entropy \cite{Dutta:2019gen}, negativity \cite{Calabrese:2012ew,Kusuki:2019zsp}, or more general multipartite measures \cite{Gadde:2022cqi,Penington:2022dhr,Yuan:2024yfg}, etc. Those topics are beyond the concept of this paper.

Here we also comment on the relationship between our upper bound of $(-1)^n I_n$ with the holographic entropy cone (HEC). The work on the holographic entropy cone analyzes the holographic entropy inequalities \cite{Bao:2015bfa,Hubeny:2018trv,Hubeny:2018ijt,He:2019ttu,HernandezCuenca:2019wgh,He:2020xuo,Avis:2021xnz,Fadel:2021urx,Bao:2024azn}. In other words, its aim is to find all the combinations of holographic entanglement entropy with zero upper (lower) bound. A significant difference in our discussion is, the number of intervals of the arbitrarily chosen region can tend to infinity, in other words, the boundary of this region might not be smooth. This violates the premise of the holographic entropy cone work \cite{He:2020xuo}, and the value of a super-balanced entropy combination can be divergent in our case. Moreover, unbalanced holographic entropy inequalities could arise from the upper bound of CMI in our work.
%During this procedure, no regions are fixed, while in our work, we might have one or several regions fixed. If we generalize our argument from \( I_n \) to any combination of HEE (as we are trying in Section 5), holographic entropy inequalities would be a special case in our context.
%Then what is the meaning of evaluating the upper bound while fixing several regions? The answer is, 
Furthermore, we can analyze the entanglement structure which involves several specific regions. We can ask questions such as ``How much tripartite entanglement can regions \( A \) and \( B \) maximally participate in, along with another region?'' ``If their tripartite entanglement could reach the information theoretical upper bound, are there any shape constraints on regions \( A \) and \( B \)?'' Those questions cannot be easily answered in holographic entropy cone program, and our ultimate goal is to find the upper bound of all combinations of holographic entanglement entropy.%Those questions cannot be answered when we only focus on the entropy inequalities without several regions fixed.

This paper is organized as follows. In Sections \ref{sec2}, \ref{sec3}, and \ref{sec4}, we investigate the upper bound of the conditional mutual information \( I(A:B|E) \) with region \( E \) arbitrarily chosen. Sections \ref{sec2} and \ref{sec3} focus on the AdS$_3$/\(\text{CFT}_{2}\) case, where the configuration with the maximum CMI is found and the divergence behavior is investigated, respectively. In Section \ref{sec4}, the result is greatly generalized, and the technical method developed there can be used to find the upper bound of \( I_4 \) and \(- I_5 \), which we discuss in Section \ref{sec5}. A brief discussion of finding the upper bound for general combinations of holographic entanglement entropy is also presented in Section \ref{sec5}. We conclude our results in Section \ref{sec6}.

\section{Maximizing CMI: the multi-entanglement phase transition rule} \label{sec2}
    \noindent To find the region $E$ which maximizes the CMI $I(A:B|E)$ with $A$ and $B$ fixed, we start from some basic analysis on the behavior of CMI with varying $E$ in AdS$_3$. As we vary $E$ from being a single interval to a set of disconnected intervals, simple calculations show that CMI could increase. Therefore, we would start from finding the maximum CMI configuration of $E$ with the number of intervals in $E$  fixed and then compare the maximum value for different numbers of intervals.  In this section, we first calculate two simple examples in Section \ref{sec2.1}: the cases where the region \( E \) has one or two intervals, and find the configuration of $E$ that maximizes CMI in those cases. We find that the geometric configurations for the region $E$ that reaches the CMI upper bound are uniquely determined by a set of phase transition conditions of RT surfaces. We will propose and prove a multi-entanglement phase transition rule utilizing an iterative method in Section \ref{sec2.2}, stating that only multi-entanglement phase transition critical points of RT surfaces could reach the upper bound of CMI, laying foundations for the analysis of the behavior of the maximum value of CMI with varying $E$ in Section \ref{sec3}.

     \subsection{Basic example: $E$ has one or two intervals}\label{sec2.1}
    
\noindent According to the RT proposal \cite{Ryu_2006} for the holographic entanglement entropy, the entanglement entropy of a subregion \( A \) in the boundary conformal field theory is proportional to the area of the bulk minimal surface \( \gamma_A \) that is homologous to \( A \)
\begin{equation}
	S_A=\frac{\text{Area}\left(\gamma_A\right)}{4G_N}.
\end{equation}
In AdS$_3$, the minimal surface for a single interval \(A\) reduces to the geodesic that connects the two endpoints of \(A\).
The length of the geodesic is \(2 R \log \frac{L}{\epsilon}\) in Poincare coordinates, where \(L\) is the length of \(A\), \(\epsilon\) is the UV cutoff and \(R\) is the AdS radius.  
The entanglement entropy of \(A\) is then given by
\begin{equation}
	S_A=\frac{c}{3}\log \frac{L}{\epsilon},
\end{equation}
where \(c\) is the central charge and we have used the identity \(\frac{L}{G_N}=\frac{2c}{3}\)\cite{Brown:1986nw}.

\subsubsection{\(E\) being one interval}
\noindent We first deal with the case that each of the subregions \(A\), \(B\) and \(E\) contains only a single interval at the boundary of $AdS_3$.
We want to find the configuration of the region $E$ (if there exists one) that maximizes the CMI \(I(A:B|E)\) under the condition that \(A\) and \(B\) are fixed. In the following of this paper, we will refer to the whole interval in between $A$ and $B$ as the {\it gap region} for convenience.

The CMI \(I(A:B|E)\) can be rewritten as  
\begin{equation}\label{CMIr}
	\begin{aligned}
		I(A:B|E)&=S_{AE}+S_{BE}-S_E-S_{ABE}\\
		        &=\left(S_{AE}-S_{A}-S_{E}\right)+\left(S_{BE}-S_{B}-S_{E}\right)+\left(S_{A}+S_{B}+S_{E}-S_{ABE}\right)\\
		        &=-I(A:E)-I(B:E)+J_3(ABE),
	\end{aligned}
\end{equation}
where we have defined \(J_3(ABE):=S_{A}+S_{B}+S_{E}-S_{ABE}\), which is the tripartite mutual information of three systems $A$, $B$ and $E$, in contrast to $I_3$, which is the tripartite information.
We will show that all the three quantities \(I(A:E)\), \(I(B:E)\) and \(J_3(ABE)\) can be represented in terms of one dimensional \footnote{The spatial dimension of the boundary is one, since we are dealing with the entanglement entropy of a time slice in AdS\(_3\).} cross ratios.

Given two disjoint intervals \(P=(x_i, x_j)\) and \(Q=(x_{k}, x_{l})\) in a line, a cross ratio is defined to be 
\begin{equation}\label{CrossratioDef}
    X(P:Q):=\frac{x_{ij} x_{kl}}{x_{jk} x_{il}},
\end{equation}
where we have denoted \(x_{mn}=x_m-x_n\).
It is straightforward to verify that the cross ratio is constructed to be invariant under one dimensional conformal transformations, i.e. the translation, dilation and special conformal transformation.
Note that in one dimension, there are \(N-3\) independent cross ratios that are constructed from \(4\) out of \(N\) points\footnote{In \(D\) dimensions, there are \(D\cdot N\) variables to build a cross ratio %\sun{cross ratios?}
and \((D+2)(D+1)/2\) constraints coming from the freedom of conformal transformations. Thus in general, the number of independent cross ratios is \(D\cdot N-(D+2)(D+1)/2\).}. 
Thus we have only \(1\) independent cross ratio given \(4\) points.
All other cross ratios constructed from the \(4\) points can be represented by \eqref{CrossratioDef}. For instance:
\begin{equation}
    \frac{x_{ij} x_{kl}}{x_{ik} x_{jl}}=\frac{X(P:Q)}{1+X(P:Q)}.
\end{equation}
The definition \eqref{CrossratioDef} is in fact the choice of the cross ratio that is directly related to the mutual information of two intervals that do not overlap, as shown in the following.

\begin{figure}[h]
	\centering
	\includegraphics[scale=0.8]{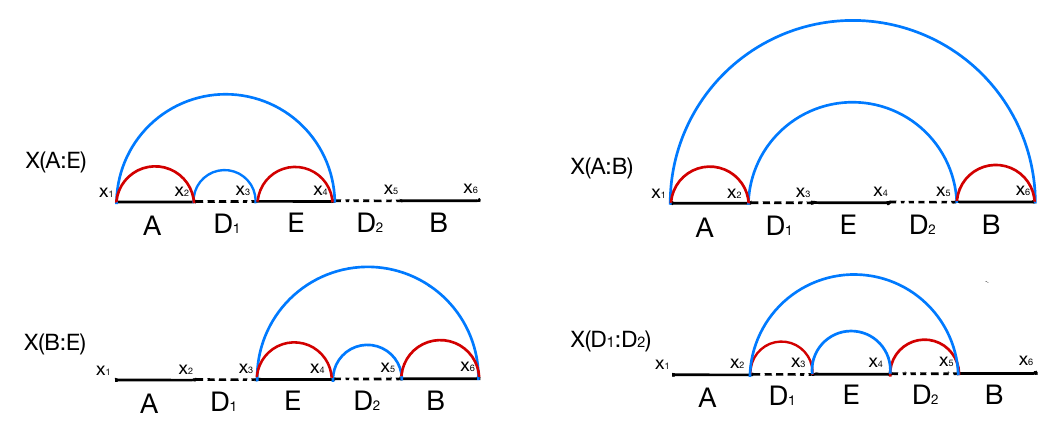}
		\caption{An illustration for the cross ratios \( X(A:E) \), \(X(B:E)\), \( X(A:B) \), \(X(D_1:D_2)\). Each of them corresponds to a mutual information and thus a geodesic configuration. The red/blue semi-circles represent the geodesics whose lengths have positive/negative signs in the calculation of the mutual information such as \eqref{IPQ}.}\label{CrossRatio}
\end{figure}

We first show that the mutual information (MI) of two disjoint intervals $P$ and $Q$ is essentially determined by the cross ratio $X(P:Q)$ defined above.
\begin{equation}\label{IPQ}
	\begin{aligned}
		I(P:Q)&=S_{P}+S_{Q}-S_{PQ}\\
		&=\frac{c}{3}\log \frac{L_P}{\epsilon}+\frac{c}{3}\log \frac{L_Q}{\epsilon}-\text{min}\left\{\frac{c}{3}\log\frac{L_P}{\epsilon}+\frac{c}{3}\log \frac{L_Q}{\epsilon},\;\; \frac{c}{3}\log\frac{L_{D}}{\epsilon}+\frac{c}{3}\log \frac{L_{PDQ}}{\epsilon}\right\}\\
		&=\text{max}\left\{0,\;\; \frac{c}{3}\log\left(\frac{L_{P}L_{Q}}{L_{D}L_{PDQ}}\right) \right\}\\
		&=\frac{c}{3}\log\left(\text{max}\left\{1,\;\; \frac{x_{ij}x_{kl}}{x_{jk}x_{il}} \right\}\right)\\
		&=\frac{c}{3}\log\left(\text{max}\left\{1,\;\; X(P:Q) \right\}\right),
	\end{aligned}
\end{equation}
where $D$ is the gap region between $P$ and $Q$, the lengths of the intervals are denoted by $L_P$, $L_Q$, $L_D$ and $L_{PDQ}$ respectively.
Since there are three intervals $A$, $B$ and $E$ to be considered in the CMI \eqref{CMIr}, we illustrate in Figure \ref{CrossRatio} our definitions of various cross ratios for the subregions: $A=(x_1, x_2)$, $E=(x_3, x_4)$, $B=(x_5, x_6)$, and intervals between them: $D_1=(x_2, x_3)$, $D_2=(x_4, x_5)$, which are
\begin{equation}\label{defX}
    X(A:E)=\frac{x_{12} x_{34}}{x_{23} x_{14}},\quad X(E:B)=\frac{x_{34} x_{56}}{x_{45} x_{36}},\quad
    X(A:B)=\frac{x_{12} x_{56}}{x_{25} x_{16}},\quad
    X(D_1:D_2)=\frac{x_{23} x_{45}}{x_{34} x_{25}}.
\end{equation}

\begin{figure}[H]
	\centering
    \includegraphics[scale=0.8]{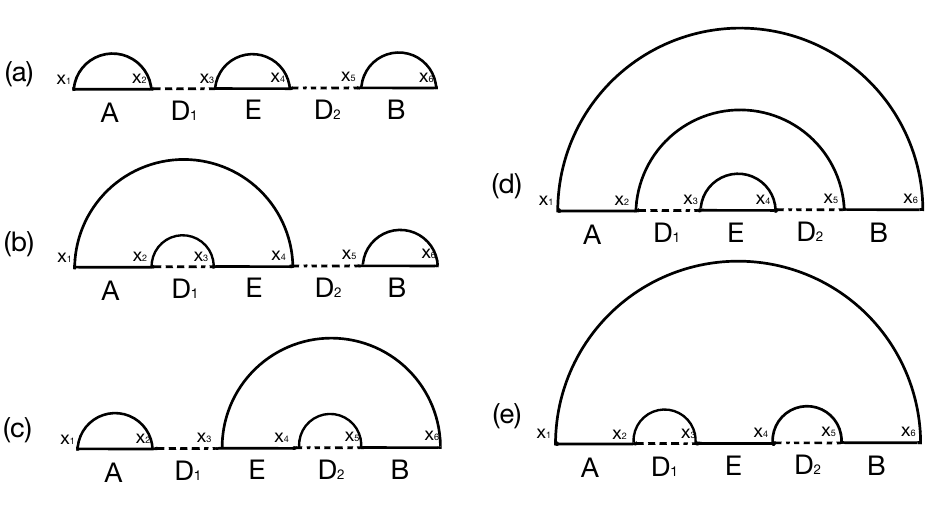}
	\caption{Five possible geodesic configurations for the calculation of \(S_{ABE}\). }\label{SABE}
\end{figure}

Next we show that \(J_3(ABE)\) is also a function of the cross ratios.
There are five {configurations} for the geodesics that compute \(S_{ABE}\), as shown in Figure \ref{SABE}.
\begin{figure}[h]
	\centering
    \includegraphics[scale=0.6]{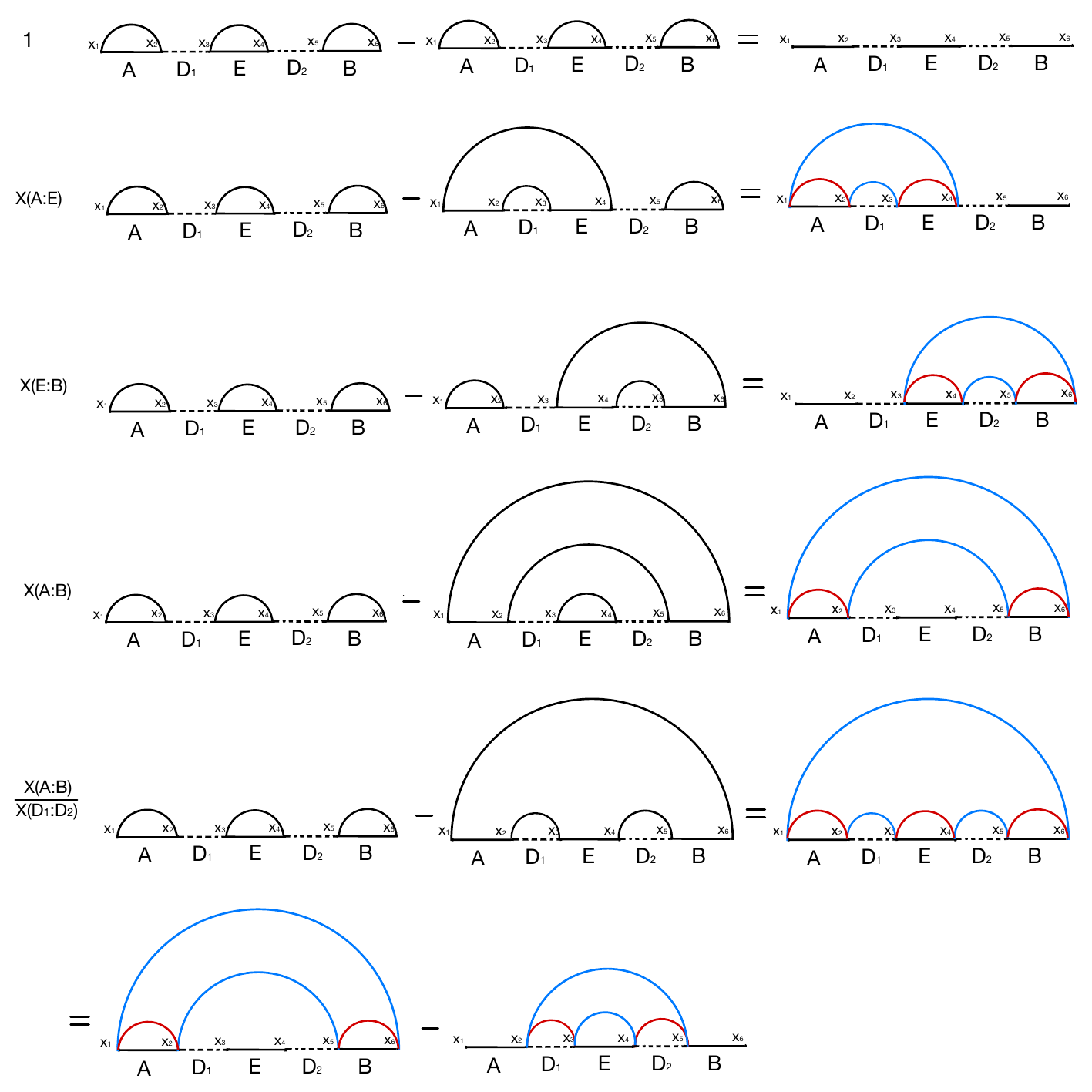}
	\caption{An illustration for the geodesic configurations that correspond to $J_3(ABE)=S_A+S_B+S_C-S_{ABE}$.
	Each of the first five lines consists of three diagrams. In each line, the first diagram shows the geodesic configuration that computes $S_A+S_B+S_C$.
	It is subtracted by the second diagram which computes $S_{ABE}$ and belongs to one of the five configurations in Figure \ref{SABE}.
	The result in each line is the configuration shown in the last diagram in which the red/blue semi-circles represent the geodesics whose lengths have positive/negative signs.
	The final result in each line corresponds to a cross ratio or a function of cross ratios.
	The first four lines correspond to $1$, $X(A:E)$, $X(B:E)$ and $X(A:B)$, respectively.
	The last diagram in the fifth line is equivalent to the difference of the two diagrams in the last line,  corresponding to the ratio between \(X(A:B)\) and \(X(D_1:D_2)\).
 }\label{XABE}
\end{figure}

\(S_{ABE}\) is proportional to the configuration which has the least total geodesic lengths among them.
Which of the five configurations is the minimal one is determined by the cross ratios of these subregions and gap regions. 
As could be seen from Figure \ref{XABE}, the configuration (a) in Figure \ref{SABE} gives an $J_3(ABE)=S_{A}+S_{B}+S_{C}-S_{ABE}$ that is proportional to $\log 1$.
The configuration (b), (c), (d) and (e) in Figure \ref{SABE} correspond to the $J_3(ABE)$ which is proportional to $\log X(A:E)$, $\log X(B:E)$, $\log X(A:B)$, and $\log (X(A:B)/X(D_1:D_2))$, respectively.
Therefore, as the quantity \(J_3(ABE)\) has a minus sign in $S_{ABE}$, the final value of $J_3(ABE)$ is determined by the maximal one among the cross ratios, i.e.
%\sun{why maximal? it is constructed from SABE so why not minimal? not maximal geodesics but maximal value of -geodesic length?} \sun{these are two different things? S is always the minimal geodesic length, -S is also the minimal geodesic, Maximum J is minimal S, but not referring to which geodesic but which configuration?} 
\begin{equation}\label{JABE}
	\begin{aligned}
		&J_3(ABE)=\frac{c}{3}\log \left(\text{max}\left\{1, X(B:E), X(A:E), X(A:B), \frac{X(A:B)}{X(D_1:D_2)}\right\}\right).
	\end{aligned}
\end{equation}
%\sun{indicate more clearly what these 12345 are, explain more on those in figure1c}.

Combining \eqref{CMIr}, \eqref{IPQ}, \eqref{defX} and \eqref{JABE}, one can write the CMI in terms of the cross ratios explicitly as
\begin{equation}\label{IABE}
	\begin{aligned}
		I(A:B|E)&=\frac{c}{3}\log\frac{\text{max}\left\{1, X(A:E), X(B:E), X(A:B), X(A:B)/X(D_1:D_2) \right\}}{\text{max}\left\{1, X(A:E) \right\}\text{max}\left\{1, X(B:E) \right\}}.
	\end{aligned}
\end{equation}
Since the cross ratios are invariant under the one dimensional conformal transformations, \(I(A:B|E)\) is also conformal invariant.

As demonstrated above, there are \(N-3\) independent cross ratios for \(N\) endpoints. 
Now that we are dealing with \(3\) intervals and \(6\) endpoints, there are \(3\) independent cross ratios.
Therefore, the CMI \eqref{IABE} can be represented by these \(3\) independent cross ratios. 
One can choose them to be \(X(A:E)\), \(X(B:E)\) and \(X(A:B)\) for simplicity and other cross ratios in \eqref{IABE} could be expressed by using these ones.
Since the regions \(A\) and \(B\) are fixed in our setup, there are only two free variables: \(X(A:E)\) and \(X(B:E)\).
The only unknown variable that needs to be represented by these independent free variables in \eqref{IABE} is \(X(D1:D2)\).  
After some calculations, it could be found that
\begin{equation}
	\frac{X(A:B)}{X(D1:D2)}= \frac{1}{2}\left(b+\sqrt{b^2+4c}\right),
\end{equation}
where \(b=X(A:E)X(B:E)+X(A:E)X(A:B)+X(A:B)X(B:E)+X(A:E)X(B:E)X(A:B)\) and \(c=X(A:E)X(B:E)X(A:B)\).
We note here that because \(b\) and \(c\) are invariant after exchanging any two of the regions \(A\), \(B\) or \(E\), the CMI \eqref{IABE} is also invariant under the exchange. This means there is no real difference in whether \(E\) locates inside the gap region in between \(A\) and \(B\) or outside of them, which is a consequence of conformal invariance of \(I(A:B|E)\).
%\sun{more clearly  as to which E and which E outside?}.\wb{There is only one interval of E here. Maybe I should analyze the two interval case in the next subsubsection.}

\begin{figure}[H]
\centering
\subfigure[]{
	\begin{minipage}[b]{.47\linewidth}
		\centering
		\includegraphics[scale=0.75]{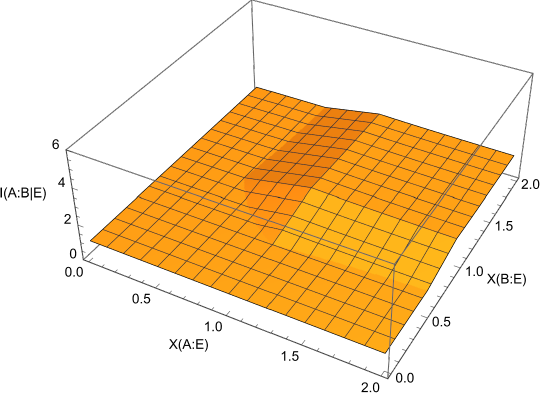}
	\end{minipage}
}
\subfigure[]{
	\begin{minipage}[b]{.47\linewidth}
		\centering
		\includegraphics[scale=0.75]{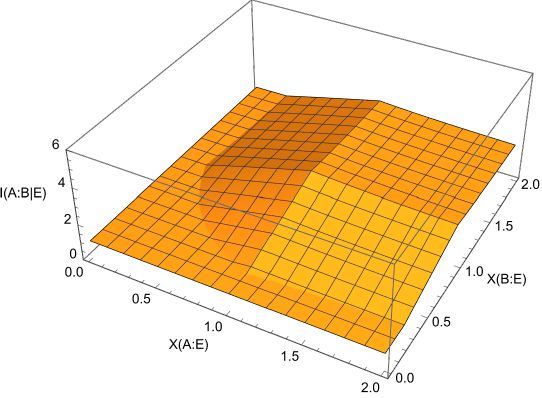}
	\end{minipage}
}\\
\subfigure[]{
	\begin{minipage}[b]{.47\linewidth}
		\centering
		\includegraphics[scale=0.75]{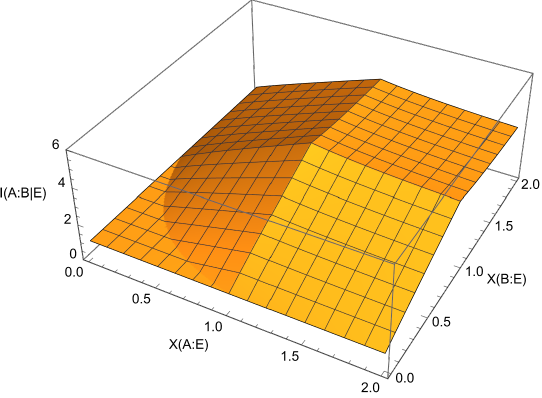}
	\end{minipage}
}
\subfigure[]{
	\begin{minipage}[b]{.47\linewidth}
		\centering
		\includegraphics[scale=0.75]{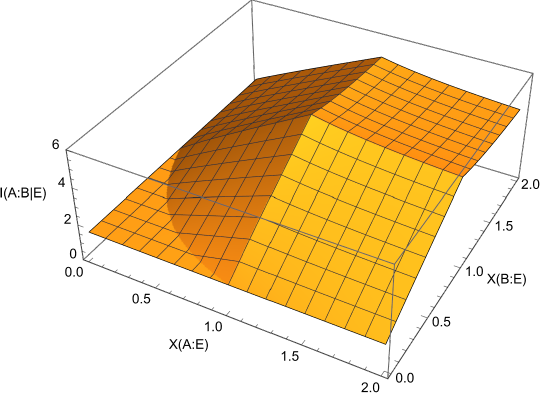}
	\end{minipage}
}
	\caption{\(I(A:B|E)\) as a function of \(X(A:E)\) and \(X(B:E)\) for different values of \(X(A:B)\): (a) \(X(A:B)=0.2\), (b) \(X(A:B)=0.5\), (c) \(X(A:B)=1\), (d) \(X(A:B)=1.5\).
	}\label{FigIABE}
\end{figure}

Then the CMI \eqref{IABE} can be computed.
Figure \ref{FigIABE} shows the value of the fraction in the logarithm of \eqref{IABE} as a function of values of \(X(A:E)\), \(X(B:E)\) at different values of \(X(A:B)\).
Through simple analysis, it could be found that there is indeed an upper bound for \(I(A:B|E)\) when varying the length of $E$, which could also be seen from Figure \ref{FigIABE}. The configuration that maximizes the CMI  has
\begin{equation}\label{maxconf1}	X(A:E)=X(B:E)=1,
\end{equation} 
which does not depend on the value of \(X(A:B)\). 
Note that the condition of a cross ratio being \(1\) is just the critical point for the entanglement phase transition between the connected and disconnected phase of the corresponding MI {as can be seen from equation \eqref{CrossratioDef}}.
Therefore, in order to reach the upper bound of CMI for the conditional region $E$ being one single interval, one should adjust the interval \(E\) to reach the critical points of both the entanglement phase transitions with \(A\) and the one with \(B\) simultaneously.
Note that it is always possible to satisfy both of these conditions, because there are two degrees of freedom of region $E$ (the two coordinates of the end points of region $E$), and those two entanglement phase transition conditions are two equations that uniquely determine them. Intuitively, one can choose the middle point of region $E$ inside the gap, then enlarge $E$ while preserving its middle point. If the middle point is near $A$ and far from $B$, $EW(AE)$ will first connect during the enlarging process and vise versa. As a result, because of the intermediate value theorem, there must exist a fine-tuned middle point of $E$ that makes $EW(AE)$ and $EW(BE)$ connect simultaneously, which is the $E$ that maximizes the CMI.

\subsubsection{\(E\) {being} two intervals}
\begin{figure}[h]
	\centering
\subfigure[]{
	\begin{minipage}[b]{1\linewidth}
		\centering
		\includegraphics[scale=0.45]{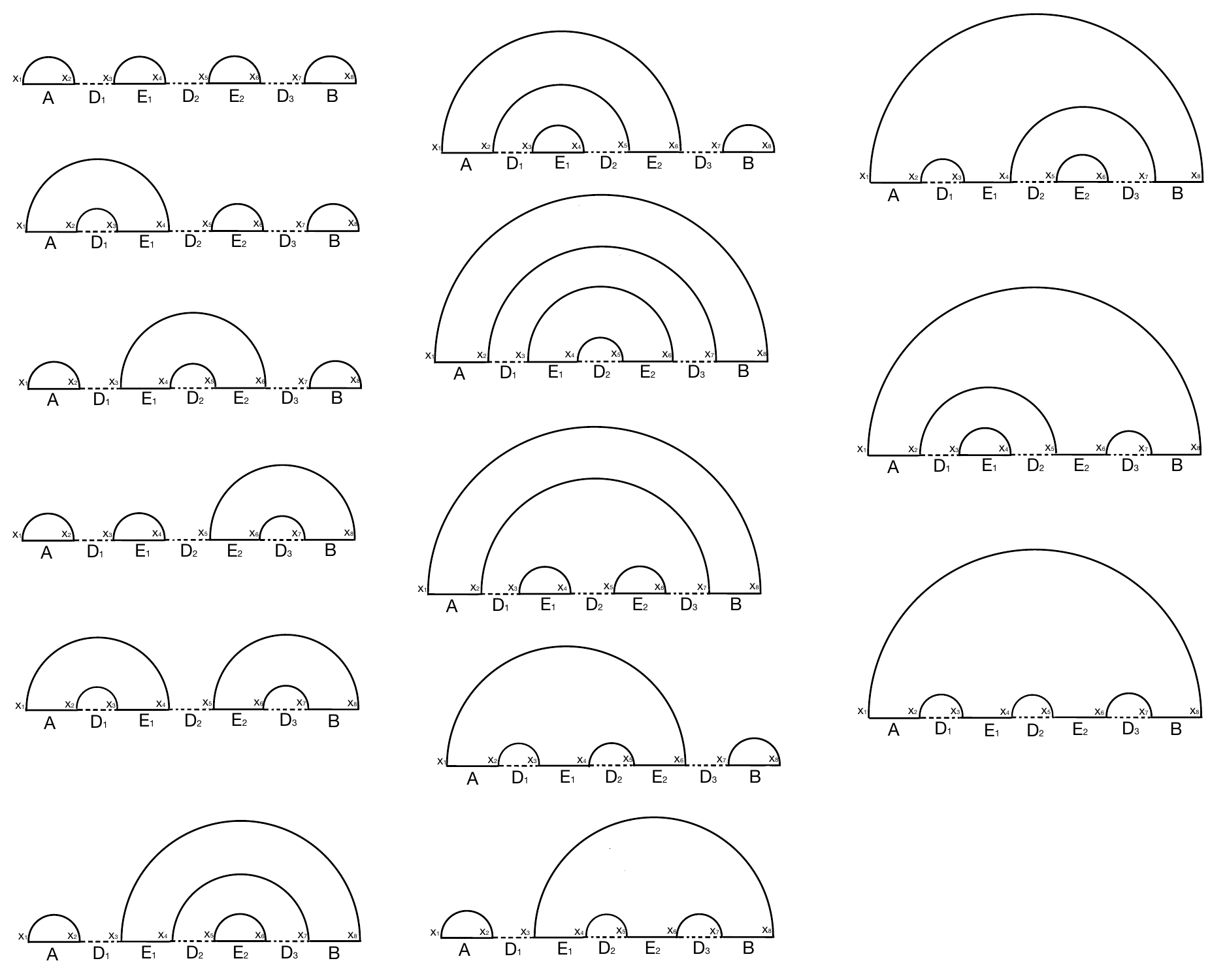}
	\end{minipage}
}\\
\subfigure[]{
	\begin{minipage}[b]{1\linewidth}
		\centering
		\includegraphics[scale=0.35]{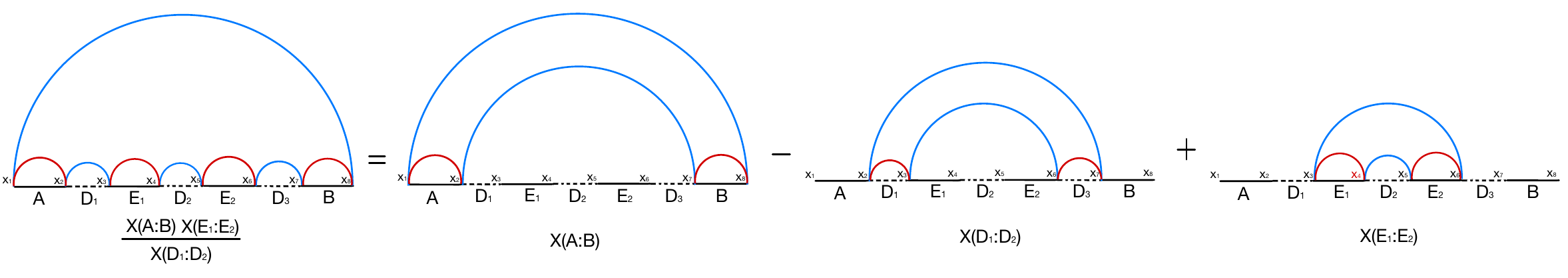}
	\end{minipage}
}
	\caption{(a) \(14\) possible configurations for the calculation of \(S_{AE_1E_2B}\). 
		(b) The totally connected phase for \(J_4(A E_1 E_2 B)\) which can be decomposed into \(3\) diagrams that each corresponds to the cross ratio \(X(A:B)\), \(X(D_1:D_3)\) and \(X(E_1:E_2)\), respectively. 
		Thus \(J_4(A E_1 E_2 B)\) is determined by the fraction \((X(A:B) X(D_1:D_3))/X(E_1:E_2) \) because of the logarithm.
		The red semi-circles signify the geodesic lengths with positive signs in the calculation of \(J_3(ABE)\) while the blue semi-circles denote the geodesic lengths with negative signs. 
	}\label{SABEE}
\end{figure}

\noindent Next, we analyze the case of $E=E_1\cup E_2$ being two disjoint intervals \(E_1\) and \(E_2\) between \(A\) and \(B\).
The CMI can be rewritten analogously as in \eqref{CMIr}
\begin{equation}\label{CMIr2}
\begin{aligned}
	I(A:B|E)=&S_{AE_1E_2}+S_{BE_1E_2}-S_{E_1E_2}-S_{AE_1E_2B}\\
	=&\left(S_{AE_1E_2}-S_{A}-S_{E_1E_2}\right)+\left(S_{BE_1E_2}-S_{B}-S_{E_1E_2}\right)+\left(S_{A}+S_{B}+S_{E_1E_2}\right.\\
	&\left.-S_{AE_1E_2B}\right)\\
	=&2\left(S_{E_1}+S_{E_2}-S_{E_1E_2}\right)-\left(S_{E_1}+S_{E_2}+S_{A}-S_{AE_1E_2}\right)-\left(S_{E_1}+S_{E_2}+S_{B}\right.\\
	&\left.-S_{BE_1E_2}\right)-\left(S_{E_1}+S_{E_2}-S_{E_1E_2}\right)+\left(S_{A}+S_{B}+S_{E_1}+S_{E_2}-S_{AE_1E_2B}\right)\\
	=&I(E_1:E_2)-J_3(AE_1E_2)-J_3(BE_1E_2)+J_4(AE_1E_2B),
\end{aligned}
\end{equation}
where \(J_4(AE_1E_2B):=S_{A}+S_{B}+S_{E_1}+S_{E_2}-S_{AE_1E_2B}\).
The geodesics responsible for the calculation of \(S_{AE_1E_2B}\) have fourteen possible different configurations which are depicted in Figure \ref{SABEE}(a).
\(S_{AE_1E_2B}\) is proportional to the minimal geodesic length among them. 
Similar to the case of mutual information, one can show that for each of the fourteen configurations, \(J_4(AE_1E_2B)\) can be represented by a function of cross ratios, which is illustrated in Figure \ref{SABEE}(b) in the example of the totally connected phase. The result for $J_4$ is

\vspace{-0.5cm}
\begin{scriptsize}
	\begin{equation}
	J_4(A E_1 E_2 B)=\frac{c}{3}\log\text{max}\left\{1, X_1, X_2, X_3, X_5, X_6, X_8, X_1 X_5, X_2 X_8, X_1 X_{10}, X_5 X_{11}, X_3/X_4, X_6/X_7, (X_2 X_8)/X_9 \right\}.
	\end{equation}
\end{scriptsize}
The CMI \eqref{CMIr2} is then given by
\begin{tiny}
	\begin{equation}\label{IABE2}
		\begin{aligned}
			I(A:B|E)&=\frac{c}{3}\log\frac{\text{max}\left\{1, X_2 \right\}\text{max}\left\{1, X_1, X_2, X_3, X_5, X_6, X_8, X_1 X_5, X_2 X_8, X_1 X_{10}, X_5 X_{11}, X_3/X_4, X_6/X_7, (X_2 X_8)/X_9 \right\}}{\text{max}\left\{1, X_1, X_2, X_3, X_3/X_4 \right\}\text{max}\left\{1, X_2, X_5, X_6, X_6/X_7 \right\}}.
		\end{aligned}
	\end{equation}
\end{tiny}
For conciseness, the cross ratios are denoted as \(X_1=X(A:E_1)\), \(X_2=X(E_1:E_2)\), \(X_3=X(A:E_2)\), \(X_4=X(D_1:D_2)\), \(X_5=X(E_2:B)\), \(X_6=X(E_1:B)\), \(X_7=X(D_2:D_3)\), \(X_8=X(A:B)\), \(X_9=X(D_1:D_3)\), \(X_{10}=X(AD_1E_1:B)\), \(X_{11}=X(A:E_2D_3B)\).

We have four intervals and eight endpoints in the two interval case.
The number of independent cross ratios is therefore five.
A convenient choice of the free variables is \(X_1\), \(X_2\), \(X_5\), \(Y_1\) and \(Y_2\), where \(Y_1:=X(E_2:A D_1 E_1)\) and \(Y_2:=X(E_1:E_2 D_3 B)\).
All the other cross ratios are functions of them:
\begin{equation}\label{XX}
	\begin{aligned}
		&X_3=\frac{X_1 Y_1(Y_1-X_2)}{X_2+U_{1}}, \quad X_4=\frac{Y_1-X_2}{X_2+U_{1}},\\
		&X_6=\frac{X_5 Y_2(Y_2-X_2)}{X_2+U_{2}}, \quad X_7=\frac{Y_2-X_2}{X_2+U_{2}},\\
		&X_8=\frac{X_1 X_5 Y_1 Y_2(1+X_2)(Y_1-X_2)(Y_2-X_2)}{U_{3}(X_2^2+(U_{1}+U_{2})X_2+Y_1 Y_2(X_1 X_2 X_5+X_1 X_5-X_2))},\\
		&X_9=\frac{X_2 (X_2-Y_1)(X_2-Y_2)}{X_2^2+(U_{1}+U_{2})X_2+Y_1 Y_2(X_1 X_2 X_5+X_1 X_5-X_2)},\\
		&X_{10}=\frac{X_5 Y_1 Y_2 (1+X_2)(Y_2-X_2)}{(X_2 + U_2) U_3}, \quad X_{11}=\frac{X_1 Y_1 Y_2 (1+X_2)(Y_2-X_2)}{(X_2 + U_2) U_3},
	\end{aligned}
\end{equation}
where \(U_{1}=(X_1+X_2+X_1 X_2)Y_1\), \(U_{2}=(X_5+X_2+X_5 X_2)Y_2\), \(U_{3}=X_2(1+Y_1+Y_2)-Y_1 Y_2\).
Considering the fact that \(X_8\) is fixed, we still need to reverse the fifth equation in \eqref{XX} to obtain \(X_2\) in terms of \(X_8=X(A:B)\) and substitute the result into all the other cross ratios.
After the substitution, there are four free variables \(X_1\), \(X_5\), \(Y_1\) and \(Y_2\).
One could further reduce the number of variables by making some assumptions according to the symmetries.
Note that by using conformal transformations, it is always possible to let \(A\) and \(B\) be mirror symmetric with respect to the vertical axis that passes through the middle point of the gap region between $A$ and $B$.
Thus, it is natural to speculate the configuration that maximizes the CMI also has the mirror symmetry, in other words we can assume \(X_1=X_5=X\), and \(Y_1=Y_2=Y\), which will also be confirmed by direct calculations.

\begin{figure}[H]
\centering
\subfigure[]{
	\begin{minipage}[b]{.47\linewidth}
		\centering
		\includegraphics[scale=0.75]{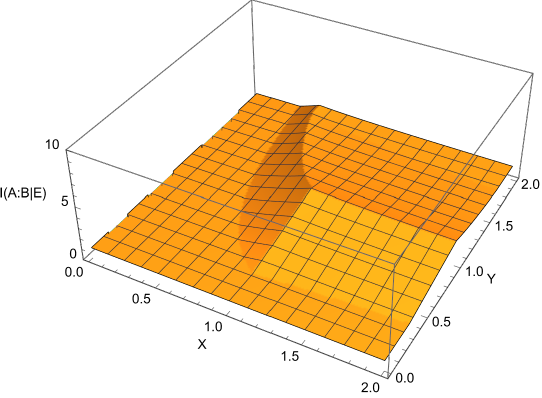}
	\end{minipage}
}
\subfigure[]{
	\begin{minipage}[b]{.47\linewidth}
		\centering
		\includegraphics[scale=0.75]{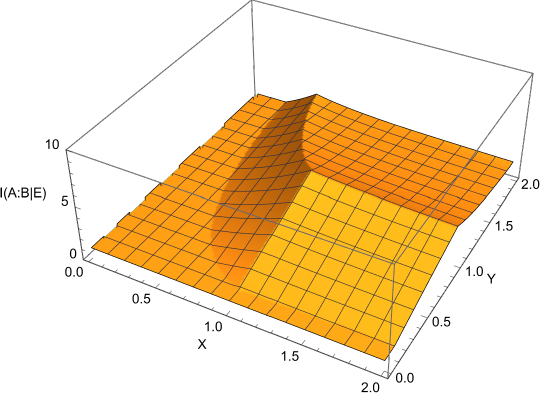}
	\end{minipage}
}\\
\subfigure[]{
	\begin{minipage}[b]{.47\linewidth}
		\centering
		\includegraphics[scale=0.75]{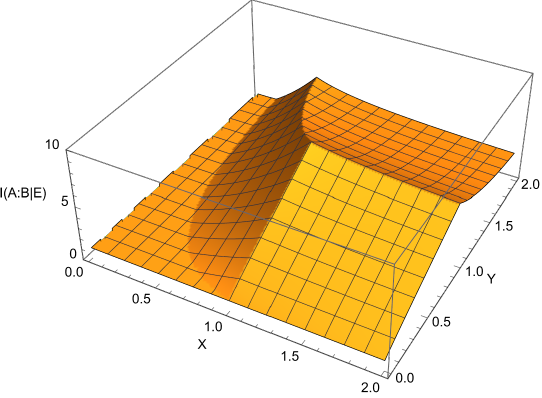}
	\end{minipage}
}
\subfigure[]{
	\begin{minipage}[b]{.47\linewidth}
		\centering
		\includegraphics[scale=0.75]{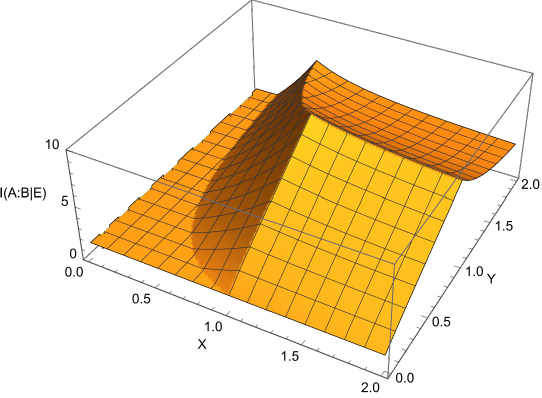}
	\end{minipage}
}
\caption{\(I(A:B|E)\) as a function of \(X=X(A:E_1)=X(B:E_2)\) and \(Y=X(E_2:AD_1E_1)=X(E_1:E_2D_3B)\) with different values of \(X_8=X(A:B)\): (a) \(X_8=0.2\), (b) \(X_8=0.5\), (c) \(X_8=1\).}\label{FigIABEE}
\end{figure}

After all the assumptions and manipulations above, the fraction in the logarithm in \eqref{IABE2} becomes a function of \(X\), \(Y\) and \(X_8\), and is shown in Figure \ref{FigIABEE}.
We find the configuration that maximizes the CMI is \(X=Y=1\) and is also independent of \(X_8=X(A:B)\).
Therefore, in order to reach the maximum, \(E_1\) and \(E_2\) should reach the critical points of four sets of phase transitions between the connected and disconnected phases of 
(1) \(E_1\) and \(A\); 
(2) \(E_1\) and \(E_2 D_3 B \);
(3) \(E_2\) and \(B\);
(4) \(E_2\) and \(A D_1 E_1 \).

    \subsection{Multi-entanglement phase transition rules}\label{sec2.2}
  \noindent From the calculations above, one can see that in the case where region $E$ is one/two intervals inside the gap region between regions $A$ and $B$, CMI reaches the upper bound when two/four entanglement phase transitions of RT surfaces occur simultaneously. Those phase transition conditions uniquely determine the position of those intervals  of $E$. This leads us to propose the following multi-entanglement phase transition rule in AdS$_3$/CFT$_2$.

    \textit{Multi-entanglement phase transition rule: given fixed boundary subregions $A$ and $B$, the conditional mutual information $I(A:B|E)$ for region $E$ being $m$ disjoint boundary intervals with $m$ fixed reaches its upper bound at the critical point where $2m$ entanglement phase transitions happen simultaneously. These $2m$ phase transition conditions uniquely determine the position of the $2m$ endpoints of intervals in $E$, i.e. the configuration of $E$ at maximum CMI.}

This multi-entanglement phase transition rule states that to find the $m$-interval region E that maximizes CMI is equivalent to finding the $2m$ multi-entanglement phase transition critical configuration. This is a generalization of what we have found in Section \ref{sec2.1} for $m=1,2$ to general $m$, and we will see that this could be proved for general $m$. Before formally proving it via an iterative method, we use an intuitive way to explain the reason why the CMI could never reach its maximum value when we move one of the $2m$ endpoints in $E$ before reaching a critical point where an entanglement phase transition of an entanglement wedge happens. 
\begin{figure}[H]
\centering
     \includegraphics[width=10.5cm]{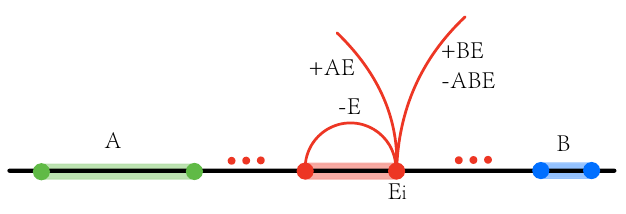}
 \caption{RT surfaces emitted from one of the $2m$ endpoints \( E_i \), marked by red arcs. The subregion that the geodesic is responsible for in the calculation of the entanglement entropy is labeled beside the geodesic and their corresponding signs in the definition of CMI (\ref{CMIformula}) are shown. Boundary subregions \( A \), \( B \), and \( E \) are shaded in green, blue, and red, respectively.
}\label{pertubei}
\end{figure}
Figure \ref{pertubei} shows the RT surfaces of $AE$, $BE$, $E$ and $ABE$ in a certain choice of configuration. Without loss of generality, for an arbitrarily chosen endpoint \( E_i \) in $E$, at most three distinct geodesics are emitted from it in these four RT surfaces. Furthermore, there are two positive signs and two negative signs in formula (\ref{CMIformula}). As a result, one can mark those three geodesics as positive and negative ones. The RT surfaces are then the left curve (labeled by \( +AE \)), the right curve (labeled by \( +BE \) and \( -ABE \)), and the one that connects to itself (labeled by \( -E \))\footnote{{This choice of labels corresponds to a specific phase for demonstration purposes, and should vary between phases.}}. By slightly modifying the position of \( E_i \) before triggering any entanglement phase transition, only the lengths of those three geodesics change, which in turn changes the value of CMI.  Through simple analysis, one can find that those four terms will become the difference of the length of at most two geodesics. As a result, as we move \( E_i \) in one direction, CMI increases or decreases monotonically (or remains constant) as long as no entanglement phase transition happens. Due to this monogamy within one phase, the maximum of CMI will only occur at a phase transition critical point as we vary \( E_i \) while leaving other endpoints unchanged. As a result, CMI must reach its maximum at critical points of entanglement phase transitions; \ie, at the local maximum of CMI, a phase transition condition of RT surfaces is associated with the configuration of \( E_i \) while other endpoints are left fixed.
 %One exception is when the lengths of those three geodesics cancel out precisely, which makes CMI constant under \( E_i \) perturbation. In this case, one can simply move \( E_i \) until a phase transition of RT surfaces occurs, so it does not violate our desired result that calculating CMI when multi-phase transitions occur is sufficient.

    \subsubsection*{An iterative method and proof of the multi-entanglement phase transition rule}

    Given a fixed number \( m \) of intervals in $E$, to reach the maximum of CMI, one can use an iterative method to modify the endpoints of each interval of \( E_i \) in $E$ repeatedly. Each modification maximizes the CMI locally, given the condition that the other endpoints are fixed. This is named the ``hill-climbing algorithm'' in programming. One can take CMI, a function of \( 2m \) variables, as a mountain (hypersurface) in \( \mathbf{R}^{2m+1} \). To climb this mountain in order to reach its peak, one can simply pick one of the $2m$ variables in order, i.e. one coordinate in the \( \mathbf{R}^{2m} \) space, and move along this coordinate to reach its maximum (the ridge). Different orders in picking the coordinates each time might result in the program finally reaching different peaks of this mountain, as we are going to see in the next section.

We choose the simple case, where \( E \) is chosen as a region of \( m \) intervals all living inside one gap region between \( A \) and \( B \) as an example. We find that the ``mountain'' of CMI has only one peak in this simple case. The procedures of this program are as follows.
\begin{itemize}
    \item 1. Set initial values of \( 2m \) endpoints \( E_{iL} \) and \( E_{iR} \), \( i \) from \( 1 \) to \( m \), which represent the left and right endpoints of the \( i \)th interval. For convenience, we can simply choose them as the points dividing the gap between \( A \) and \( B \) into \( 2m+1 \) equal length sections.
    \item 2. Vary the coordinate of \( E_{iL} \) from \( i=1 \) to \( m \) in order. Each new $E_{iL}$ maximizes the CMI locally as we mentioned above.
    \item 3. Vary the coordinate of \( E_{iR} \) from \( i=m \) to \( 1 \) in order. Each new $E_{iR}$ maximizes the CMI locally.
    \item 4. Repeat procedure 2 and 3 over and over again.
\end{itemize}

The order of moving endpoints in steps 2 and 3 in this procedure guarantees the convergence of this iteration in practice. According to the arguments above, each step in 2 or 3 leads to a critical point of the phase transition of a certain RT surface. As a result, the iterative method will lead the construction of those \( m \) intervals to the ``peak" of the mountain where \( 2m \) phase transitions occur simultaneously, \ie, CMI reaches its upper bound at the multi-entanglement phase transition critical point of the RT surfaces. This result naturally holds in practice, so the multi-entanglement phase transition rule is proved.

\section{Varying the number of intervals in E: divergence behavior of CMI} \label{sec3}
\noindent Utilizing the multi-entanglement phase transition rule proposed in Section \ref{sec2}, we could obtain the region $E$ with fixed number of intervals $m$ that maximizes the CMI. For fixed $m$, there are multiple but finite many configurations which reach the critical point of $2m$ phase transitions. Only one among them has the global maximum CMI, so we need to compare all of them to find that configuration. In this section, to perform this procedure for general $m$ we introduce a multi-entanglement phase transition (MPT) diagram which could depict the phase transition conditions in a concise way at the multi-entanglement phase transition critical point. We then exhibit the results for the maximum value of the CMI for general $m$ from the iterative method and analyze the relationship between the number of intervals \( m \) and the maximum value of CMI both analytically and numerically. %In subsection 3.3, we introduce another method to analyze the upper bound of CMI in more complex cases such as the wormhole geometry, and conclude a universal result.\sun{why not move to sec 4?}

    \subsection{MPT diagram}\label{sec3.1}
    \noindent At the beginning of Section \ref{sec2.1}, we obtained the phase transition {configurations} for the subregion $E$ at the critical point when CMI reaches its maximum for \( m=1 \) and \( m=2 \). Those conditions are
\begin{itemize}
    \item \( m = 1 \):  an entanglement phase transition for region $AE$ from being disconnected to being connected and an entanglement phase transition for region $BE$ occur simultaneously. We denote this condition as \( A/E \) and \( B/E \).
    \item \( m = 2 \): phase transitions \( A/E_1 \), \( B/E_2 \), \( AE_1/E_2\footnote{In Section \ref{sec2}, we write the entanglement phase transition condition as $AD_1E_1/E_2$, while we omit the gap region $D_1$ in between region $A$ and $E_1$ here. Technically, this entanglement phase transition condition only constrains the four coordinates of the endpoints $\partial AD_1E_1$ and $\partial E_2$, so as long as the entanglement wedge of $AE_1$ is connected, the exact position of $\partial D_1$ is irrelevant in condition $AD_1E_1/E_2$. Therefore, the two conditions $AD_1E_1/E_2$ and $AE_1/E_2$ are in fact equivalent and as a result, for convenience, we can simply write $AE_1/E_2$ instead.} \), and \( BE_2/E_1 \) occur simultaneously.
\end{itemize}
We introduce a simple diagram to illustrate these phase transition configurations, which we name the multi-entanglement phase transition diagram. 
\begin{figure}[H]
\centering
     \includegraphics[width=13cm]{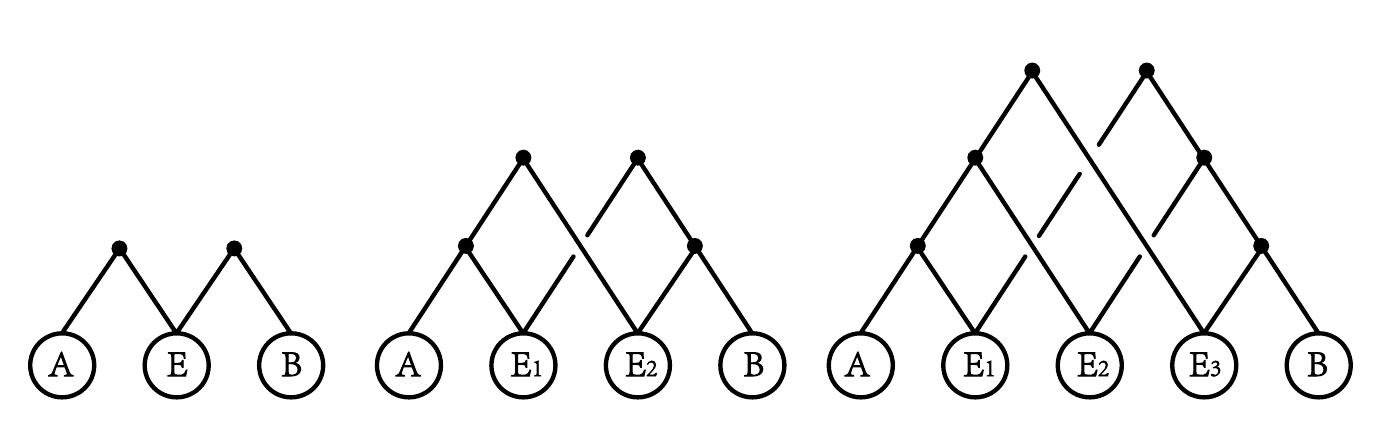}
 \caption{Multi-entanglement phase transition diagrams for illustration of the critical phase transition conditions in the maximum CMI configuration for $E$ being \( m = 1, 2, 3 \) intervals. The whole region of E stays inside the gap region between $A$ and $B$.}
\label{MPT1}
\end{figure}
As shown in Figure \ref{MPT1}, there are three elements in an MPT diagram: circles, legs, and dots. A circle represents a single interval with its name labeled inside it. A dot represents a phase transition critical point which imposes a constraint condition between two regions associated with the two legs. For each leg, the region associated with this leg is the combination of all regions which connect to this leg from downward. For diagrams with more than one level, the dot on the top level can have a leg connecting to another dot below it, which indicates a phase transition between a single interval and the combination of some intervals. For example, the left top dot in the middle figure of Figure \ref{MPT1} indicates \( AE_1/E_2 \).

There are some fundamental rules for drawing MPT diagrams. First of all, there exist \( 2m \) dots in total, and each interval of region \( E \) can trace upward by legs to at least two dots. This rule ensures that the phase transition conditions uniquely determine the $2$ endpoints of each intervals. Second, any dot can connect directly to at most one dot below it. If it connects to two dots, the phase transition conditions would overdetermine the system. For example, if we have \( A/E_1 \) and \( B/E_1 \), the entanglement wedge of \( ABE_1 \) is fully connected due to the strong subadditivity, so a condition like \( AE_1/B \) cannot be satisfied. Thirdly, any dot must connect with either region \( A \) or \( B \). Otherwise, this dot represents a phase transition condition purely inside region \( E \), like \( E_1/E_2 \). As we will see later, those diagrams, though they exist, are not the ones that might reach the maximum of CMI.

From Figure \ref{MPT1}, it is easy to generalize and draw the MPT diagram with larger \( m \) intervals, where \( m \) is a finite integer. However, when we consider placing the intervals of region \( E \) in both of the gaps between regions \( A \) and \( B \), the MPT diagrams cannot be uniquely determined for \( m \geq 2 \). We have to find other ways to constrain the diagrams and perform specific calculations to find the one with maximum CMI.

    \subsubsection*{MPT diagram for E living in two gap regions and the disconnectivity  condition}
    Now we analyze the case where the $m$ intervals of region $E$ could live in both the two gap regions between region $A$ and $B$ and outside $A$ and $B$. 
       {We take the region $A$ to be the interval $[A_L,A_R]$, and  region $B$ the interval $[B_L,B_R]$, with $A_L<A_R<B_L<B_R$ without loss of generality. }
     %Without loss of generality, given region $A$ the interval $[A_L,A_R]$, region $B$ the interval $[B_L,B_R]$, with $A_L<A_R<B_L<B_R$. 
     The two gaps between $A$ and $B$ and outside them are $[A_R,B_L]$ and $(-\infty,A_L]\cup[B_R,\infty)$, separately. Note that one of the intervals of region $E$ could be chosen as $(-\infty,E_{iR}]\cup[E_{iL},\infty)$.
     %Note that due to special conformal transformation, one can always make region $A$ the complement of a single region $(-\infty,m]\bigcup[n,\infty)$, and region $B$ is $[p,q]$ with $m<p<q<n$. In which case, two gaps between $A$ and $B$ are $(m,p)$ and $(q,n)$.
    
\begin{figure}[H]
\centering
     \includegraphics[width=15cm]{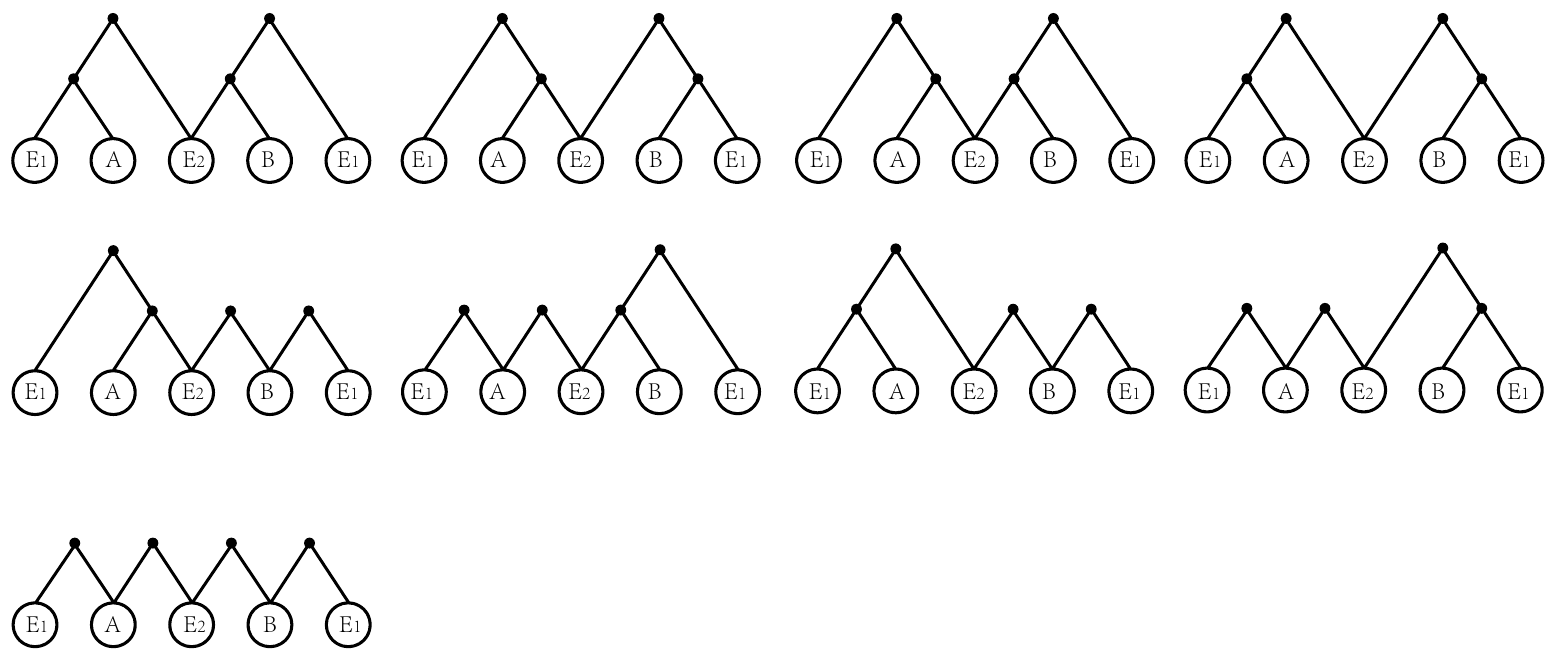}
 \caption{$m=2$ MPT diagrams for the case of $E$ living in both gap regions inside and outside of $A$ and $B$.}
\label{MPT2}
\end{figure}

{In this case, there are more than one consistent MPT diagrams for each $m>1$ and we need to first write out all possible MPT diagrams and then compare the maximum values in the diagrams to pick the one that maximizes the CMI.} In fact, there are nine MPT diagrams in the simplest \( m = 2 \) case, as shown in Figure \ref{MPT2}. Utilizing a special conformal transformation, one can always make the length of region \( A \) equal to region \( B \). As a result, the configuration has a reflection symmetry under \( x \to -x \), and diagrams that can transform to each other under the reflection symmetry have the same CMI values. Another symmetry of the MPT diagram is the special conformal transformation \( x \to 1/x \), which exchanges the positions of \( E_1 \) and \( E_2 \). Taking these two symmetries into consideration, the nine diagrams in Figure \ref{MPT2} have only four distinct local maximum values. Specifically, diagrams (1.1) and (1.2) share the same value, written as \( 2\log(I_{(1)}) \); diagrams (1.3) and (1.4) share the same value, written as \( 2\log(I_{(2)}) \); diagrams (2.1), (2.2), (2.3), and (2.4) share the same value, written as \( 2\log(I_{(3)}) \); diagram (3.1) has a distinct value, written as \( 2\log(I_{(4)}) \).

\begin{figure}[H]
\centering
     \includegraphics[width=12cm]{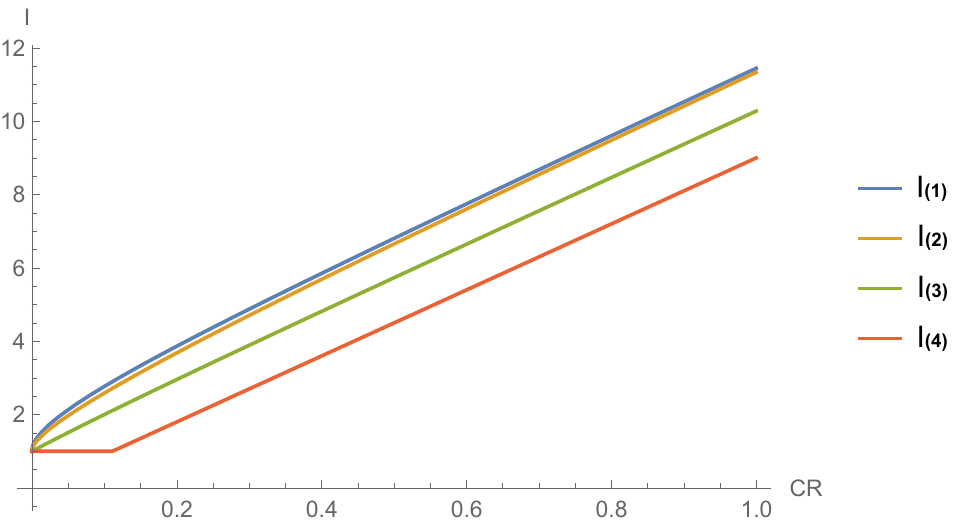}
 \caption{$I_{(1)}$, $I_{(2)}$, $I_{(3)}$, and $I_{(4)}$ defined from the seven MPT diagrams for the $m=2$ two-gap case as functions of the cross-ratio between \( A \) and \( B \).}
\label{I1234}
\end{figure}

Due to the conformal symmetry, \( I_{(1)} \), \( I_{(2)} \), \( I_{(3)} \), and \( I_{(4)} \) are functions only of the cross-ratio between regions \( A \) and \( B \). They could be directly calculated and the graph of these functions is plotted in Figure \ref{I1234}. One can find that though $I_{(1)}$ and $I_{(2)}$ are very close to each other, $I_{(1)}$ is always the maximum among the four values, while $I_{(3)}$ and $I_{(4)}$ are much smaller than the former two. 

{We could continue to perform this procedure for $m=3$, where one interval of $E$ stays in one gap region while the other two intervals stay in the other gap region in between or outside $A$ and $B$.}  Though the total number of consistent MPT diagrams is finite, it is so large—about fifty—that drawing them one by one would be a formidable task. As a result, we have to find stricter constraints for MPT diagrams. These new constraints concern the connectivity of entanglement wedges in the configuration that maximizes the CMI as follows.

%{\bf Connectivity conditions for the maximal CMI configuration \sun{list as 3 items?}}
\begin{itemize}
    \item Constraint I: the entanglement wedge of \( ABE \) must be totally connected.
    \item Constraint II: the entanglement wedge of \( E \) inside a gap must be totally disconnected.
    \item Constraint III, \textbf{disconnectivity condition}: the mutual information between region $A(B)$ and $E$ vanishes.
    \begin{equation}\label{disconnect}
        I(A:E) = I(B:E) = 0.
    \end{equation} 
\end{itemize}

We explain these constraints one by one. In the MPT diagram with maximum CMI, if there exists a single interval \( E_i \) where, in \( ABE \)'s entanglement wedge, \( E_i \) is disconnected—which means \( I(ABE\backslash E_i : E_i) = 0 \)—one can find
\begin{equation}\label{ABEcon}
\begin{aligned}
    I(A:B|E) &= S_{AE} + S_{BE} - S_E - S_{ABE} \\
             &= S_{AE\backslash E_i} + S_{E_i} + S_{BE\backslash E_i} + S_{E_i} - S_{E\backslash E_i} - S_{E_i} - S_{ABE\backslash E_i} - S_{E_i} \\
             &= I(A:B|E\backslash E_i),
\end{aligned}
\end{equation}
i.e., the CMI will be the same if we delete \( E_i \) from \( E \). We can do this procedure over and over again until all disconnect $E_i$ are deleted so that constraint 1 is satisfied. Therefore, we only need to consider configurations that satisfy constraint I.

If the entanglement wedge of region \( E \) is partially connected—let us say, the entanglement wedges of \( E_i \) and \( E_{i+1} \) are connected {in $E$'s entanglement wedge,}—we denote the gap between \( E_i \) and \( E_{i+1} \) as \( G_{i,i+1} \). Then we can ``merge'' \( E_i \), \( G_{i,i+1} \), and \( E_{i+1} \) into a single interval to replace the former \( E_i \) and \( E_{i+1} \). Denoting the new region \( E \) as \( E_{\text{new}} \), we have
\begin{equation}
\begin{aligned}
    I(A:B|E)&=S_{AE}+S_{BE}-S_E-S_{ABE}    \\
            &=S_{AE_{new}}+S_{G_{i,i+1}}+S_{BE_{new}}+S_{G_{i,i+1}}-S_{E_{new}}-S_{G_{i,i+1}}-S_{ABE_{new}}-S_{G_{i,i+1}}\\
            &=I(A:B|E_{new}),
\end{aligned}
\end{equation}
i.e., the CMI will be the same if we {merge} \( E_i \), \( G_{i,i+1} \), and \( E_{i+1} \) together. We can do this procedure over and over again until all connect $E_i$ are merged together so that constraint II is satisfied. Therefore, we only need to consider configurations that satisfy constraint II.

The third constraint, \ie, the disconnectivity condition, which states the vanishing of mutual information \( I(A:E) \) and \( I(B:E) \) when CMI reaches its upper bound, is a very powerful constraint. We will revisit it in Section \ref{sec4} as a core statement. For now, we can intuitively understand this statement as follows. In Section \ref{sec2.2}, when we try to prove the multi-entanglement phase transition rule, we only need to {analyze} the phase transition critical conditions for RT surfaces of regions that appear in formula \ref{CMIformula}, \eg, \( AE \), \( BE \), \( E \), and \( ABE \). The entanglement wedge of region \( AE \) will not be affected by the phase transition of \( A/E_i \);\footnote{Note that we are saying that the connectivity of $A$ and $E_i$ in the entanglement wedge of $AE_i$ does not matter, while the connectivity of $A$ and $E_i$ in the entanglement wedge of the whole $AE$ still matters.} only when the whole entanglement wedge \( AE \) reaches its phase transition critical point could CMI be affected. In other words, MPT diagrams with conditions like \( A/E_i \) and \( A/E_j \) will not be the diagrams with maximum CMI. Only conditions like \( AE_{i_1}E_{i_2}.../E_{i_m} \) are valid. In Figure \ref{MPT2}, only diagrams on the first row satisfy this disconnectivity condition. In practice, one can indeed find that \( I_1 \) and \( I_2 \), which are much larger than \( I_3 \) and \( I_4 \), are the only two rightful candidates.

\begin{figure}[H]
\centering
     \includegraphics[width=15cm]{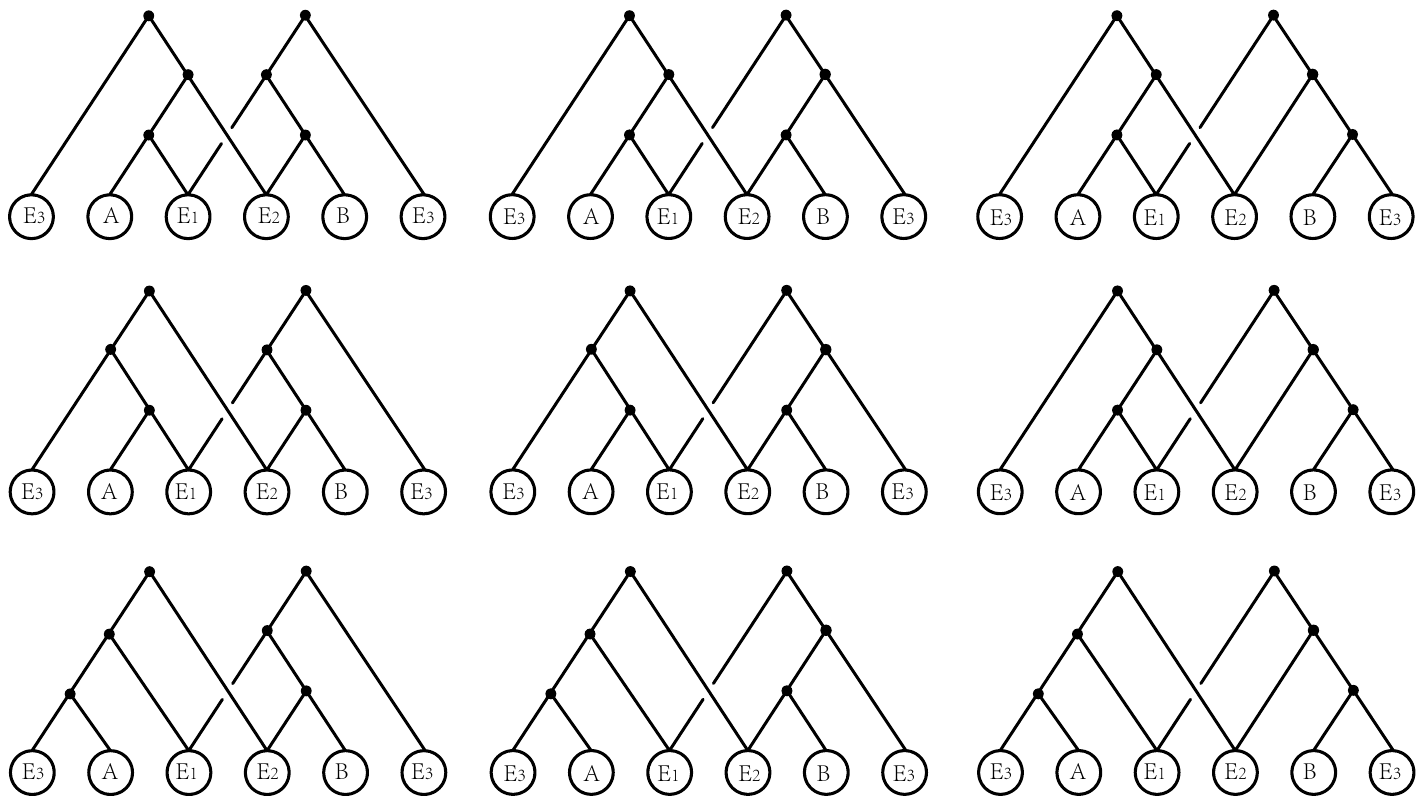}
 \caption{$m=3$ MPT diagrams for $E$ living in two gap regions under the constraint of disconnection conditions. }
\label{MPT3}
\end{figure}

With these three additional constraints taken into consideration, the number of MPT diagrams is greatly reduced.
For example, when \( m = 3 \) {with $E$ split into one and two intervals each living in one of the gap regions}, let \( E_1 \) and \( E_2 \) be two intervals inside the gap region between regions \( A \) and \( B \), while \( E_3 \) is an interval in the other gap region outside $A$ and $B$. There are only nine {valid} %possible orders, each corresponding to an 
MPT diagrams as shown in Figure \ref{MPT3}.
{The corresponding phase transition conditions of each diagram are listed as follows, from left to right and from top to bottom.}
\begin{equation*}
    \begin{aligned}
        &(A_{123};B_{213}): A/E_1, AE_1/E_2, AE_1E_2/E_3; B/E_2, BE_2/E_1, BE_2E_1/E_3\\
        &(A_{123};B_{231}): A/E_1, AE_1/E_2, AE_1E_2/E_3; B/E_2, BE_2/E_3, BE_2E_3/E_1\\                
        &(A_{123};B_{321}): A/E_1, AE_1/E_2, AE_1E_2/E_3; B/E_3, BE_3/E_2, BE_3E_2/E_1\\
        &(A_{132};B_{213}): A/E_1, AE_1/E_3, AE_1E_3/E_2; B/E_2, BE_2/E_1, BE_2E_1/E_3\\
        &(A_{132};B_{231}): A/E_1, AE_1/E_3, AE_1E_3/E_2; B/E_2, BE_2/E_3, BE_2E_3/E_1\\                
        &(A_{132};B_{321}): A/E_1, AE_1/E_3, AE_1E_3/E_2; B/E_3, BE_3/E_2, BE_3E_2/E_1\\        
        &(A_{312};B_{213}): A/E_3, AE_3/E_1, AE_3E_1/E_2; B/E_2, BE_2/E_1, BE_2E_1/E_3\\
        &(A_{312};B_{231}): A/E_3, AE_3/E_1, AE_3E_1/E_2; B/E_2, BE_2/E_3, BE_2E_3/E_1\\                
        &(A_{312};B_{321}): A/E_3, AE_3/E_1, AE_3E_1/E_2; B/E_3, BE_3/E_2, BE_3E_2/E_1,\\     
    \end{aligned}
\end{equation*}
%\begin{equation*}
%    \begin{aligned}
%        (A E_1 E_2 E_3, B E_2 E_1 E_3): A\backslash E_1, AE_1\backslash E_2, AE_1E_2\backslash E_3, B\backslash E_2, BE_2\backslash E_1, BE_2E_1\backslash E_3\\
%        (A E_1 E_2 E_3, B E_2 E_3 E_1): A\backslash E_1, AE_1\backslash E_2, AE_1E_2\backslash E_3, B\backslash E_2, BE_2\backslash E_3, BE_2E_3\backslash E_1\\                
%        (A E_1 E_2 E_3, B E_3 E_2 E_1): A\backslash E_1, AE_1\backslash E_2, AE_1E_2\backslash E_3, B\backslash E_3, BE_3\backslash E_2, BE_3E_2\backslash E_1\\
%        (A E_1 E_3 E_2, B E_2 E_1 E_3): A\backslash E_1, AE_1\backslash E_3, AE_1E_3\backslash E_2, B\backslash E_2, BE_2\backslash E_1, BE_2E_1\backslash E_3\\
%        (A E_1 E_3 E_2, B E_2 E_3 E_1): A\backslash E_1, AE_1\backslash E_3, AE_1E_3\backslash E_2, B\backslash E_2, BE_2\backslash E_3, BE_2E_3\backslash E_1\\                
%        (A E_1 E_3 E_2, B E_3 E_2 E_1): A\backslash E_1, AE_1\backslash E_3, AE_1E_3\backslash E_2, B\backslash E_3, BE_3\backslash E_2, BE_3E_2\backslash E_1\\        
%        (A E_3 E_1 E_2, B E_2 E_1 E_3): A\backslash E_3, AE_3\backslash E_1, AE_3E_1\backslash E_2, B\backslash E_2, BE_2\backslash E_1, BE_2E_1\backslash E_3\\
%        (A E_3 E_1 E_2, B E_2 E_3 E_1): A\backslash E_3, AE_3\backslash E_1, AE_3E_1\backslash E_2, B\backslash E_2, BE_2\backslash E_3, BE_2E_3\backslash E_1\\                
%        (A E_3 E_1 E_2, B E_3 E_2 E_1): A\backslash E_3, AE_3\backslash E_1, AE_3E_1\backslash E_2, B\backslash E_3, BE_3\backslash E_2, BE_3E_2\backslash E_1\\     
%    \end{aligned}
%\end{equation*}
{The terms inside the parentheses on the left are acronyms indicating the phase transition configuration written on the right side. The numbers in the lower indexes correspond to the order of phase transition conditions on the right side. Note that for each line, the multi-entanglement phase transitions happen simultaneously and the order in the sequence in each phase transition configuration is only an order in presentation, though it may be connected with the graphic representation of these phase transition configurations. For example, \( A_{123} \) represents the sequence of phase transition conditions of
$A/E_1,  AE_1/E_2,  AE_1 E_2/E_3$, while \( A_{132} \) represents the sequence of phase transition conditions of
$A/E_1, AE_1/E_3, AE_1E_3/E_2.$ Here, it must be easier for \( A \) to have phase transitions with the intervals near it, so the number \( 1 \) must precede the number \( 2 \). Overall, the order of numbers in the index indicates different phase transition configurations.
}
 %from left to right, from top to bottom in order, we have figure \ref{MPT3}. 

\begin{figure}[H]
\centering
     \includegraphics[width=14cm]{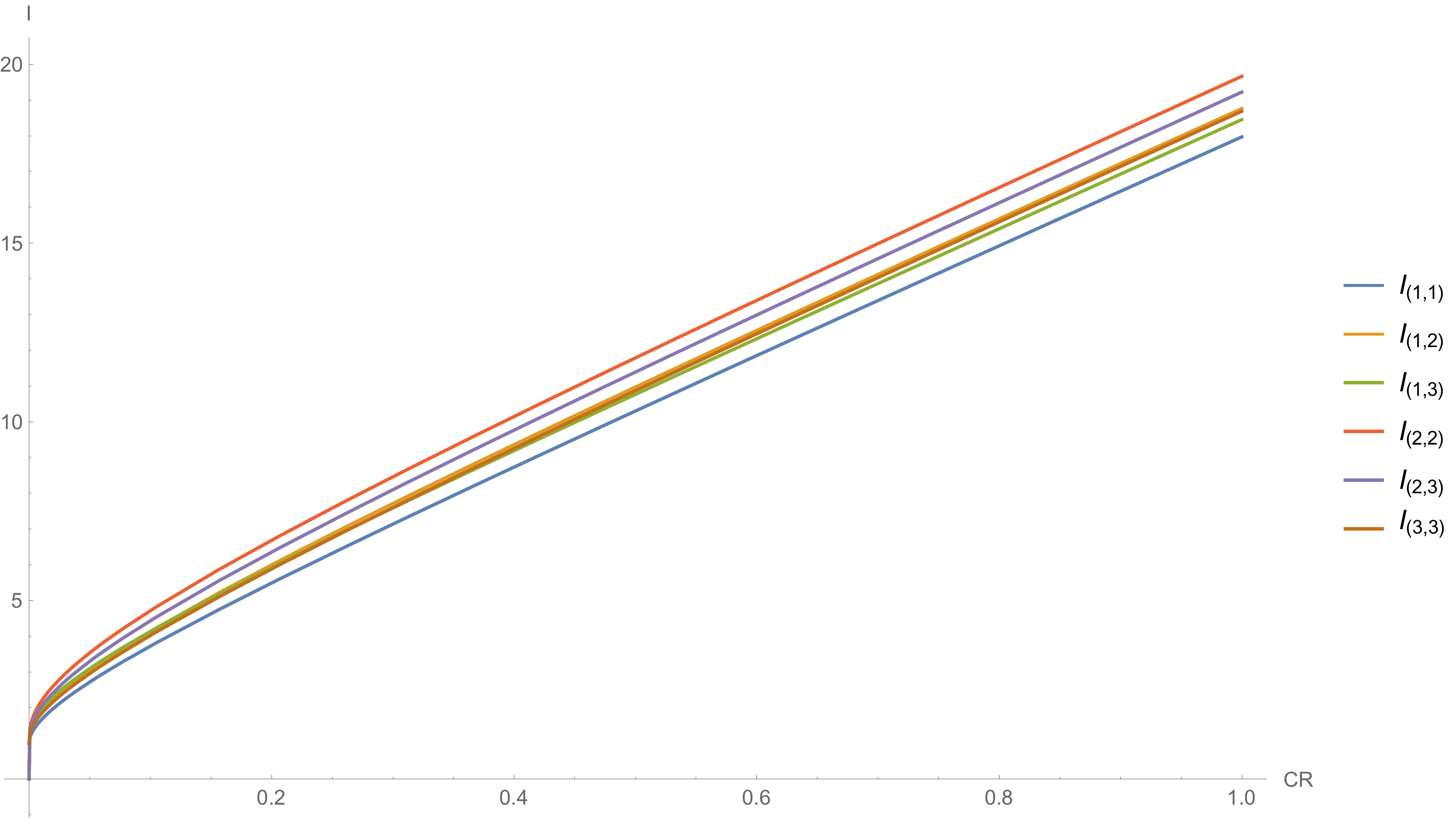}
 \caption{$I_{(p,q)} = \exp(\frac{I(A:B|E)}{2})$ as functions of the cross ratio between $A$ and $B$. The index $(p,q)$ refers to the $(p,q)$th entanglement phase configuration in Figure \ref{MPT3}.}
\label{I123456}
\end{figure}

It is worth noting that due to {the reflection} symmetry, diagram \( (p,q) \) is symmetric to diagram \( (q,p) \) leading to the same value of CMI, making the number of distinct local maximum values of CMI values six. Evaluating the CMI of each diagram as we did before, we plot \( I_{(1,1)} \), \( I_{(1,2)} \), \( I_{(1,3)} \), \( I_{(2,2)} \), \( I_{(2,3)} \), and \( I_{(3,3)} \) as functions of the cross ratio between \( A \) and \( B \) in Figure \ref{I123456}. As we expect, although the values of CMI corresponding to each diagram are very close to each other, there still exists a diagram larger than any other diagrams for any value of cross ratios, which is diagram \( (2,2) \) in Figure \ref{MPT3}.

%In fact, one has \sun{what does has to mean here} to establish an order from \( E_1 \) to \( E_n \), so that the phase transition conditions between region \( A \) and them happen in order. For example, in diagram (1.1)\sun{explain what 1.1} in figure \ref{MPT2}, we have \( A\backslash E_1 \) and \( AE_1\backslash E_2 \), while in diagram (1.2), we have \( A\backslash E_2 \) and \( AE_2\backslash E_1 \)\sun{explain more words}. Generalizing to the \( n > 2 \) case, the diagram could be labeled by two orders of \( E_1, \ldots, E_n \), each representing the order of phase transitions with \( A \) and with \( B \), respectively.\sun{explain more clearly}

\begin{figure}[H]
\centering
     \includegraphics[width=15cm]{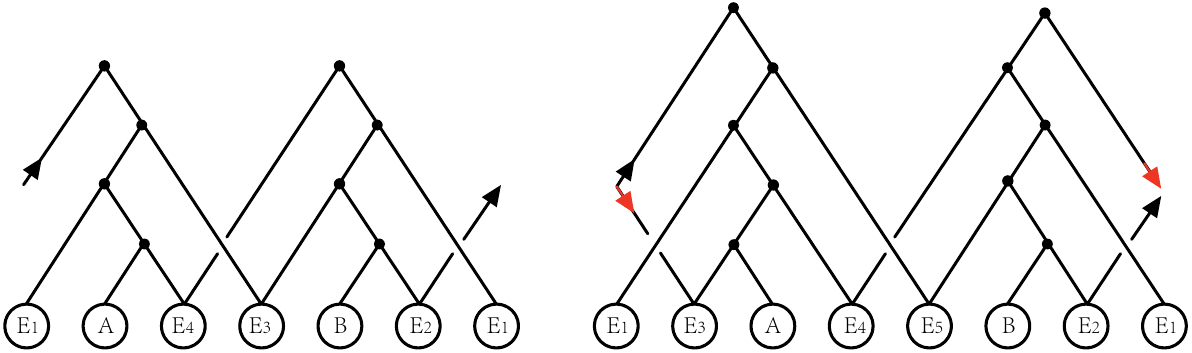}
 \caption{$m=4$ and $m=5$ MPT diagrams for two gap region cases with maximum CMI value. The left and right arrows indicate that the legs with the same color of arrows in each diagram are connected.}
\label{MPT45}
\end{figure}

We can generalize this procedure to the $m = 4$ and $m = 5$ cases in the same vein. {One only have to list all orders corresponding to each MPT diagrams, and compare their CMI to find the maximum one.} Here, we skip the technical details and present the numerical results for $m=4$ in Table \ref{n4}. One can find that this table is symmetric with respect to both the {principle} diagonal and the {auxiliary} diagonal. As a result, the number of distinct values of CMI is twelve. The configuration with the maximum value is labeled $(A_{4132}, B_{2314})$, \ie, the multi-entanglement phase transition configuration of $A/E_4, AE_4/E_1, AE_4E_1/E_3,AE_4E_1E_3/E_2; B/E_2, BE_2/E_3$, $ BE_2E_3/E_1,BE_2E_3E_1/E_4$. Its corresponding MPT graph is depicted in the left figure of Figure \ref{MPT45}. When \( m = 5 \), there are \( 10 \times 10 = 100 \) candidate diagrams, and the right figure in Figure \ref{MPT45} presents the diagram with the largest CMI. 

\begin{table}[H]
    \centering
    \begin{tabular}{c|c|c|c|c|c|c}
         & $A_{1243}$    & $A_{1423}$ & $A_{1432}$ & $A_{4123}$ & $A_{4132}$ & $A_{4312}$\\
    $B_{2134}$ & 12.2212 & 13.0777 & 13.1206 & 13.2314 & 13.3684 & 12.4431 \\
    $B_{2314}$ & 13.0777 & 14.1521 & 14.2872 & 14.2080 & 14.4255 & 13.3684 \\
    $B_{2341}$ & 13.1206 & 14.2872 & 14.2256 & 14.1836 & 14.2080 & 13.2314 \\
    $B_{3214}$ & 13.2314 & 14.2080 & 14.1836 & 14.2256 & 14.2872 & 13.1206 \\ 
    $B_{3241}$ & 13.3684 & 14.4255 & 14.2080 & 14.2872 & 14.1521 & 13.0777 \\
    $B_{3421}$ & 12.4431 & 13.3684 & 13.2314 & 13.1206 & 13.0777 & 12.2212
    \end{tabular}
    \caption{$\exp(\frac{I(A:B|E)}{2})$ for $m = 4$ in the two gap region case. The labels on the top and on the left stipulate the configuration of the phase transitions for $A$ and $B$, respectively. The values correspond to the case where the cross ratio between $A$ and $B$ is set to be $\frac{1}{3}$.}
    \label{n4}
\end{table}

After analyzing diagrams with maximum CMI in different \( m \), we find that the MPT diagram which reaches the maximum value has two features as follows.

\begin{itemize}
    \item {The dots are ``zigzag'' arranged.}
    \item {The order of the phase transition between \( E_i \) and \( A \) is the exact opposite of the order of the phase transition between \( E_i \) and \( B \).} {For example, if we have $A_{ijkl}$, $B_{lkji}$ has the exact opposite order}.
\end{itemize}

The first feature means that in the phase transition configuration, $A$ or $B$ has phase transitions with those subregions in $E$ that each transition includes one new interval $E_i$ into the subregion from the left or right side respectively. Being ``zigzag" refers to the pattern where if one new interval in $E$ is added from the left side in one phase transition condition, the subsequent subregion in the phase transition condition would include the nearest new interval in $E$ from the right side, and vice versa. For example, in the left figure of Figure \ref{MPT45}, the phase transitions happen for subregions of $A/E_4$, $AE_4/E_1$, $AE_4E_1/E_3$, and so on. Due to the second feature, when \( m \) is an odd number, the graph is symmetric, which is consistent with the figures we draw when \( m = 2, 3, 4, \) and \( 5 \).

    \subsection{Numerical and analytical calculations}
\noindent As we have found the desired MPT diagram for any \( m \) in the one gap region or two gap region cases, we can perform the direct calculation for CMI and present the numerical and analytical results—the relationship between the maximum value of CMI and \( m \). We will discuss the behavior of the maximum value of CMI with $E$ being a region of $m$ disjoint intervals for the following several cases: the case where fixed regions $A$ and $B$ are adjacent subregions, the case where $A$ and $B$ are not adjacent while all intervals in $E$ live in one gap region between $A$ and $B$, the case where $m$ intervals of $E$ live in both gap regions between and outside $A$ and $B$, and more general cases.

%\subsubsection{Numerical Results and Findings}
\subsubsection*{$A$ and $B$ are adjacent}
The easiest case is when two regions \( A \) and \( B \) are adjacent at a point, so there is only one gap {outside} \( A \) and \( B \). We can simply use a conformal transformation to push this adjacent point to infinity in the Poincaré coordinate system, and \( A \) and \( B \) become \( (-\infty, 0] \) and \( [1, \infty) \) respectively. In this case, using Figure \ref{MPT1}, one can easily find that the \( 2m \) endpoints of \( E_i \) are simply the \( 2m + 1 \) equal division points of the gap region \( (0, 1) \). The value of CMI is infinite due to the divergence of the mutual information (MI) between regions \( A \) and \( B \), so we calculate their difference, which is the tripartite information, in the maximal configuration of an $m$ interval $E$
\begin{equation}\label{I3cal}
\begin{aligned}
    -I_3 &= I(A:B|E) - I(A:B)\\
         &= 2\log\frac{1}{\epsilon}+S_{E}-S_{ABE}\\
         &=2\log(2m+1).
\end{aligned}
\end{equation}
In the second line, the disconnectivity condition (\ref{disconnect}) is {utilized}. \( S_E \) is the summation of the entropy of \( m \) intervals {from constraint II}, while \( S_{ABE} \) is the entropy of the gap regions in between and beside them {from constraint I}. Due to the fact that the endpoints of \( E \) are chosen to be the \( 2m + 1 \) equal division points of \( (0, 1) \), only the entropy of one single interval \( -2\log\frac{1}{(2m+1)\epsilon} \) remains after the cancellation, resulting in \( \exp\left( \frac{I(A:B|E)}{2} \right) \), which is linearly diverging when \( m \) tends to infinity.

A surprising result emerges: no finite upper bound of (minus) tripartite information exists; it will diverge if \( m \) goes to infinity. One might wonder if this divergence results from the fact that \( A \) and \( B \) are adjacent. To answer this question, let us analyze the more complex case where \( A \) and \( B \) are non-adjacent single intervals with the cross ratio \( CR \).

\subsubsection*{The case where $m$ intervals of $E$ live in one gap region between $A$ and $B$}

\begin{figure}[H]
\centering
     \includegraphics[width=10cm]{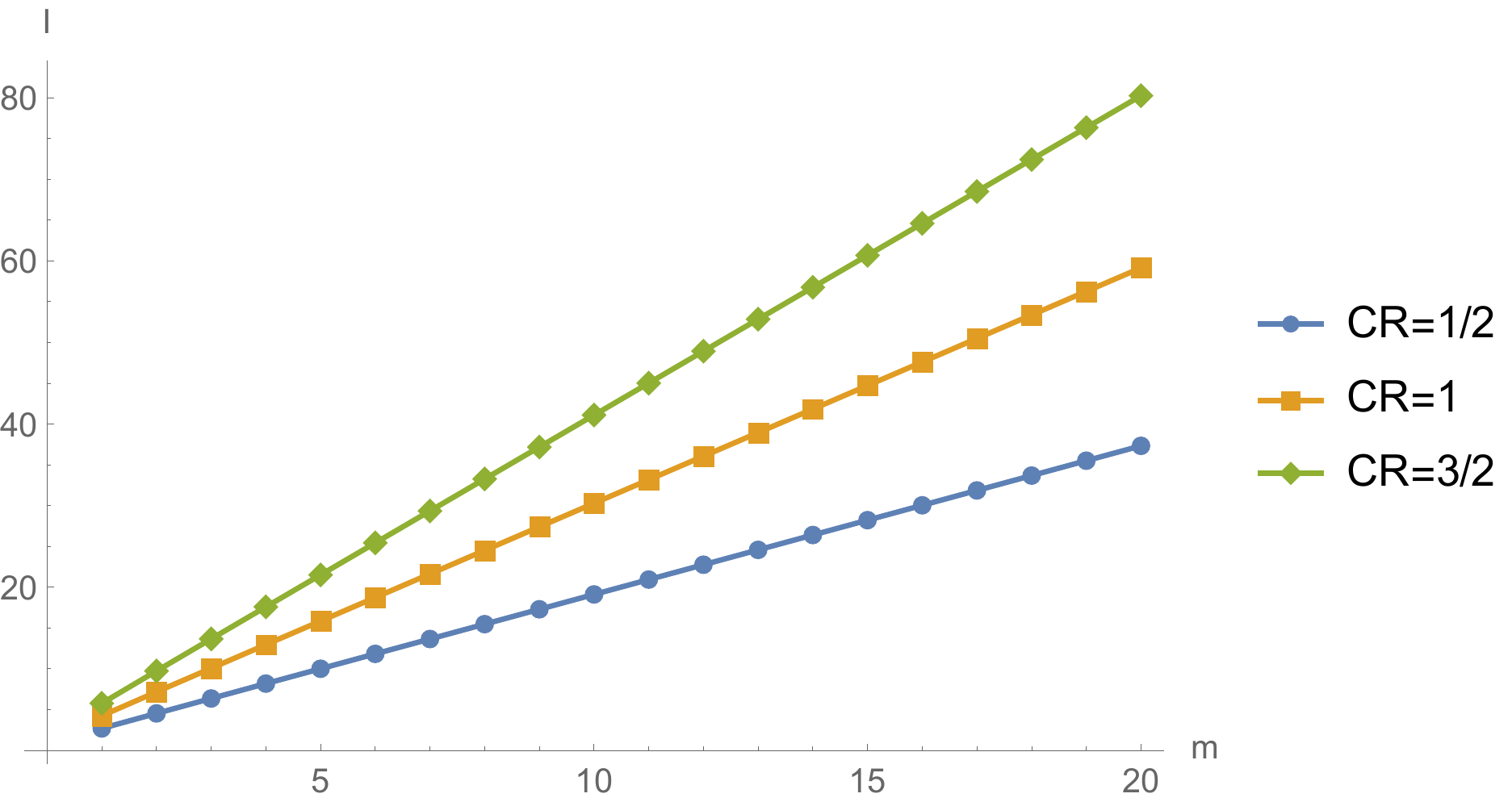}
 \caption{\( I = \exp\left( \frac{I(A:B|E)}{2} \right) \) as a function of \( m \) when all $m$ intervals of region $E$ live in only one gap between region $A$ and $B$, with the cross ratio between \( A \) and \( B \) chosen as different values.}
\label{I_n1}
\end{figure}

First, we only place intervals of \( E \) inside one gap region between \( A \) and \( B \). According to the MPT diagrams in Figure \ref{MPT1}, one can numerically solve the equations of critical conditions for phase transition configurations. Each critical condition is a quadratic equation. Together, we have a system of quadratic equations with \( 2m \) variables. Utilizing symmetry, it becomes a system of quadratic equations with \( m \) variables. It is worth noting that the ``NSolve'' function in \(\mathsf{Mathematica}\) cannot handle this equation for \( m > 8 \) in this case. As a result, for \( m > 8 \), we have to use the iterative method in Section \ref{sec2.2} to approximate the solution. Ultimately, the {numerical results of the} relationship between \( \exp\left( \frac{I(A:B|E)}{2} \right) \) and \( m \) for different values of cross ratios \( CR \) is plotted in Figure \ref{I_n1}.

One can observe that \( \exp\left( \frac{I(A:B|E)}{2} \right) \) is nearly an exact linear function of \( m \). As it grows with $m$, we are most concerned about its divergence behavior, but since one cannot precisely solve the equations when \( m \) tends to infinity (which takes polynomial time with respect to \( m \)), we cannot be certain about the divergence behavior of \( I_3 \) with respect to \( m \). The phase transition configuration conditions are a list of quadratic equations of the length of intervals, so that the length of those intervals in the configuration of maximum $I_3$ should be algebraic numbers. \( \exp\left( \frac{I(A:B|E)}{2} \right) \) as a rational polynomial of those length should also be an algebraic number (we take $CR$ as a rational coefficient here), which is the zero of a polynomial with integer coefficients. 

Finding this polynomial whose zero point is $\exp\left( \frac{I(A:B|E)}{2} \right)$ could give us the maximum value of $\exp\left( \frac{I(A:B|E)}{2} \right)$ and determine its divergence behavior when $m$ tends to infinity. In appendix A, we obtain this polynomial in both the case where $E$ lives in one gap region between $A$ and $B$, and the case of $E$ living in two gap regions through quite tedious calculations. Here we summarize the main results for this polynomial and details for the calculations are in appendix A.

%Below, we will show that the integer coefficients of this polynomial (whose zero point is \( \exp\left( \frac{I(A:B|E)}{2} \right) \)) are simple functions of \( n \).

%\wb{WB: The zero points of the polynomials do not have to be algebraic numbers, since \(CR\) may not be a rational number. }

For the case where $m$ intervals of $E$ live in one gap between $A$ and $B$, the polynomial whose zero point gives maximum \( \exp\left( \frac{I(A:B|E)}{2} \right) \) is found to be 
\begin{equation}
	P_m(CR)=(1+CR)x(x-1)^m-CR (x+1)^{m+1}.
\end{equation}
According to the corresponding polynomial equation \( P_m(CR)=0 \), the upper bound of \( \exp\left( \frac{I(A:B|E)}{2} \right) \) is the solution $x$ of the simple equation as follows
\begin{equation}\label{algebraic1}
    \frac{x}{1+x} \left( \frac{x - 1}{x + 1} \right)^m = \frac{CR}{1 + CR}.
\end{equation}
One can check that this result for the maximum value of \( \exp\left( \frac{I(A:B|E)}{2} \right) \)) is consistent with numerical results in Figure \ref{I_n1} that we have obtained previously.

Now we study the divergence behavior of the maximum value of \( \exp\left( \frac{I(A:B|E)}{2} \right) \)) at large $m$, which is the $x$ in (\ref{algebraic1}). As \( m \) tends to infinity, we make an ansatz for {the asymptotic behavior of $x$}
\begin{equation}
    \lim_{m \to \infty} x = a m + b + \frac{c}{m} + O\left( \frac{1}{m^2} \right),
\end{equation}
{where $a$, $b$ and $c$ are corresponding coefficients to be determined from the equation.} The left-hand side of equation (\ref{algebraic1}) tends to 
\begin{equation}
\begin{aligned}
    \lim_{m \to \infty} \frac{x}{1 + x} \left( \frac{x - 1}{x + 1} \right)^m =& e^{ - \frac{2}{a} } - \frac{ ( a - 2b ) e^{ - \frac{2}{a} } }{ a^2 m } \\
    +& \frac{(6b^2-2a(1+3b+3b^2)+3a^2(1+b+2c))e^{ - \frac{2}{a}}}{3a^4 m^2}+O(m^{-3}).
\end{aligned}
\end{equation} As it equals the right-hand side of equation (\ref{algebraic1}), \( \frac{CR}{1 + CR} \), which is a constant, we have

\begin{equation}
    a = \frac{2}{ \log\left( 1 + \frac{1}{CR} \right) }, \quad b = \frac{a}{2} = \frac{1}{ \log\left( 1 + \frac{1}{CR} \right) }, \quad c = \frac{1}{12}\left(-3+2\log(1+\frac{1}{CR}) \right).
\end{equation}

Finally, we {have proved} that the leading order divergence behavior of \( \exp\left( \frac{I(A:B|E)}{2} \right) \) with respect to \( m \) is indeed linear. The cross ratio between \( A \) and \( B \) only affects the coefficient of this linear term, rather than affecting the linear divergence behavior.

\subsubsection*{The case where $m$ intervals of $E$ live in both gap regions}
\begin{figure}[H]
\centering
     \includegraphics[width=14cm]{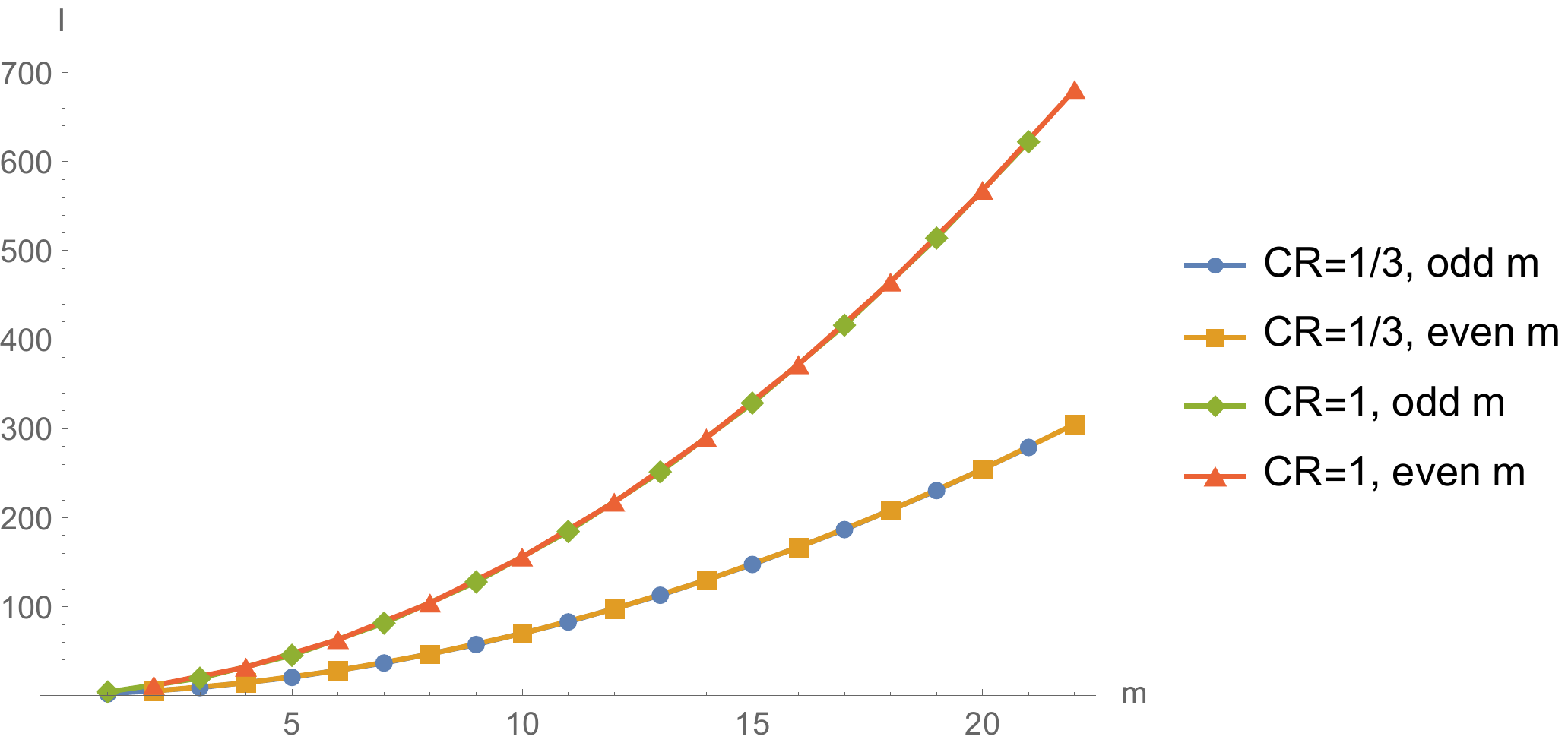}
 \caption{\( I = \exp\left( \frac{I(A:B|E)}{2} \right) \) as a function of \( m \), with the cross ratio between \( A \) and \( B \) chosen as different values. The values of \( I \) in even \( m \) and odd \( m \), though close, satisfy different rules.}
\label{I_n2}
\end{figure}

For the case where $m$ intervals of $E$ live in both gap regions between and outside $A$ and $B$, the method is basically the same. The difference is that the system of quadratic equations that determine the lengths of intervals at the maximum configuration is now obtained in Figure \ref{MPT45}. Because the numbers of the chosen intervals in the two gaps are the same when \( m \) is even and different when \( m \) is odd, one should analyze the case for even \( m \) and odd \( m \) respectively. The iterative method is basically the same as in Section \ref{sec2}, except that the order of modifying the endpoints of \( E \) will change. For example, when \( m = 5 \), the order of modifying the endpoints $E_{iR}$ and $E_{iL}$ would become
\begin{equation}
    \begin{aligned}
        E_{3R}, E_{1R}, E_{4L}, E_{2R}, E_{5L}, \\
        E_{5R}, E_{2L}, E_{4R}, E_{1L}, E_{3L}.
    \end{aligned}
\end{equation}
In fact, this order in the first line is exactly the same as the order of phase transitions between \( E_i \) and \( A \), and the second line is the order of phase transitions between \( E_i \) and \( B \). In practice, this order will guarantee the convergence of this iterative method to the desired diagram {of} Figure \ref{MPT45}. Utilizing this method, we specifically calculate \( I_3 \) in various \( m \) both odd and even, and plot \( \exp\left( \frac{I(A:B|E)}{2} \right) \) as a function of \( m \) with different cross ratios between \( A \) and \( B \) respectively, as shown in Figure \ref{I_n2}.

From Figure \ref{I_n2}, it seems that, different from the one-gap cases, \(I=\exp\left( \frac{I(A:B|E)}{2} \right)\) seems like a perfect quadratic function of \( m \). We could determine the exact large $m$ divergence behavior of \(\exp\left( \frac{I(A:B|E)}{2} \right)\) through the same method utilizing a polynomial with integer coefficients as for the one gap region case. The details in finding this polynomial function could be found in Appendix A.  We find that in this case, \( I \) is also generated by a polynomial. The two polynomials \eqref{Pn2o} and \eqref{Pn2e} that govern the upper bound \(I\) in the odd $m$ and even $m$ case can be written together as
{\begin{equation}\label{algebraic2}
		x(x - 1)^m - \frac{CR}{4} \left[ (\sqrt{x} + 1)^{m + 1} - (\sqrt{x} - 1)^{m + 1} \right]^2 + CR(x - 1)^m \delta,
\end{equation}}
where \( \delta = 0 \) for even \( m \) and \( \delta = 1 \) for odd \( m \), and the zero point $x$ of this polynomial gives the upper bound value of \(I= \exp\left( \frac{I(A:B|E)}{2} \right)\).

When \( m \) tends to infinity, we make an ansatz\footnote{One can find that only by making the ansatz with the highest degree of \( m \) being \( 2 \) can this polynomial reach zero when \( m \) tends to infinity.} {for the asymptotic behavior of $x$} that
\begin{equation}
    \lim_{m \to \infty} x = a m^2 + b m + c + \frac{d}{m} + O\left( \frac{1}{m^2} \right).
\end{equation}
Substituting this ansatz into polynomial (\ref{algebraic2}) and setting it to zero, we can get the values of the coefficients as follows
{\begin{equation}
		\begin{aligned}
		&a = \frac{1}{\operatorname{arccsch}^2(\sqrt{CR})}, \quad b = \frac{2}{\operatorname{arccsch}^2(\sqrt{CR})}, \\
		&c = \frac{2}{3}+\frac{1}{\operatorname{arccsch}^2(\sqrt{CR})}-\frac{2(1+\delta \; CR)\operatorname{csch}\left[2\operatorname{arccsch}(\sqrt{CR})\right]}{CR\operatorname{arccsch}(\sqrt{CR})}, \quad d = 0.
		\end{aligned}
\end{equation}}
At the end, we derive the exact divergence behavior of \(I=\exp\left( \frac{I(A:B|E)}{2} \right)\) with respect to \( m \), and most importantly, it is quadratic in $m$.

\subsubsection*{More general cases}
\begin{figure}[H]
\centering
     \includegraphics[width=10cm]{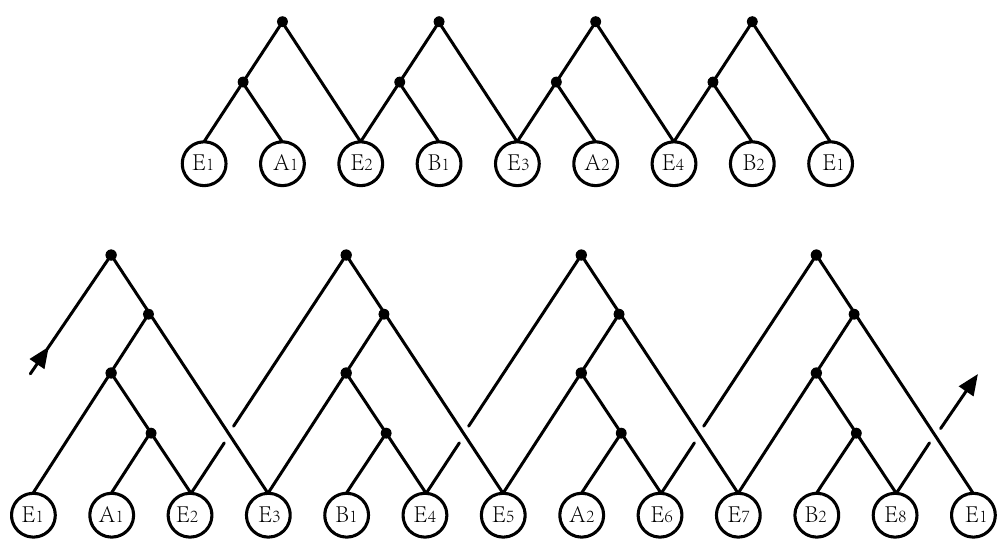}
 \caption{$m=4$ and $m=8$ MPT diagrams for the maximum CMI configuration in the case where $E$ could reside in four gap regions.}
\label{MPTG4}
\end{figure}

There are more complex cases with more gap regions between $A$ and $B$; \eg, when regions $A$ and $B$ are disconnected and they ``interlace'' with each other in space like $A_1B_1A_2B_2$, in which case there are four gap regions between $A$ and $B$. It is easy to guess that $\exp\left( \frac{I(A:B|E)}{2} \right)$, as a function of $m$, has a fourth power divergent behavior $m^4$. In practice, this generalization holds true. We give a quick overview here. The MPT diagrams for the maximum CMI configuration in the case where $E$ could reside in four gap regions are shown in Figure \ref{MPTG4} for $m=4,8$.

Figure \ref{MPTG4} is basically a replica of the two gap region cases for $m=2$ and $m=4$ in Figures \ref{MPT2} and \ref{MPT45}. The calculation of $I_3$ is basically the same as in formula (\ref{I3cal}). Due to the replica configuration, terms of entropy of single intervals in $S_E$ and terms of entropy of gaps in between those interval $E_i$ are twice those of the two-gap cases. As a result, $I_3$ in the four-gap cases with $4m$ intervals should approximately be twice that of the two-gap cases with $2m$ intervals. Therefore, we determine that the divergence behavior of $\exp\left( \frac{I(A:B|E)}{2} \right)$ for four gap region cases is of order $m^4$.

\subsubsection*{Physical implications}

Let us analyze the physics behind the divergence behaviour of $\exp\left( \frac{I(A:B|E)}{2} \right)$ when region $A$ and $B$ are not adjacent. In this case, the mutual information $I(A:B)$ is finite so that $-I_3=I(A:B|E)-I(A:B)$ diverges because $I(A:B|E)$ diverges at large $m$. As an IR quantity that directly cancels out the possible UV divergence in mutual information, $-I_3$ should not be expected to have UV divergence in any definite configuration of $E$ with a finite number of $m$. {Here, what happens is that} it diverges when we make the number of intervals that $E$ contains tend to infinity. Is this divergence related to the UV divergence we have {in some entanglement measures including the entanglement entropy and mutual information}? The answer is yes. When $m$ tends to infinity, the length of the smallest interval inside $E$ will tend to $O\left( \frac{1}{m} \right)$. As a result, we have to choose the UV cutoff $\epsilon < L_{AB} \frac{1}{m}$. Conversely, given the UV cutoff $\epsilon$ fixed, the largest $m$ is chosen to be proportional to $\frac{1}{\epsilon}$. On the other hand, we have
\begin{equation}\label{divergence}
\begin{aligned}
    -I_3 &\sim 2\log(2m+1) \propto 2\log\frac{1}{\epsilon} + O(1),\quad\quad\quad\,\,\, (\text{one gap}),\\
    -I_3 &\sim 2\log(am^2+bm+c) \propto 4\log \frac{1}{\epsilon} + O(1),\,\,\, (\text{two gaps}),\\
    -I_3 &\sim 2\log(am^4+\dots) \propto 8\log \frac{1}{\epsilon} + O(1),\quad\quad \,(\text{four gaps}).\\
\end{aligned}
\end{equation}
One can find that after substituting $m$ by $\frac{1}{\epsilon}$, the divergence behaviour of $-I_3$ is exactly twice the UV divergence of the entanglement entropy of region $A$ (or $B$). %The divergence behaviour of $-I_3$ is exactly twice the UV divergence of the entanglement entropy of $A(B)$.

Let us compare this result in holography with the upper bound of CMI ($-I_3$) in quantum information theory. Using the Araki-Lieb inequality twice, one can easily get
\begin{equation}\label{cmiineq}
    I(A:B|E) \leq 2\min(S_A, S_B).
\end{equation}
We assume $S_A<S_B$ without loss of generality, and this inequality (\ref{cmiineq}) is saturated, \ie{}, $I(A:B|E)=2S_A$ iff
\begin{equation}\label{mivanish}
    I(A:B) = I(A:E) = I\left( A:(ABE)^c \right) = 0, \quad I(A:BE)=2S_A.
\end{equation}
Due to the vanishing of the first two mutual information above in (\ref{mivanish}), subsystem \( A \) is not correlated or entangled with either \( B \) or \( E \) in both the classical or quantum senses. Meanwhile, the vanishing of mutual information \( I\left( A:(ABE)^c \right) \) together with the fact that $I(A:BE)=2S_A$ indicates that \( A \) contributes all its degrees of freedom to the entanglement with the union region \( BE \). This correlation between $A$ and $BE$ is purely quantum, as classical correlations can contribute at most the amount of \( S_A \) to the mutual information of $A$ with any other subsystem, while quantum entanglement can contribute up to the amount of \( 2S_A \) to the mutual information. We can further confirm this point by directly calculating the squashed entanglement—a nearly perfect pure quantum entanglement measure \cite{Li_2018,Li_2014,Wilde_2016,Avis_2008,Yang_2009,Christandl_2004,Brand_o_2011,Ju:2023dzo}—\( E_{sq}(A:BE) \) between \( A \) and \( BE \) as follows.
\begin{equation}
\begin{aligned}
    E_{sq}(A:BE) &= \frac{1}{2} \inf_{\rho_{ABEF}} I(A:BE|F), \\
    &= \frac{1}{2} \inf_{\rho_{ABEF}} \left( S_{AF} + S_{BEF} - S_{ABEF} - S_F \right), \\
    &= \frac{1}{2} \inf_{\rho_{ABEF}} \left( S_A + S_F + S_{BEF} - S_{ABEF} - S_F \right), \\
    &= \frac{1}{2} \inf_{\rho_{ABEF}} I(A:BEF), \\
    &= \frac{1}{2} I(A:BE)= S_A.
\end{aligned}
\end{equation}

Here, \( \rho_{ABEF} \) is any possible extension of \( \rho_{ABE} \) in a mathematically consistent way, and the vanishing of mutual information \( I(A:F) = 0 \)\footnote{Due to the purification theorem, there always could exist a system $G$ that purifies $ABEF$, and we have
\begin{equation*}
    S_{BE}-S_A=S_{ABE}\,\,
    \Rightarrow\,\, S_{AFG}-S_A=S_{FG}
    \,\,\Rightarrow \,\,I(A:FG)=0
    \,\,\Rightarrow \,\,I(A:F)=0
\end{equation*}} is used in simplifying the second line. Since \(2S_A=I(A:BE)\leq I(A:BEF) \leq 2S_A \) due to the Araki-Lieb inequality, any extension \( \rho_{ABEF} \) yields \( I(A:BE|F) = 2S_A \), and thus the squashed entanglement $E_{sq}(A:BE)$ is definitively \( S_A \). This indicates that the correlation between \( A \) and \( BE \) is purely quantum as the squashed entanglement is a measure of pure quantum entanglement \footnote{Note that the $1/2$ coefficient difference between the squashed entanglement and mutual information is due to convention in definition.}.

In quantum information theory~\cite{bengtsson2016brief}, the tripartite entanglement \( E_3(A:B:E) \) is often defined as
\begin{equation}\label{tripartiteE3}
    E_3(A:B:E) = Ent(A:BE) - Ent(A:B) - Ent(A:E),
\end{equation}
where \( Ent \) denotes a measure of bipartite quantum entanglement in mixed states. When the right hand side of (\ref{tripartiteE3}) are all genuine quantum entanglement measures, the left hand side should be genuine tripartite entanglement. It depicts the global entanglement among the tripartite subsystems \( A \), \( B \), and \( E \), independent of the bipartite entanglements among them. Here, if we choose the mutual information to serve as this measure, then \( -I_3 \) corresponds to \( E_3(A:B:E) \) by definition. The fact that the mutual information \( I(A:BE) \) coincides with the squashed entanglement \( E_{sq}(A:BE) \), as shown earlier, justifies this choice.

We emphasize here that usually $I_3$ is not a good measure for tripartite entanglement, however, here due to the properties that this maximum configuration has, $I_3$ serves as a valid measure to capture the amount of tripartite entanglement among $A$, $B$ and $E$.
In our case, when the CMI tends to the upper bound, the fact that formula (\ref{divergence}) exhibits the same divergent behavior as \( 2S_{A(B)} \) implies that nearly all UV degrees of freedom of \( A \) and \( B \) are participating in the tripartite entanglement with $E$.

Formally, when discussing the divergence behaviour of the entanglement entropy of region $A$ in CFT, people always intuitively imagine that the local Bell pairs near the boundary of region $A$ contribute to it. However, Bell pairs cannot have non-zero tripartite information with another region; in our calculation, we reveal that the degrees of freedom localized near the boundary of $A$, rather than participating in the bipartite entanglement, are participating in the tripartite entanglement.

Using the method in this subsection, we have only been able to show this conclusion for the $UV$ behavior of the entanglement. In the next section, we will develop a more universal method and obtain the same conclusion for the whole entanglement behavior including IR degrees of freedom.

\section{The upper bound of CMI: a more universal method to determine E} \label{sec4}
\noindent In the previous section, we analyze the divergence behaviour of $I_3$ and come to the conclusion that nearly all UV degrees of freedom of $A$ and $B$ are participating in the tripartite entanglement. However, as we cannot determine the exact relationship between the number of intervals $m$ and the UV cutoff $\epsilon$, the $O(1)$ term in formula (\ref{divergence}) is still unknown. In this section, we are trying to analyze whether all the IR degrees of freedom are participating in tripartite entanglement as well. We are going to investigate more complicated cases, such as the {two sided black hole} and higher-dimensional cases. In higher dimensions, $\partial E$ has infinitely many degrees of freedom, so any finite number of entanglement phase transition conditions cannot fully determine the shape of region $E$. In this sense, the multi-entanglement phase transition rule which we state in Section 2.2 is not useful any more; however, the disconnectivity condition is still valid in a weaker form. We first introduce this weaker form of the disconnectivity condition and prove it. Then we utilize this condition to give a direct calculation of the upper bound of CMI in various general cases. At last, we give a general and concise formula for the upper bound of CMI.

\subsection{Disconnectivity condition}%more gaps, higher dim and general geometry

\noindent When $E$ has multiple intervals, analyzing CMI for general geometries and complicated regions of $A$ and $B$ is a formidable task; instead, it is easy to only analyze the disconnected configuration that satisfies the disconnectivity condition (\ref{disconnect}). As a result, it would be great to prove that only analyzing the disconnected cases is enough for calculating the upper bound of CMI, and we have the disconnection condition in a weaker form as follows.

\textbf{The weaker form of disconnectivity condition.} \textit{Given a configuration $E$ with a non-vanishing $I(A:E)$ or $I(B:E)$, there always exists a disconnected configuration of $E$ with vanishing $I(A:E)$ and $I(B:E)$ whose CMI $I(A:B|E)$ is not less than the former one.}

This statement is weaker than the disconnectivity condition stated in the last section, as we do not demand that the disconnected configuration has the same $m$ as the connected configuration. One can choose a configuration with $m$ many times larger than that of the connected configuration, as long as its CMI is larger. Though the theorem is weaker, it is easy to prove and it is enough for us to only analyze the disconnected configurations for finding maximum CMI. {Note that the previous stronger version of disconnectivity condition still holds for the calculation of maximum configuration of $-I_3$, however, as we will show later, this weaker form of disconnectivity condition will remain valid for the calcualtion of maximum $I_4$ and $-I_5$ while the stronger form does not hold there.}

\begin{figure}[H]
\centering
     \includegraphics[width=10cm]{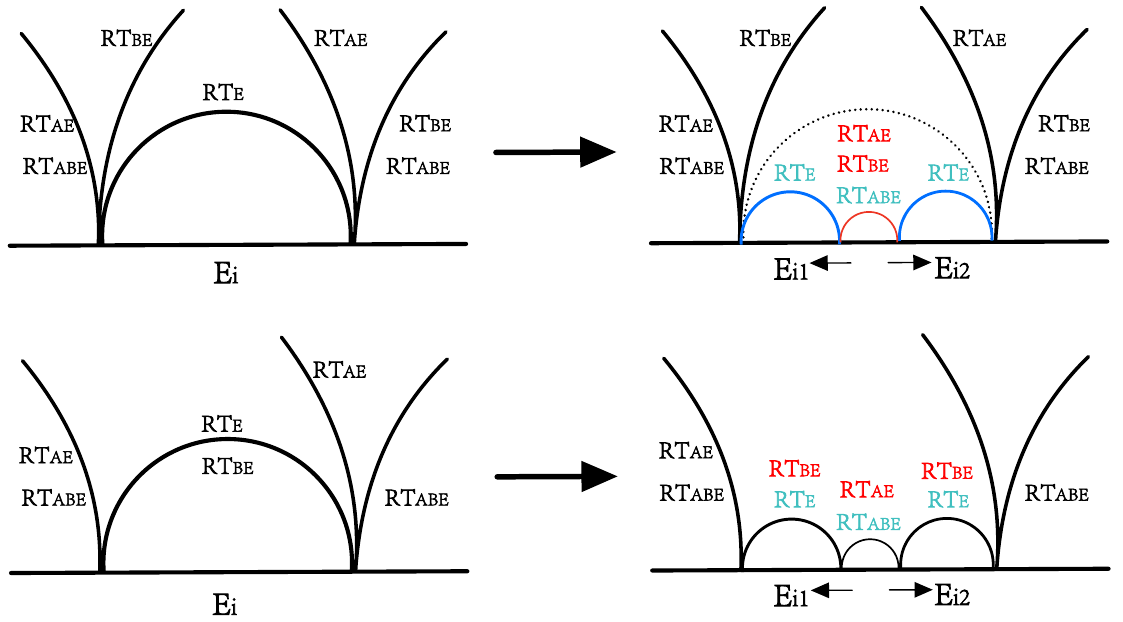}
 \caption{Proof of the weaker form disconnectivity condition. Each RT surface of region $M$ is labeled as $RT_{M}$. Only the surfaces whose area will change under modifying the endpoints of the gap are marked by blue and red, which represent their signature in formula (\ref{CMIformula}).}
\label{CMIDISPRO}
\end{figure}

Let us prove it in the general situation. There are four steps to prove this theorem. According to Figure \ref{CMIDISPRO}, we start from {one of the intervals} of $E$, written as $E_i$, whose entanglement wedge is connected with both $A$ and $B$ in the configuration of the entanglement wedge of regions $AE$ and $BE$. We split $E_i$ from the middle into two subregions $E_{i1}$ and $E_{i2}$ by adding a gap inside while preserving the outer boundary of $E_i$ unchanged. When the gap is very small, the entanglement wedge of region $E_i$ is fully connected, and as we have argued in the last section, the value of CMI does not change after opening this small gap.

The second step: when we continuously enlarge the gap, the first phase transition of the entanglement wedge occurs, which makes $E_{i1}$ and $E_{i2}$ disconnected in the entanglement wedge of $E$, while preserving the connectivity of $E_{i1}$ and $E_{i2}$ in the entanglement wedge of $AE$ and $BE$. This is shown in Figure \ref{CMIDISPRO}.
 
The third step: we continuously enlarge the gap inside $E$. One can find that $S_{E_{i1}}$ and $S_{E_{i2}}$ both decrease, while the entropy of the gap region in-between $S_{gap}$ increases. These three quantities are the only ones that affect the value of CMI because the outer boundary of $E_i$ remains unchanged. The sign of them inside formula (\ref{CMIformula}) is drawn as the different colors of those RT surfaces in Figure \ref{CMIDISPRO}; red represents positive and blue represents negative. It is easy to observe that when we enlarge the gap, CMI increases until another phase transition occurs. In principle, the phase transition could occur on one of $E_{i1}$ or $E_{i2}$, \ie  the entanglement wedge of $AE$,\footnote{Here, we assume that $E_{i1}$ and $E_{i2}$ become disconnected in the entanglement wedge of $AE$ before $E_{i1}$ and $E_{i2}$ become disconnected in the entanglement wedge of $BE$, without loss of generality.} suddenly becomes disconnected between $A$ and $E_{i1(2)}$ while remaining the connectivity between $A$ and $E_{i2(1)}$. However, we can make the phase transition occur on $E_{i1}$ and $E_{i2}$ simultaneously by adjusting the middle point of this gap\footnote{If the middle point of the gap region between $E_{i1}$ and $E_{i2}$ is far left inside the original region $E$, $E_{i1}$ disconnects with $A$ first; otherwise, if the middle point is far right inside the original region $E$, $E_{i2}$ disconnects with $A$ first. Therefore, there must exist a fine-tuned middle point which makes the phase transition occur on $E_{i1}$ and $E_{i2}$ simultaneously.}. After the phase transition, $E_{i1}$ and $E_{i2}$ disconnect with $A$ in the entanglement wedge of $AE$. Remember that when we enlarge the gap, the CMI increases, so we prove that:
\textit{for a configuration whose intervals of $E$ connect with both $A$ and $B$ inside the entanglement wedge of $AE$ and $BE$ respectively, one can always find another configuration whose intervals connect to at most one of $A$ and $B$ inside the entanglement wedge of $AE$ and $BE$, with larger CMI.}

After performing the third step for all the intervals of $E$, we are left with a configuration where all intervals of $E$ only connect with $A$ in $EW(A)$ or only connect with $B$ in the $EW(BE)$. The fourth step is to consider an interval $E_j$ which is only connected with $B$ in the entanglement wedge of $BE$, and find another disconnected configuration with CMI not less than it. To achieve this final goal, we have to split $E_j$ again. In the second line of Figure \ref{CMIDISPRO}, this process is illustrated. The only difference is that the CMI will not change no matter how we enlarge the gap, as long as $BE_{j1}$ and $BE_{j2}$ are connected in the entanglement wedge of $BE$. Again, we can make the phase transition occur between $E_{j1}$ and $B$ and between $E_{j2}$ and $B$ simultaneously. At the end, both $E_{j1}$ and $E_{j2}$ are disconnected from $B$ and the value of CMI stays the same. Combined with the last three steps, in the end, we find a disconnected configuration with CMI not less than that of the connected configuration, and the disconnectivity condition is proven\footnote{Note that this proof is valid for various geometries and in higher-dimensional cases. When \( E_i \) is a disk, the ``gap'' inside region \( E_i \) would be a disk; when \( E_i \) is an annulus, the gap would be an annulus that splits \( E_i \) into two annuli, etc.}.

{Therefore, to search for maximum configurations of $-I_3$, we could calculate the values of $-I_3$ only in configurations satisfying the following constraints, and pick the largest one among them. The constraints are: I, the entanglement wedge of $ABE$ is fully connected; II, the entanglement wedge of $E$ is fully disconnected; III, the disconnectivity condition. Note that there could be configurations that do not satisfy these conditions while at the same time have the same maximum value of $-I_3$. However, in these cases, as we argued, because $I(A:E)$ or $I(B:E)$ is not zero, $-I_3$ may not fully represent tripartite entanglement.}

   \subsection{Direct calculation of maximum CMI in various cases}
\noindent The disconnectivity condition stipulates that analyzing disconnected configurations is enough for finding the upper bound of CMI. As the number of intervals in $E$ is not fixed, there could be infinitely many configurations that satisfy the disconnectivity condition, \ie. $I(A:E)=I(B:E)=0$. However, after this condition is satisfied, the formula for CMI or $I_3$ will get much simplified and we only need to pick the maximum CMI configuration from all configurations that satisfy the disconnectivity condition. In this subsection, we calculate the maximum CMI configurations for two types of geometries: the asymptotic $AdS_3$ geometry and the two sided black hole geometry, under the limit that \( m \) goes to infinity.

\subsubsection*{Asymptotic $AdS_3$} 

\begin{figure}[H]
\centering
\includegraphics[width=13cm]{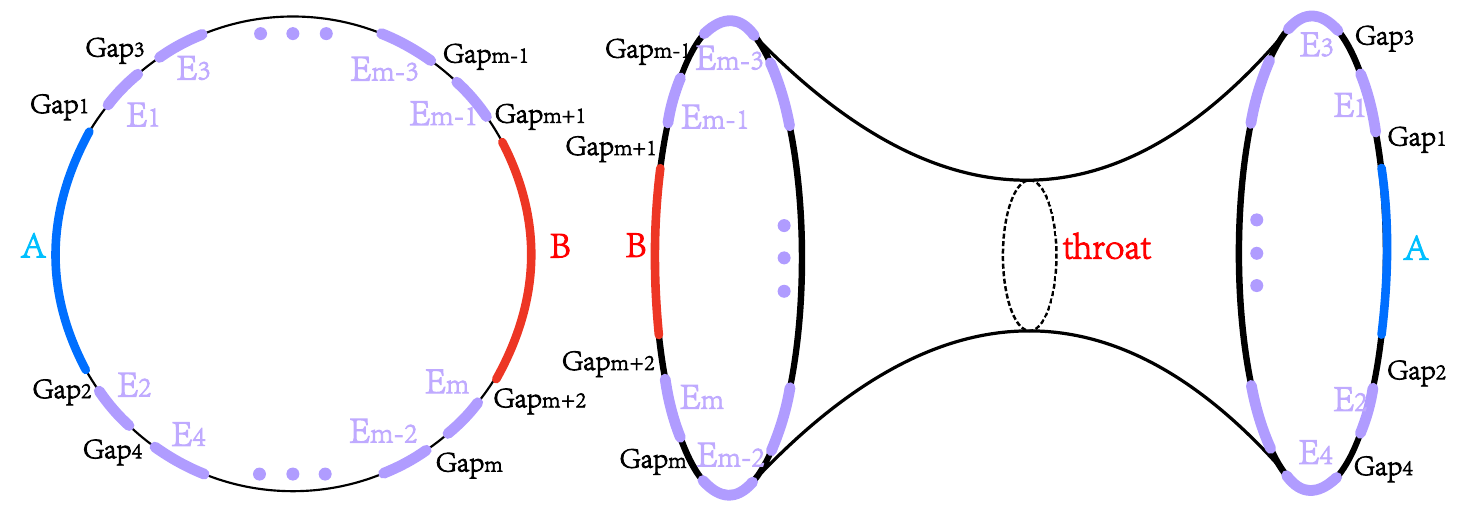}
\caption{Illustration for finding the maximum CMI configurations when intervals of $E$ could live anywhere that has no overlap with $A$ or $B$ for $AdS_3$ (left) and the two sided black hole (right). Regions \( A \), \( B \), and \( E \) are marked by blue, red, and purple intervals, respectively. The \( m + 2 \) gap regions between \( E_i \), \( A \), and \( B \) are labeled as \( \text{Gap}_i \).}
\label{GENCMI}
\end{figure}

The maximum configuration for $E$ should only be when intervals of $E$ live both in-between and outside $A$ and $B$. The left side of Figure \ref{GENCMI} depicts the configuration of such an $E$ in global coordinates. Using the disconnectivity condition, we can calculate the value of CMI as follows.
\begin{equation}\label{generalCMI}
\begin{aligned}
    I(A:B|E) &= S_{AE} + S_{BE} - S_E - S_{ABE} \\
             &= S_A + S_E + S_B + S_E - S_E - S_{ABE} \\
             &= S_A + S_B + \sum_{i=1}^m S_{E_i} - \sum_{i=1}^{m+2} S_{\text{Gap}_i},
\end{aligned} 
\end{equation}
{where $S_{Gap_i}$ denotes the entanglement entropy of the $i$-th gap region between $A$, $B$ and $E$ as shown in Figure \ref{GENCMI}.}
In the second line, we have used the disconnectivity condition and in the third {line, we have used the fact that $S_{ABE}$ is fully connected as we argued in section \ref{sec3} and $S_{ABE}$ then is equivalent to the entanglement entropy of its fully disconnected complementary gap regions. }
One can find that if we want to make the right side of this equality as large as possible, we have to make \( S_{E_i} \) very large and \( S_{\text{Gap}_i} \) very small. However, if \( \sum_{i=1}^m S_{E_i} \gg \sum_{i=1}^{m+2} S_{\text{Gap}_i} \) %\zy{ if \( S_{E_i} \gg  S_{\text{Gap}_i} \) for every \(i\)},
the entanglement wedges of \( AE \) and \( BE \) will be connected {as the gap region is too small}, which violates the disconnectivity condition. Note that this means that (\ref{generalCMI}) does not hold for this case and disconnected configurations should always have larger values of CMI due to the disconnectivity condition. Therefore, we cannot make \( \sum_{i=1}^m S_{E_i}-\sum_{i=1}^{m+2} S_{\text{Gap}_i} \) extremely large and there should be an upper bound on \( \sum_{i=1}^m S_{E_i}-\sum_{i=1}^{m+2} S_{\text{Gap}_i} \) coming from the constraint of the disconnectivity condition. This constraint should be the following. For the entanglement wedges of \( AE \) and \( BE \) which satisfy the disconnectivity condition, the lengths of the connected geodesics must be larger than that of the disconnected geodesics as the latter are the real RT surfaces. This gives the following constraint on the values of these entanglement entropies.
\begin{equation}
\begin{aligned}
    S_A + \sum_{i=1}^m S_{E_i} &\leq \sum_{i=1}^{m+2} S_{\text{Gap}_i} - S_{\text{Gap}_{m+1}} - S_{\text{Gap}_{m+2}} + S_{\text{Gap}_{m+1} B \text{Gap}_{m+2}}, 
\end{aligned}
\end{equation}
so that 
\begin{equation}\label{upperbound}
\begin{aligned}
    S_A + S_B + \sum_{i=1}^m S_{E_i} - \sum_{i=1}^{m+2} S_{\text{Gap}_i} &\leq S_{\text{Gap}_{m+1} B \text{Gap}_{m+2}} + S_B - S_{\text{Gap}_{m+1}} - S_{\text{Gap}_{m+2}}, \\ \ie,
    I(A:B|E) &\leq S_{\text{Gap}_{m+1} B \text{Gap}_{m+2}} + S_B - S_{\text{Gap}_{m+1}} - S_{\text{Gap}_{m+2}}.
\end{aligned}
\end{equation}

Overall, we find that CMI has an upper bound due to this constraint. We hope to find the maximum value on the right hand side of (4.3) by tuning $E$, \ie the configuration of $E$ that makes the value of CMI as large as possible. This happens when \( m \to \infty\), where we could have
\begin{equation}\label{limit}
    \lim_{m \to \infty} S_{\text{Gap}_{m+1}} = \lim_{m \to \infty}S_{\text{Gap}_{m+2}} = 0, \quad \lim_{m \to \infty} S_{\text{Gap}_{m+1} B \text{Gap}_{m+2}} = S_B.
\end{equation}
Substituting (\ref{limit}) into (\ref{upperbound}), we finally get
\begin{equation}
    I(A:B|E) \leq 2S_B.
\end{equation}
The system is symmetric when exchanging \( A \) and \( B \), so we have
\begin{equation}\label{twogapresult}
    I(A:B|E) \leq 2\min(S_A, S_B).
\end{equation}
However, this is what we have already known in quantum information theory, where it is derived using the Araki-Lieb inequality as we mentioned earlier. The crucial point is whether we could reach this upper bound in holography. In other words, can this upper bound be saturated in holography?

The answer is yes. Taking \( \min(S_A, S_B) = S_B \) without loss of generality, the inequality (\ref{twogapresult}) would be saturated when (\ref{upperbound}) is saturated, i.e. when the system reaches the phase transition point of the RT surface of region \( AE \). Given a sufficiently small UV cutoff \( \epsilon \) so that we could make $m$ very large, (\ref{limit}) could be approached arbitrarily close. Therefore, the
inequality (\ref{twogapresult}) could be saturated with sufficiently large \( m \). As a result, CMI would be infinitely close to its upper bound when \( \epsilon \) tends to zero. Combined with the argument in the last section and the facts $I(B:E)=I(B:A)=0$, all degrees of freedom in boundary subregion \( B \), including the IR degrees of freedom, contribute to the tripartite entanglement.

\noindent \textbf{ Two sided black hole geometry}

We discuss another example here, the two sided black hole, as shown on the right side of Figure \ref{GENCMI}. \( A \) and \( B \) are two subregions whose upper bound of $-I_3$ we are studying. $A$ and $B$ could both stay on the same side of the black hole, in which case the result would be the same as the previous asymptotic AdS$_3$ case. Therefore, here we consider the case where $A$ and $B$ live on two different boundary CFTs with their dual bulk spacetime connected by a wormhole.

Now we calculate the upper bound of \( I(A:B|E) \). The disconnectivity condition still applies in this case and formula (\ref{generalCMI}) is still valid. The key step is still calculating the maximum value of \( \sum_{i=1}^m S_{E_i} - \sum_{i=1}^{m+2} S_{\text{Gap}_i} \), which is determined by the disconnectivity condition of the entanglement wedge of \( AE \). However, in the two sided black hole case, there are two ways for the entanglement wedge of \( AE \) to be connected, each posing a different constraint on the upper bound of the disconnected configuration. We have to choose the tighter one to be the final upper bound, which could be saturated.

The first way for the entanglement wedge of \( AE \) to be connected is the same as the asymptotic AdS$_3$ case, \ie, the entanglement wedge of \( AE \) connects every single interval of \( E \). This will give a constraint the same as inequality (\ref{upperbound}), which results in \( I(A:B|E) \leq S_{\text{Gap}_{m+1} B \text{Gap}_{m+2}} + S_B - S_{\text{Gap}_{m+1}} - S_{\text{Gap}_{m+2}} \). However, in the two sided black hole case, there exists another connected configuration of the entanglement wedge of \( AE \); that is, intervals of \( E \) on the same boundary (right boundary in Figure \ref{GENCMI}) as region \( A \) connect with \( A \), while the intervals of \( E \) on the other boundary (left boundary in Figure \ref{GENCMI}) disconnect with \( A \). The surface area of this connected phase would be composed of three parts: \( \sum_{i=1}^{\frac{m}{2}+1} S_{\text{Gap}} + \sum_{i=\frac{m}{2}+1}^{m} S_{E_i} + S_{\text{throat}} \)\footnote{Suppose \( m \) is an even number without loss of generality, with \( \frac{m}{2} \) intervals placed in each boundary.}. Due to the disconnectivity condition, the entanglement wedge of \( AE \) is disconnected {at the maximum configuration}; we have
\begin{equation}
\begin{aligned}
    & S_A + \sum_{i=1}^{m} S_{E_i} \leq \sum_{i=1}^{\frac{m}{2}+1} S_{\text{Gap}_i} + \sum_{i=\frac{m}{2}+1}^{m} S_{E_i} + S_{\text{throat}}, \\ 
    & S_A - \sum_{i=1}^{\frac{m}{2}+1} S_{\text{Gap}_i} + \sum_{i=1}^{\frac{m}{2}} S_{E_i} \leq S_{\text{throat}}.
\end{aligned}
\end{equation}
Using the symmetry of exchanging \( A \) and \( B \), we have
\begin{equation}
    S_B - \sum_{i=\frac{m}{2}+2}^{m+2} S_{\text{Gap}_i} + \sum_{i=\frac{m}{2}+1}^{m} S_{E_i} \leq S_{\text{throat}}.
\end{equation}
Adding these two inequalities together, we have
\begin{equation}
\begin{aligned}
    S_A + S_B - \sum_{i=1}^{m+2} S_{\text{Gap}_i} + \sum_{i=1}^{m} S_{E_i} &\leq 2S_{\text{throat}}, \\\ie,\quad
    I(A:B|E) &\leq 2S_{\text{throat}}.
\end{aligned}
\end{equation}

Now we have two upper bounds for the value of CMI: $S_{\text{Gap}_{m+1} B \text{Gap}_{m+2}} + S_B - S_{\text{Gap}_{m+1}} - S_{\text{Gap}_{m+2}}$ and $2S_{\text{throat}}$. We need to check which of these two upper bounds is the tighter one, which is the correct one to pick. Let us analyze the values of these two upper bounds as follows. When \( m \) is relatively small, the entropy of a single gap is relatively large, resulting in \( S_{\text{Gap}_{m+1} B \text{Gap}_{m+2}} + S_B - S_{\text{Gap}_{m+1}} - S_{\text{Gap}_{m+2}} \) being relatively small. This implies that at small $n$, the tighter inequality (\ref{upperbound}) should be the true upper bound of CMI in disconnected configurations. When \( m \) is large, the situation changes: \( 2S_{\text{throat}} \), an IR term without UV divergence, becomes the evidently tighter upper bound. For a disconnected configuration with region \( E \) being \( m \) intervals, one can always construct \( E' \) with \( m+1 \) intervals to replace \( E \), whose CMI is larger than that of \( E \). As a result, we shall take the upper bound with larger \( m \) as the actual upper bound of CMI, \ie,
\begin{equation}
    -I_3 = I(A:B|E) - I(A:B) \leq 2S_{\text{throat}} - I(A:B).
\end{equation}

This result indicates that when $I(A:B)=I(A:E)=0$ while $-I_3$ reaches its upper bound, $-I_3=2S_{\text{throat}}$ fully characterizes the tripartite entanglement of $A$, $B$ and $E$. 
This is a surprising result, as we always take the throat as a symbol of bipartite entanglement between two CFT boundaries. However, we found that the throat area contributes to the tripartite entanglement between arbitrarily chosen sufficiently small regions \( A \) and \( B \) with another region \( E \). Compared with the $AdS_3$ result, this indicates the long-range nature in the tripartite global entanglement in $ABE$.

At the maximum configuration where $-I_3$ reaches the information theoretical upper bound, we could see from the disconnectivity condition that $I(A:B)=I(A:E)=0$ while $I(A:BE)=2S_A$. This indicates that $A$ is fully tripartite globally entangled with $B$ and $E$. As $A$ and $B$ are two small, distant, arbitrarily chosen subregions, this implies that in holographic systems, any two small distant subregions are highly tripartite entangled with a third region. All bipartite entanglement between $A$ and $BE$ emerges from the tripartite global entanglement among $A$, $B$ and $E$. 

To reinforce this conclusion that all bipartite entanglement emerges from tripartite global entanglement, we provide another piece of evidence here. What we have shown above is that $-I_3(A:B:E)$ for distant small subregions $A$ and $B$ whose mutual information is zero could reach an upper bound of $2S_A$. Now we show that for two arbitrary disjoint intervals $A$ and $B$ which have a nonzero mutual information, $I(A:B)\neq 0$ in fact quantifies the tripartite entanglement that $a\subset A$ and $b\subset B$ with vanishing $I(a:b)$ are participating, together with a region $e\subset AB$. Therefore, the mutual information $I(A:B)$ comes from the tripartite global entanglement among three subsystems spanning $A$ and $B$.
The proof is as follows.

\noindent {\textbf{A constrained upper bound: given $A$, $B$ with $I(A:B)\neq 0$, $a\subset A$ and $b\subset B$, constrain $e$ to be within $AB$, $-I_3(a:b:e)\leq I(A:B)$.}

Regions $a$ and $b$ are small subregions with $a\subset A$ and $b\subset B$ with vanishing mutual information $I(a:b)=0$. $A$ and $B$ are two disjointed intervals with $I(A:B)\neq 0$. Region $e$ is constrained to be subregions inside $AB$. Now, we evaluate the uppper bound of $-I_3(a:b:e)$. Compared with the cases we analyzed before, the difference is that we further add constraints on $e$ to be inside $AB$. Due to the disconnectivity condition of $EW(ae)$, and $EW(be)$, we have 
\begin{equation}\label{ConstraintECMIDIS}
\begin{aligned}
    S_a+\sum_{i\in A}S_{e_i}&\leq \sum_{i\in A}S_{gap_i}+S_A,\\
    S_b+\sum_{i\in B}S_{e_i}&\leq \sum_{i\in B}S_{gap_i}+S_B,
\end{aligned}
\end{equation}
respectively. As $S_{abe}=\sum_{i} S_{\text{gap}_i}+S_{AB}$, formula (\ref{generalCMI}) should be slightly modified as 
\begin{equation}
    I(a:b|e)=S_a+S_b+\sum_{i} S_{e_i} - \sum_{i} S_{\text{gap}_i}-S_{AB}.
\end{equation}
Combined with (\ref{ConstraintECMIDIS}), we can finally get the upper bound of $I(a:b|e)$ to be exactly the mutual information $I(A:B)$ between $A$ and $B$. This result indicates that the mutual information between $A$ and $B$ is the upper bound of tripartite entanglement for two sufficiently small subregions $a\subset A$ and $b\subset B$ to participate along with another region $e\subset AB$. As a result, we provide another piece of evidence that bipartite entanglement emerges form tripartite global entanglement.}

Therefore, based on all the facts above, we can make the following bold statement. 
\begin{itemize}
    \item No Bell pairs exist in holographic states; all bipartite entanglement emerges from tripartite entanglement; any two small distant subsystems are highly tripartite entangling with another system.
\end{itemize}

\subsection{Entanglement of state-constrained purification}
\noindent We have successfully derived the upper bound of CMI in the asymptotic AdS$_3$ background and the two sided black hole. These two cases lead to two different ends that seem unrelated. One may wonder what will happen if we consider more complicated cases, such as when $A$ and $B$ are partially interlaced intervals or in higher-dimensional wormhole geometry, etc. Analyzing them one by one would be a tedious (but not difficult) task. However, a keen observation can be made that $\min(S_A, S_B)$ in the asymptotic AdS$_3$ case and $S_{\text{throat}}$ in the wormhole case are both the minimal area of the surface that divides the entanglement wedge of $A$ and $B$. Could this rule be valid in more general cases? We find that inequality (\ref{generalCMI}) always holds true. The nontrivial part in different cases is: due to the disconnectivity condition, how could we constrain the value of $\sum_{i=1}^{m} S_{E_i} - \sum_{i=1}^{m+2} S_{\text{Gap}_i}$?

The disconnectivity condition stipulates that for the configuration of $AE$ with a connected RT surface, its area must be larger than the disconnected configuration (real RT surface). Therefore, it imposes constraints on the value of $-I_3$ in the disconnected configuration that all connected configurations should have a larger value of $\sum_{i=1}^{m} S_{E_i} - \sum_{i=1}^{m+2} S_{\text{Gap}_i}$ than the disconnected configuration. The problem is, which connected configuration of $AE$ should give a minimum value of $\sum_{i=1}^{m} S_{E_i} - \sum_{i=1}^{m+2} S_{\text{Gap}_i}$? This question could easily be answered by reformulating the configuration in the following way. $E$ can be divided into two parts: the part that connects with $A$ in the RT surface of $AE$ and the part that does not which is disconnected by itself\footnote{In CFT$_2$, these two parts are collections of complement intervals. However, in higher-dimensional cases, as $n$ tends to infinity, any strip of $E$ and the gap regions between them are thin enough \cite{Czech:2014wka,Bhattacharjee:2024ceb} (of order $O\left( \frac{1}{m} \right)$). A sufficiently thin strip of $E$ can have different connectivity at different locations, such as the upper half of this thin strip being connected with $A$ while the lower half being disconnected. We need a line (codimension-2 object) $\partial D$ to divide these two parts.}. We define a region $D\supseteq A$ where intervals of region $E$ inside $D$ connect with $A$ in the connected configuration of $AE$, while regions outside $D$ disconnect with $A$. We have
\begin{equation}
\begin{aligned}
    &S_A + \sum_{i=1}^n S_{E_i} \leq \sum_{i \in D} S_{\text{Gap}_i} + \sum_{i \notin D} S_{E_i} + S_D, \\
    &S_A + \sum_{i \in D} S_{E_i} - \sum_{i \in D} S_{\text{Gap}_i} \leq S_D.
\end{aligned}
\end{equation}
Here we also require the complement of $D$, $D^c$, to be the region where intervals inside $D^c$ connect with $B$. Then we can simply substitute $A$ and $D$ into $B$ and $D^c$, respectively. We have
\begin{equation}
    S_B + \sum_{i \in D^c} S_{E_i} - \sum_{i \in D^c} S_{\text{Gap}_i} \leq S_D.
\end{equation}
Adding those two inequalities together, we have
\begin{equation}\label{FinalCMI}
    I(A:B|E) \leq 2S_D \quad\quad (D \supseteq A, \ D^c \supseteq B).
\end{equation}

To make this upper bound as tight as possible, we can simply require $S_D$ to be the minimal surface dividing the entanglement wedge of $A$ from $B$. As before, this upper bound can be infinitely approached when $\epsilon \to 0$ and $m \to \infty$.

\begin{figure}[H]
\centering
\includegraphics[width=13cm]{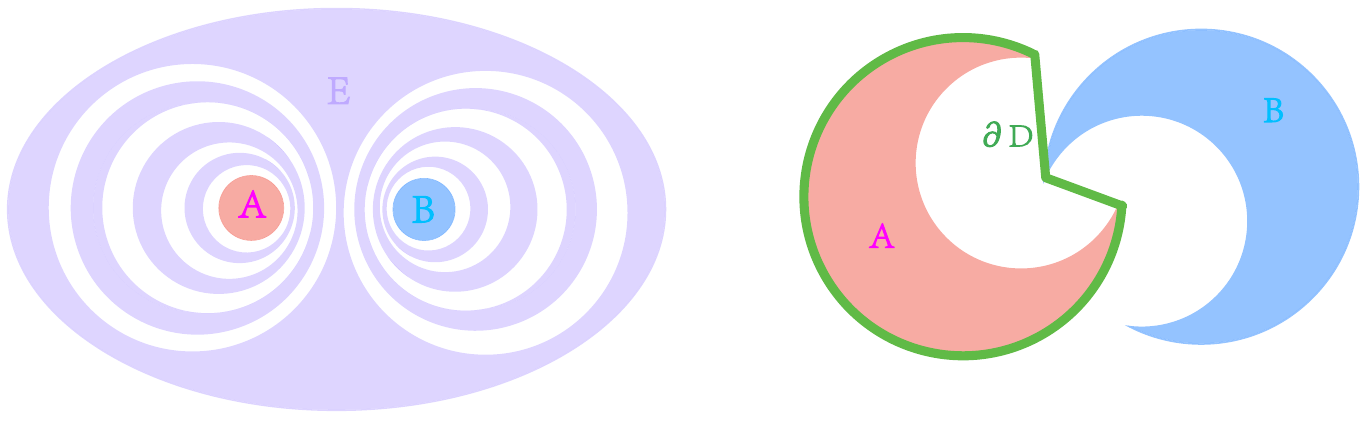}
\caption{CMI in higher dimensions. Left: an example of the configuration of region $E$ with $m=7$. Right: another example where part of $B$ is inside the convex hull of $A$. $\partial D$ is the shortest line that divides $A$ and $B$; the upper bound of CMI is given by $S_{D}$.}
\label{Highdim}
\end{figure}

Let us analyze the consequence of (\ref{FinalCMI}) in higher dimensions. In higher-dimensional cases, the shape of region $E$ can be arbitrarily chosen. No finite phase transition condition can determine the entire shape of region $E$, and the MPT diagram loses its power. In this case, we can choose region $E$ as a collection of curved strips (corresponding to the intervals in low dimension) as shown in Figure \ref{Highdim}. Formula (\ref{FinalCMI}) is still valid, \ie, CMI will tend to $2\min\{S_D\}$ when $m$ tends to infinity. However, as we can use a very thin bridge to connect those strips into a single region, and CMI would nearly not change during this procedure, as a result, CMI could be arbitrarily large even when $m$ is finite \eg, region $E$ is composed of a large amount of strips and thin bridges connecting them. Moreover, when $A$ and $B$ are concave regions, the minimal surface that divides $A$ and $B$ will be the RT surface of the convex hull of $A$ (or $B$), instead of the RT surface of $A$ (or $B$) itself. When the convex hull of $A$ partially contains $B$, the situation will be more complicated, as shown in the right side of Figure \ref{Highdim}. \textit{This reveals the relationship between the convexity of the region and the tripartite entanglement it participates in.} Specifically speaking, the UV degrees of freedom of region $A$ near the concave part (the part which is near $\partial A$ but not near the boundary of the convex hull of $A$) cannot fully participate in the tripartite entanglement with a region outside the convex hull of $A$. However, when $B$ is partially inside the convex hull of region $A$, the degrees of freedom near the concave part of $A$ could participate in the tripartite entanglement along with $B$ and $E$.

We would like to impart the area of the minimal surface $S_D$ a physical meaning in the sense of quantum information theory. The definition of the RT surface of $D$ is very similar to the definition of the entanglement wedge cross section (EWCS). In \cite{Takayanagi:2017knl,Mori:2024gwe}, EWCS between $A$ and $B$ is defined as the minimal surface that divides the entanglement wedge of $AB$ into two parts, with its area being conjectured as the entanglement of purification (EoP) between $A$ and $B$. When the entanglement wedge of $AB$ is disconnected, no surface is needed to divide it into two parts. The area of EWCS vanishes in this case \cite{Jain:2022csf}, as does the EoP when $I(A:B) = 0$ to leading order in $\frac{1}{G_N}$.

However, in our case, $S_D$ does not vanish when $I(A:B) = 0$. Instead, it is a large value. Thus it is not the traditional EoP. We define it as the entanglement of state-constrained purification (EoSP) as follows:
\begin{equation}
    \text{EoSP}(A:B) = \min (S_{AA'}), \quad (\text{where } AA'BB' \text{ is the entire boundary(ies)}).
\end{equation}
The only difference between EoP and EoSP is that $A'$ and $B'$ are constrained to be boundary subregions in this holographic theory instead of some arbitrary extensions in quantum information theory. This constraint makes EoSP not the optimal choice from the viewpoint of EoP, resulting in it being greatly larger than EoP. We can easily prove that the value of EoSP we define is exactly $S_D$, as region $AA'$ is region $D$.

Finally, we find the true upper bound of CMI, which can be written as
\begin{equation}\label{CMIEoHP}
    \sup_{E} I(A:B|E) = 2\text{EoSP}(A:B).
\end{equation}

\section{Four and five partite information}\label{sec5}

\noindent To analyze the multiparite entanglement structure, we investigate the upper bound of the holographic four and five partite information in this section. We give the reason why we should investigate them in the first subsection, prove the disconnectivity condition for \( I_4 \) in the second subsection, and calculate the upper bound of \( I_4 \) in holographic CFT\(_{1+1}\) and higher-dimensional cases in the third subsection. The conclusion is that the upper bound of \( I_4 \) is finite in holographic CFT\(_{1+1}\) when we fix three subregions\footnote{If we fix two subregions with the other two subregions arbitrarily chosen, the divergence behaviour of $I_4$ is exactly the same as $I_3$ and could be UV divergent.}, but infinite in higher dimensions. This reveals the fundamental difference between the multipartite entanglement structure in holographic CFT\(_{1+1}\) and higher dimensions.

\subsection{Overview}

\noindent We have derived the upper bound of holographic conditional mutual information in general cases. {As we have proposed in Section \ref{sec1}}, the ultimate goal is to find the upper bound of any linear combination of the holographic entanglement entropy, with \( p \) regions fixed and \( q \) regions arbitrarily chosen, {to understand the underlying multipartite entanglement structure}. We will give this question a more detailed answer in an upcoming work \cite{Ju:2025}. For now, we have only studied the upper bound of $-I_3$ with two subregions fixed and analyzed the tripartite entanglement structure. A natural generalization would be analyzing four-partite or five-partite entanglement structure, and the first step is to find a measure of multipartite entanglement which we can analyze using similar methods. A first choice should be $I_4$ and $I_5$. In the following we will give a comprehensive analysis showing why they are the best choices here. 

The properties that this measure should have are listed as follows.
\begin{itemize}
    \item 1. It must be a linear combination of entanglement entropies so that the RT formula is still valid, and we can analyze it in a similar (but not the same) method as in the previous section.
    \item 2. It must characterize the \( n \)-partite global entanglement, which means, it is nonzero if and only if the entanglement wedge of these \( n \) regions is fully connected.
    \item 3. It must have a divergence behavior that we have not investigated for now.
\end{itemize}

Only property 3 needs explanation. One can always write down a combination of entanglement entropy such as, in the simplest case, \( S_E \). However, its divergence behavior is known (trivial), which is proportional to the length of \( \partial E \). To cancel this divergence, we can use the mutual information \( I(A:E) \). The divergence behavior of \( I(A:E) \) will not be enlarged when we add the length of \( \partial E \) (endpoints of \( E \) in CFT\(_{1+1}\)). However, the divergence behavior of mutual information is trivial as well—it will diverge when \( E \) and \( A \) have a conjunct boundary, and the divergent term will be proportional to the length (in CFT\(_{2+1}\)) of this conjunct boundary. To cancel this divergence, we have to use \( -I_3 = I(E:AB) - I(E:A) - I(E:B) \), the tripartite information as the remaining part after canceling the UV divergent term in MI. However, as we have found in the last section, \( I_3 \) would diverge as well when the number of intervals of \( E \) tends to infinity. For now, to us, \( I_3 \) also has a known (trivial) divergent behavior. Another example of the combination of entanglement entropy with known divergent behaviour is as follows 
\begin{equation}
I(A:B:C|E) - J_3(ABC) = -I_3(A:B:E) - I_3(AB:C:E),
\end{equation}
where
\begin{equation}
\begin{aligned}
    I(A:B:C|E) &= S_{AE} + S_{BE} + S_{CE} - 2S_E - S_{ABCE}, \\
    J_3(ABC) &= S_A + S_B + S_C - S_{ABC},
\end{aligned}
\end{equation}
are tripartite conditional mutual information \cite{Yang_2009} and tripartite mutual information, respectively. This combination looks like a reasonable generalization of \( I_3 \); however, it has a known divergent behavior the same as \( I_3 \), so we will not investigate it.

The third property strongly implies that we should expand the combination of entanglement entropy in the \( I \) basis \cite{Hubeny:2018ijt,He:2019ttu}:
\begin{equation}
{\mathbf{Q} = \sum_{\mathcal{I} \subset [N]} \nu_{\mathcal{I}} {\mathsf{I}}_{\mathcal{I}},}
\end{equation}
which uses the combination of entanglement entropy of single regions, mutual information between two single regions, and \( I_n \) among \( n \) single regions as the basis to represent the combination \( \mathcal{Q} \). If the first-order term inside this combination is \( S_E \) or \( I(E:\dots) \) or \( I_3(E:\dots:\dots) \), then this combination has known divergent behavior. As a result, the combination with nontrivial behavior for \( n = 4 \) cases is \( I_4(E:A:B:C) \), which is defined as follows.
\begin{equation}\label{I4def}
\begin{aligned}
    I_4(E:A:B:C) &= I_3(E:A:B) + I_3(E:A:C) - I_3(E:A:BC) \\
    &= S_A + S_B + S_C + S_E - S_{AB} - S_{AC} - S_{BC} - S_{AE} - S_{BE} - S_{CE} \\
    &\quad + S_{ABC} + S_{ABE} + S_{ACE} + S_{BCE} - S_{ABCE} \\
    &= I_3(A:B:C) - I(A:B|E) + I(A:B|CE).
\end{aligned}
\end{equation}

From this definition, one can easily verify that \( I_4 \) is symmetric with respect to regions \( A \), \( B \), \( C \), and \( E \), and if the entanglement wedge of region \( ABCE \) is not fully connected, we have \( I_4 = 0 \), which satisfies property 2. Also, the divergence of $CMI$ cancels out so $I_4$ indeed has a divergence behaviour that we have not known yet, which property 3 demands.

We are going to analyze the case where regions \( A \), \( B \), and \( C \) are fixed, while region \( E \) is arbitrarily chosen. It is known in holography \cite{Erdmenger:2017gdk} that \( I_4 \) has no definite sign. As a result, \( I_4 \) has both nontrivial positive upper bound and negative lower bound. However, we can prove that their addition is \( 2I_3(A:B:C) \). A keen observation is to substitute region \( E \) with \( F \), where \( F \) purifies \( ABCE \), \ie, \( ABCEF \) is the entire boundary. We have

\begin{equation}\label{I4min}
\begin{aligned}
    I_4(E:A:B:C) &= I_3(A:B:C) - I(A:B|E) + I(A:B|CE) \\
                 &= I_3(A:B:C) - I(A:B|CF) + I(A:B|F) \\
                 &= 2I_3(A:B:C) - I_4(F:A:B:C).
\end{aligned}
\end{equation}

As a result, for an \( E \) which reaches the upper bound of \( I_4 \), the corresponding \( F \) will reach the lower bound, which is \( 2I_3(A:B:C) - I_4 \). When \( I_3(A:B:C) = 0 \), they are opposite numbers. Thus obtaining only one of the upper and lower bounds of \( I_4 \) is sufficient.

Can we draw MPT diagrams as we did in Section \ref{sec3} to analyze \( I_4 \)? Unfortunately, the answer is no, because given a finite \( m \), the configuration with maximum \( I_4 \) is not unique. Unlike the case we presented in Section \ref{sec2.1}, the maximum value of \( I_4 \) is not a peak but a plateau on the mountain, which means \( I_4 \) remains maximum in a set of configurations rather than in a unique MPT diagram. The multi-entanglement phase transition rule does not hold true in the \( I_4 \) case. Therefore we have to employ the disconnectivity condition to find the upper bound configuration for $I_4$.

\subsection{Disconnectivity condition}

\noindent The disconnectivity condition in \( I_4 \) is a stronger generalization of that in \( I_3 \), which states that only analyzing the disconnected case which satisfies
\begin{equation}\label{I4disconnect}
    I(E:AB) = I(E:AC) = I(E:BC) = 0,
\end{equation}
is enough to evaluate the upper bound of \( I_4 \). In other words, for a connected configuration of \( E \) which does not satisfy formula (\ref{I4disconnect}), there always exists a disconnected configuration \( E \) whose \( I_4 \) is larger.

\begin{figure}[h]
\centering
\includegraphics[width=15cm]{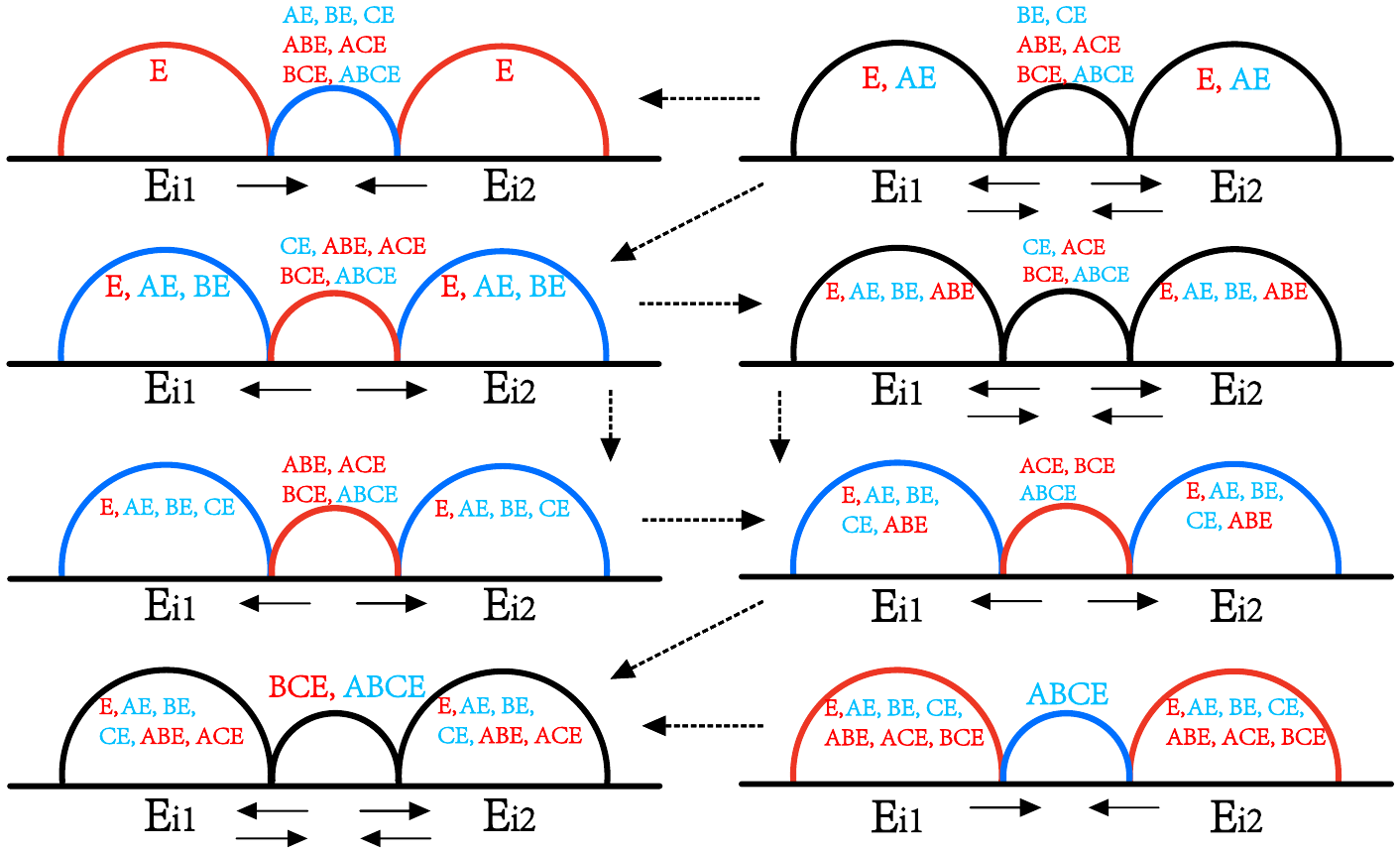}
\caption{Proof of the \( I_4 \) version of the disconnectivity condition. The labels of subregions $AE$, $BE$, etc. near each curve indicate the subregions whose RT surface involves this curve. The colors red and blue in the labels of the subregions $AE$, $BE$, etc. represent the positivity of this term in \( I_4 \), and the color of the semicircle represents the positivity of the summation of all RT surfaces which include the length of this semicircle (black means that the length of this circle cancels out in \( I_4 \)). The black arrows below the endpoints of gaps between \( E_{i1} \) and \( E_{i2} \) represent the direction of the modification of those endpoints that enlarges the \( I_4 \). The dashed arrows between each diagram represent the direction of increasing \( I_4 \).}
\label{DISI4}
\end{figure}

To prove this theorem, we split the interval \( E_i \) which connects to at least one of \( A \), \( B \), and \( C \) in the entanglement wedge of \( ABE \) or \( BCE \) or \( ACE \) into two pieces, \( E_{i1} \) and \( E_{i2} \), as we did in the last section. Figure \ref{DISI4} presents this whole procedure.

In this figure, each of the eight diagrams shows a configuration where interval \( E_i \) is split into two intervals \( E_{i1} \) and \( E_{i2} \) with a gap region in between. When the position of the endpoints of the gap is modified, only the area of three minimal surfaces (semicircles in the figure) changes to affect \( I_4 \). According to formula (\ref{I4def}), there are eight terms in \( I_4 \) which contain \( E \): \( S_{E} \), \( S_{AE} \), \( S_{BE} \), \( S_{CE} \), \( S_{ABE} \), \( S_{BCE} \), \( S_{ACE} \), and \( S_{ABCE} \), respectively. Each of those RT surfaces contains the semicircles in those figures and are labeled on those semicircles. The colors red and blue in the labels of $AE$, $BE$, etc. represent the positivity of this term in \( I_4 \), and the color of the semicircle represents the positivity of the summation of all RT surfaces which include the length of this semicircle. The color black means that the length of this circle cancels out in \( I_4 \). The black arrows below the endpoints of gap regions between \( E_{i1} \) and \( E_{i2} \) represent the direction of the modification of those endpoints that enlarges the \( I_4 \). The dashed arrows between each diagram represent the direction of increasing \( I_4 \).

 Let us analyze these configurations in Figure \ref{DISI4} one by one. The first configuration is when \( E_i \) connects with \( A \) or \( B \) or \( C \) in all entanglement wedges except the entanglement wedge of \( E \). Now we use diagram (a.b) to denote the subfigure in the a-th line and b-th row in Figure \ref{DISI4}. As shown in diagram (1.1), splitting \( E_i \) into \( E_{i1} \) and \( E_{i2} \) with a gap region between them will decrease \( I_4 \), as one should decrease the length of the gap to make the length of the blue curve decrease and the length of red curve increase in order to increase \( I_4 \). The second configuration is that \( E_i \) only disconnects with \( A \) in the entanglement wedge \( AE \) while connecting with \( A \) or \( B \) or \( C \) in all other entanglement wedges, as shown in diagram (1.2). In this case, moving the endpoints of the gap between \( E_{i1} \) and \( E_{i2} \) will not modify \( I_4 \), so we can enlarge the gap between \( E_{i1} \) and \( E_{i2} \) until the phase transition of the entanglement wedge of \( BE \) occurs to \( E_{i1} \) and \( E_{i2} \) simultaneously and make the entanglement wedge \( BE \) disconnected, which is shown in diagram (1.3). We can split \( E_i \), which disconnects with \( A \) and \( B \) in the entanglement wedges of \( AE \) and \( BE \) while connecting with other entanglement wedges, again. This time, splitting will increase \( I_4 \) to diagram (2.2) or diagram (3.1). Splitting again, diagram (2.2) or diagram (3.1) will lead to diagram (3.2). Splitting diagram (3.2) will lead to diagram (4.1). However, splitting diagram (4.1) will decrease \( I_4 \). As a result, the maximum of \( I_4 \) might be chosen as the configuration between diagram (4.1) and (4.2), which is the phase transition point. In this case, \( E \) disconnects with \( A \), \( B \), and \( C \) in the entanglement wedges of \( ABE \), \( BCE \), and \( ACE \), which satisfies the disconnectivity condition.

From the argument above, we can see that \( I_4 \) in the disconnected configuration reaches the maximum among all diagrams except diagram (1.1), so we are one step away from proving the disconnectivity condition. This step is to rule out the possibility of diagram (1.1) being the diagram with maximum \( I_4 \).  However, this task seems difficult and we will use another way instead. We will try to find the configuration with the minimal value of \( I_4 \) and based on equation (\ref{I4min}), use \( F \) to substitute \( E \) to find the configuration with maximum \( I_4 \).

The procedure of finding the minimal value is exactly the opposite of finding the maximum value. One just has to reverse all arrows in Figure \ref{DISI4}. As we already analyzed before, it is easy to find that there are two configurations which might reach the minimal value: diagram (1.2) and diagram (4.2). Let us analyze diagram (4.2) first. One should enlarge the gap between \( E_{i1} \) and \( E_{i2} \) in order to decrease \( I_4 \). At last, the entanglement wedge of \( ABCE \) could be disconnected and \( I_4 \) reaches its minimal value. However, in this case, intervals \( E_{i1} \) and \( E_{i2} \) disconnect with all entanglement wedges. In formula (\ref{ABEcon}) of Section \ref{sec3}, we have argued that this case will lead to the result that eliminating \( E_{i1} \) and \( E_{i2} \) will not affect the value of CMI. This argument is still valid in the \( I_4 \) case. Until now, we understand that diagram (4.2) leads to a configuration that has zero $I_4$, which would be one of the local minimal values of \( I_4 \).  However, the real lower bound of $I_4$ must be negative. Thus only diagram (2.1) has the right to become the configuration with minimal \( I_4 \). As adjusting the length of the gap will not change \( I_4 \), we can shorten it until a phase transition between diagram (1.1) and (1.2) happens, in which case, \( E \) connects with \( A \) or \( B \) or \( C \) in all entanglement wedges. Therefore, this could be the configuration of minimum $I_4$.

Then we could produce a configuration with maximum $I_4$ from this minimum $I_4$ configuration using equation (\ref{I4min}). We need to substitute \( E \) by the complementary region $F$ into (\ref{I4min}) to find the configuration with maximum \( I_4 \). In this configuration, we denote \( F \) as the region that purifies \( ABCE \); then \( F \) is the collection of gap regions between \( E_i \) and \( E_{i+1} \), and the connectivity of \( AE \), \( BE \), and \( CE \) is equivalent to the disconnectivity of \( BCF \), \( ACF \), and \( ABF \), respectively, as could be checked. From equation (\ref{I4min}), \( F \) must be the region that maximizes \( I_4 \), and the disconnectivity condition (\ref{I4disconnect}) is proven.

\subsection{\( I_4 \) in \(\text{AdS}_3/\text{CFT}_{2}\) and higher-dimensional cases}

\begin{figure}[H]
\centering
\includegraphics[width=13cm]{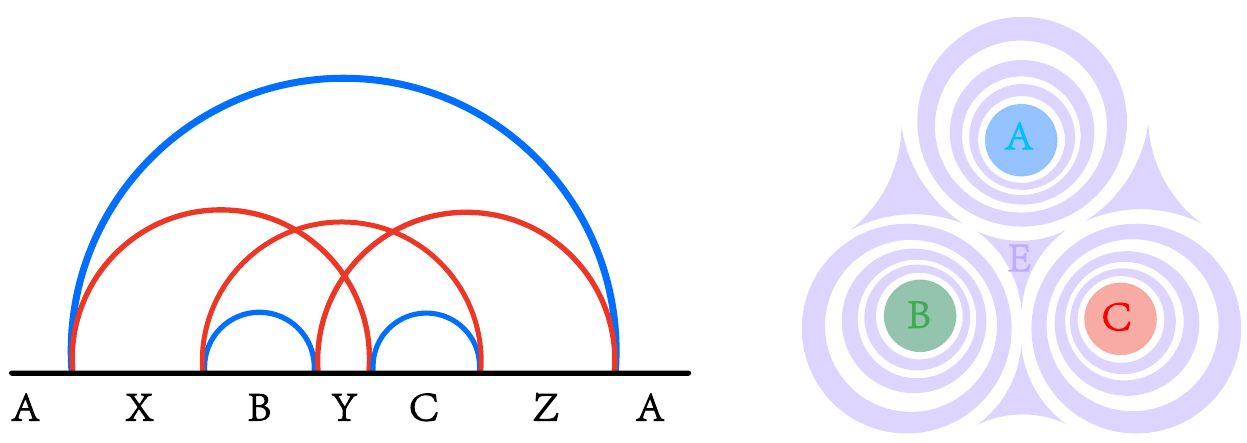}
\caption{Calculation of \( I_4 \) in \(\text{AdS}_3/\text{CFT}_{2}\) and \(\text{AdS}_4/\text{CFT}_{3}\). \( A \), \( B \), and \( C \) are intervals (left) and disks (right), respectively.}
\label{I4calculation}
\end{figure}

\noindent With the disconnectivity condition proved, let us calculate the upper bound of \( I_4 \) in holographic \(\text{CFT}_{1+1}\) and \(\text{CFT}_{2+1}\) directly. Figure \ref{I4calculation} shows the case where \( A \), \( B \), and \( C \) are intervals (disks).

\noindent{\textbf {Upper bound of $I_4$ in \(\text{AdS}_3/\text{CFT}_{2}\).}}

In the case of \(\text{AdS}_3/\text{CFT}_{2}\), there are three gap regions between regions \( A \) and \( B \), \( B \) and \( C \), and \( C \) and \( A \), which we denote as \( X \), \( Y \), and \( Z \), respectively, as shown in Figure \ref{I4calculation}. We are going to put intervals of the region $E$ in these gap regions. Due to the disconnectivity condition (\ref{I4disconnect}), we have 
\begin{equation}\label{I4indiscon}
\begin{aligned}
    I_4(E:A:B:C)&=S_{ABC}+S_E-S_{ABCE}\\
    &=S_{ABC}+\sum_{i \in D^c} S_{E_i} - \sum_{i \in D^c} S_{\text{Gap}_i}.
\end{aligned}
\end{equation}
In the first line of this equation, the disconnectivity condition is used so that only three terms are left. As a result, calculating the upper bound of \( I_4 \) is still a problem of calculating the maximum value of \( \sum_{i} S_{E_i} - \sum_{i} S_{\text{Gap}_i} \), and the disconnectivity condition constrains its value from being too large.

Due to the fact that \( I(E:AB)=0 \) from the disconnectivity condition, the area of the disconnected RT surface (true RT surface) of region \( ABE \) must be smaller than the area of the connected ``RT surface'' (fake RT surface) of region \( ABE \) as follows
\begin{equation}
    S_{AB}+\sum_{E_i\in X} S_{E_i} \leq \sum_{ \text{Gap}_i \in X} S_{\text{Gap}_i} + S_{AXB},
\end{equation}
where the RT surface of region \( AXB \) is marked by the one of the red semicircle in the left of Figure \ref{I4calculation}. This inequality prevents region \( AB \) from connecting with the intervals inside the gap region \( X \).

Using \( I(E:AC)=I(E:BC)=0 \) from the disconnectivity condition, we can get the other two inequalities as follows
\begin{equation}
    S_{AC}+\sum_{E_i\subset Z} S_{E_i} \leq \sum_{ \text{Gap}_i \subset Z} S_{\text{Gap}_i} + S_{CZA},
\end{equation}
\begin{equation}
    S_{BC}+\sum_{E_i\subset Y} S_{E_i} \leq \sum_{ \text{Gap}_i \subset Y} S_{\text{Gap}_i} + S_{BYC}.
\end{equation}
Adding these three inequalities together, we can get
\begin{equation}
\begin{aligned}
    &S_{AB}+S_{BC}+S_{AC}+\sum_{i} S_{E_i} - \sum_{i} S_{\text{Gap}_i} \leq S_{AXB} + S_{CZA} + S_{BYC}, \\
    &I_4 \leq S_{AXB} + S_{CZA} + S_{BYC} - (S_{AB} + S_{BC} + S_{AC}) + S_{ABC},
\end{aligned}
\label{FinI4AdS3}
\end{equation}
where we have used formula (\ref{I4indiscon}) in the second line. Finally, we get the upper bound of \( I_4 \). Specifically, on the left side of Figure \ref{I4calculation}, this upper bound is the summation of red curves minus the summation of blue curves plus $I_3(A:B:C)$. We can find that the UV divergent terms of the red curves and the blue curves cancel out. As $I_3(A:B:C)$ is an IR term, it will only finitely affect the value of \( I_4 \). As a consequence, the upper bound of \( I_4 \) is definitely a finite value without UV divergence.

\noindent{\textbf {Upper bound of $I_4$ in higher dimensions.}}

Let us calculate the maximum value of \( I_4 \) in higher dimensions. As shown on the right side of Figure \ref{I4calculation}, similar to Figure \ref{Highdim}, \( E \) is chosen as strips (annuli) which surround regions \( A \), \( B \), and \( C \). Formula (\ref{I4indiscon}) is still valid. The only difference is the constraint on \( \sum_{i} S_{E_i} - \sum_{i} S_{\text{Gap}_i} \) imposed by the disconnectivity condition, which is
\begin{equation}\label{I4disconstraint}
    S_{AB}+\sum_{i} S_{E_i} \leq \sum_{i} S_{\text{Gap}_i} - S_{\text{Gap}_m} + S_{C \text{Gap}_m},
\end{equation}
where \( S_{\text{Gap}_m} \) is the entanglement entropy of the gap region adjacent to \( C \), which will be infinitely thin when \( m \) tends to infinity. Rewriting the last equation as
\begin{equation}\label{I4disconstraint2}
\begin{aligned}
    S_{ABC}+\sum_{i} S_{E_i} - \sum_{i} S_{\text{Gap}_i} &\leq S_{C} - S_{AB} + S_{ABC}, \\
    I_4 &\leq 2S_C - I(AB:C),
\end{aligned}
\end{equation}
along with the other two disconnectivity conditions $I(E:BC)=I(E:AC)=0$, we have
\begin{equation}
    I_4 \leq \min\left( 2S_C - I(AB:C),\ 2S_B - I(AC:B),\ 2S_A - I(BC:A) \right).
\end{equation}
This upper bound can be saturated in a way similar to the saturation of $-I_3$. The key thing is to make the strips of $E$ surrounding \( A \) and \( B \) not too wide so that the entanglement wedge $EW(ABE)$ is disconnected, but further adding \( C \) will make $EW(ABCE)$ fully connected. Therefore, we have to make those strips of $E$ infinitely close to \( C \). To ensure that $I_4$ diverges at the upper bound configuration, we have to make the number of strips in $E$ approach infinity, and at least one of the strips is infinitely close to \( C \). To achieve this, the very first premise is that there exists a gap region which is adjacent to regions \( A \), \( B \), and \( C \) simultaneously, where we can put region \( E \) inside it. %This could be easily achieved in higher dimensional holography. 
However, in the \(\text{AdS}_{3}/\text{CFT}_{2}\) case, a gap region has only two endpoints, which makes it adjacent to at most \( 2 \) other regions. That is the technical reason why the upper bound of \( I_4 \) is finite in \(\text{AdS}_{3}/\text{CFT}_{2}\). In Section \ref{sec2}, we also find that the divergence behavior of \( \exp(-I_3/2) \) is of order \( m^g \), where \( g \) is the number of the gap regions adjacent to both \( A \) and \( B \). Here, this feature is generalized to the requirement that the gap regions need to be adjacent to all three regions $A$, $B$ and $C$.

How about in higher dimensions? Of course, there are special cases in higher-dimensional holography where there is no gap region adjacent to all of the boundary regions \( A \), \( B \), and \( C \) at the same time, such as the case where \( A \) surrounds \( B \) and isolates \( B \) from \( C \). In those cases, \( I_4 \) will have a finite upper bound. However, for three small convex regions \( A \), \( B \), and \( C \) which are far from each other in higher dimensions, the upper bound of \( I_4 \) could approach \( 2\min(S_A, S_B, S_C) \), which reaches the maximum value of \( I_4 \) in quantum information theory.

This finding reveals the fundamental difference of the four-partite entanglement structure in \(\text{AdS}_{3}/\text{CFT}_{2}\) and higher-dimensional dual CFTs. This is different form the holographic entropy cone \cite{Bao:2015bfa} works, where it is found that the HEC in \(\text{AdS}_{3}/\text{CFT}_{2}\) is the same as in higher-dimensional cases, as long as subregions on the boundary can be chosen as a collection of disconnected intervals. While in our work, regions \( A \), \( B \), and \( C \) can be chosen as disconnected intervals, but as long as they are a collection of a finite number of intervals, \( I_4 \) will always have a finite upper bound in \(\text{AdS}_{3}/\text{CFT}_{2}\), unlike in higher dimensions.

As in the upper bound configuration of $I_4$, we have all the tripartite and bipartite entanglement among any two or three subsystems in $A,B,C,$ and $E$ vanishing due to the disconnectivity conditions. Therefore, \( I_4 \) could serve as the measure of four-partite global entanglement. We can see that in \(\text{AdS}_{4}/\text{CFT}_{3}\), any degrees of freedom localized in three different locations are participating in the four-partite entanglement, while in \(\text{AdS}_{3}/\text{CFT}_{2}\), the four-partite entanglement these three fixed subregions participate in is sparse\footnote{In \(\text{AdS}_{3}/\text{CFT}_{2}\), if we fix two small and distant subregions with the other two unfixed, we could still find a configuration of these two unfixed subregions that makes $I_4$ reach the quantum information theoretical upper bound $2\min (S_A,S_B)$.}. This different behavior in different dimensions reflects the underlying entanglement structures in the dual boundary systems.
\subsection{Upper bound of $-I_5$}

\noindent After we have finished the discussion of \( I_4 \), let us talk about \( I_5 \) briefly. \( I_5 \) is defined as 
\begin{equation}\label{I5formula}
\begin{aligned}
    I_5(A:B:C:D:E) &= I_4(A:B:C:D) + I_4(A:B:C:E) - I_4(A:B:C:DE) \\
    &= -( I(A:BCDE) - I(A:E) - I(A:C|BE) - I(A:B|DE) \\&\,\,\,\,\,\,- I(A:D|CE) )+I_4(A:B:C:D) \\
    &\geq -(I(A:BCDE) - I(A:E)-I_4(A:B:C:D) ) \\
    &\geq -2S_A+I_4(A:B:C:D),
\end{aligned}
\end{equation} where in the fourth line we have used the strong subadditivity inequality and in the fifth line, Araki-Lieb inequality is utilized.
We care about the upper bound of \( -I_5 \) most, and it is \( 2S_A \) in quantum information theory (we consider the case with vanishing $I_4(A:B:C:D)$ for simplicity).

\begin{figure}[H]
\centering
\includegraphics[width=13cm]{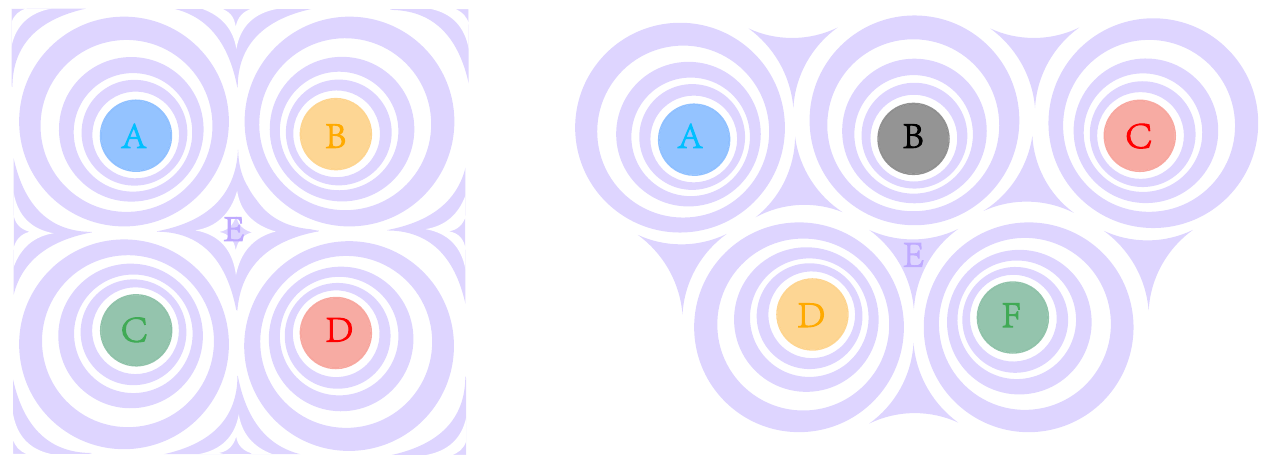}
\caption{Configurations of $E$ which make $I_5(A:B:C:D:E)$ (left) and $I_6(A:B:C:D:E:F)$ (right) approach $2\min(S_A,S_B,S_C,S_D)$ and $2\min(S_A,S_B,S_C,S_D,S_F)$ respectively.}
\label{I5I6}
\end{figure}

In \(\text{AdS}_{3}/\text{CFT}_{2}\), the upper bound of $-I_5$ is finite with three or four subregions kept fixed because it is a linear combination of $I_4$ as could be seen from the first line of (\ref{I5formula}).  Can holographic \( -I_5 \) in higher dimensions reach this quantum information theoretical upper bound of $2S_A$? The answer is yes. One could draw five disks ($A$ is the smallest one) with \( E \) being curved strips surrounding them, as shown in Figure \ref{I5I6}, making the mutual information between \( E \) and any combination of three regions among regions \( A,~B,~C,~D \) vanish. Repeat the procedure (\ref{I4disconstraint}) and (\ref{I4disconstraint2}) for the calculation of $I_5$, and one can get the upper bound of \( -I_5 \) being exactly \( 2S_A \), which is the same as in quantum information theory. Though the disconnectivity condition might not be easily proven in the \( I_5 \) case, as we have already found the case where \( I_5 \) reaches its theoretical upper bound in quantum information theory, we do not have to prove it anymore. The generalization of this procedure to the calculation of the upper bound of $I_6$ will be left to future work.

Finally, let us make some remarks on the divergence behavior of $(-1)^n I_n$ when more than one subregion could be tuned.
One can make \( I_4 \) diverge in \(\text{AdS}_{3}/\text{CFT}_{2}\) if we only fix regions \( A \) and \( B \) and arbitrarily choose regions \( C \) and \( E \). One can easily find that due to formula (\ref{I4def}), its divergence behavior will be equivalent to the divergence behavior of CMI in \(\text{AdS}_{3}/\text{CFT}_{2}\) with \( AB \) fixed and \( E \) arbitrarily chosen. This is just an example for analyzing the divergence behavior of the linear combination of holographic entanglement entropy. We can summarize the divergence behaviour of $I_n$ in a table as follows.

\begin{table}[H]
    \centering
    \begin{tabular}{|c|c|c|c|}
 \hline
 \text{Combination}    & tuning \( E \) & tuning \( E \), \( A \) & tuning \( E \), \( A \), \( B \) \\
 \hline
    \( S_E/2 \) & \( \int \partial E \log\frac{1}{\epsilon} \) & - & - \\
    \( I(E:A)/2 \) & \( (\int \partial A) \log\frac{1}{\epsilon} \) & \(\int \partial E \log\frac{1}{\epsilon} \) & - \\
    \( -I_3(E:A:B)/2 \) & \( (\int \partial G_{AB}) \log n_E \) & \( (\int \partial B) \log n_E \) & \( \int \partial E \log n_E \) \\
    \( I_4(E:A:B:C)/2 \) & \((\int\partial G_{ABC}) \log n_E\) & \( (\int\partial G_{BC}) \log n_E \) & \( (\int \partial C) \log n_E \) \\
    \( -I_5(E:A:B:C:D)/2 \) & \((\int\partial G_{ABCD}) \log n_E\) &\( (\int\partial G_{BCD}) \log n_E \)& \( (\int\partial G_{CD}) \log n_E \) \\
 \hline
    \end{tabular}
    \caption{Divergence behavior of different combinations of holographic entanglement entropy. \( n_E \) is the number of the intervals of the tuning region. \( \int \partial A \) represents the measure of \( \partial A \). \( G_{AB} \) represents the number of gap regions adjacent to both \( A \) and \( B \). $G_{ABC}$ represents the gap region adjacent to $ABC$, which does not exist in \(\text{AdS}_{3}/\text{CFT}_{2}\), resulting in the upper bound being finite.}
    \label{cc}
\end{table}

We will further investigate this table in future work \cite{Ju:2025}. We only make some comments here. First, if we could tune regions \( E \) and \( A \), the divergence behavior of \( I_{n} \) will be the same as the divergence behavior of \( I_{n-1} \) in the case where we only tune region \( E \), and it remains the same if we could tune \( E \), \( A \), and \( B \) or even more regions. As a result, we can come to the conclusion in \(\text{AdS}_{3}/\text{CFT}_{2}\) that any two distant small subregions are highly multipartite entangled while any three distant small subregions are not. Moreover, as both the upper bound and lower bound of \( I_4 \) with three fixed subregions are finite in \(\text{AdS}_{3}/\text{CFT}_{2}\), the upper bound and lower bound of \( I_{n>4} \) where three subregions are fixed will also be finite as they can be written as a linear combination of \( I_4 \).

\section{Conclusion and discussion}\label{sec6}
\noindent In this paper, we calculate the upper bound of \( n \)-partite information in holographic theory for the cases \( n = 3, 4, 5 \) with one region arbitrarily chosen and the other regions fixed, in order to analyze the holographic multipartite entanglement structure. Though \( n \)-partite entanglement might not serve as a perfect measure for multipartite entanglement in all circumstances, the case where \( I_n \) reaches its information theoretical upper bound, as we have seen in this paper, strongly suggests the presence of a large amount of multipartite entanglement as all $k$-partite entanglement among $k$ of the $n$ subregions with $k<n$ vanishes at this upper bound configuration. It is necessary to review our results and conclusions as follows.

In Sections \ref{sec2} and \ref{sec3}, we evaluate the upper bound of \( I(A:B|E) \) or $-I_3(A:B:E)$ with \( A \), \( B \) fixed and \( E \) arbitrarily chosen; the final result is $-I_3(A:B|E)\leq 2EoSP(A:B)-I(A:B)$, where $EoSP(A:B)$ is the entanglement of state-constrained  purification that we have defined. \( I_3 \) is considered an IR term, which diverges when the number of intervals in region \( E \) tends to infinity, with its upper bound approaching its maximum value in quantum information theory. This reveals that any two local subregions \( A \) and \( B \), even far apart from each other, can fully participate in tripartite entanglement along with another region \( E \). From the analysis in the wormhole case, we find that the throat of the wormhole encodes the tripartite entanglement information among any two small subregions each sitting on a different boundary and another chosen region. 

We conclude that there is no pure bipartite entanglement such as Bell pairs existing in holographic theory; all bipartite entanglement between two regions emerges from the tripartite entanglement among their subregions. Furthermore, in the higher-dimensional case, it is found that the tripartite entanglement structure is deeply related to the convexity of the region: a concave region with negative \( K \) (the trace of the extrinsic curvature of its boundary) cannot contribute all its degrees of freedom to the participation in the tripartite entanglement with another region outside its convex hull.

For \( n = 4 \), we have proved that analyzing only the disconnectivity configurations which satisfy (\ref{I4disconnect}) is sufficient for evaluating the upper bound of \( I_4 \). Using this theorem, we showed that the upper bound of \( I_4 \) is finite in \(\text{AdS}_{3}/\text{CFT}_{2}\) but infinite in higher dimensions. This reveals the fundamental difference in the multipartite entanglement structure between \(\text{AdS}_{3}/\text{CFT}_{2}\) and higher-dimensional cases. Specifically, four-partite entanglement is sparse in \(\text{AdS}_{3}/\text{CFT}_{2}\) for three fixed small and distant subregions, but in higher dimensions, any three local subregions far apart from each other are highly four-partite entangled; any tripartite entanglement among three regions emerges from four-partite entanglement among four subregions inside them.

For \( n = 5 \), the situation is quite similar to the \( I_4 \) case. One difference is that its upper bound remains finite in \(\text{AdS}_{3}/\text{CFT}_{2}\) when we fix three regions and arbitrarily choose the other two regions. However, when \( n \geq 6 \), we do not know the exact upper bound of \( I_n \) in both quantum information theory and holography; the disconnectivity condition might need to be analyzed in detail. However, we can always construct a configuration which make $I_{n\geq 6}$ approach $2\min (S_A,S_B,...)$, as shown in right side of Figure \ref{I5I6}. This indicates that in higher dimensions, $n-1$ local subregions (with vanishing $n-1$-partite entanglement) are highly $n$-partite globally entangling. 

Our ultimate goal is to find the upper (lower) bound of any combinations of holographic entanglement entropy given \( n \) regions fixed and other regions arbitrarily chosen. Analyzing the upper bound of the combinations with several regions fixed, as we did, could reveal more information about holographic entanglement structures, such as the differences between the entanglement structures in different dimensions.
To achieve this, we can rewrite this combination in the \( I \)-basis \cite{Hubeny:2018ijt,He:2019ttu}, and the divergence behavior of it is the same as that of the leading-order term. However, the cases where the divergences of the leading-order terms cancel out might lead to nontrivial behavior of the upper bound, as a result, any 3-balanced \cite{Hubeny:2018ijt,He:2019ttu,Hubeny:2018trv} combination of holographic entanglement entropy will be finite when we fix at least three subregions $A,~B,~C$. 
%In the special case where the combination we choose is one of the inequalities in holographic entropy cone works \cite{}, this bound will be exactly zero. Moreover, inequalities which HEC works investigate are cases where no regions are fixed and all regions can be arbitrarily chosen, so the problem of finding HEC can be viewed as a special situation in our case. 

We can further rewrite the upper bound we find in (\ref{FinalCMI}) and (\ref{FinI4AdS3}) as a form of \textit{unbalanced holographic entropy} inequality. Normally, holographic entropy inequality is always balanced \cite{Hubeny:2018trv,Hubeny:2018ijt}, however, as we can choose some region $E$ as a combination of infinite many intervals, unbalanced holographic entropy inequalities could emerge. We will further investigate it in future work \cite{Ju:2025}.

\section*{Acknowledgement}

We would like to thank Bart{\l}omiej Czech and Joydeep Naskar for theirhis valuable discussions. This work was supported by Project 12347183, 12035016 and 12275275 supported by the National Natural Science Foundation of China.

\appendix
\section{The derivation for the polynomial that generates the maximum value of CMI}
\noindent A method to obtain the maximum value of \( I(A:B|E) \) for each integer number \( n \) (the number of intervals of \( E \)) analytically is presented in the following.
We suppose that the fixed regions \( A \) and \( B \) are non-adjacent single intervals as considered in Section \ref{sec3.1}.
Without lost of generality, we only consider symmetric configurations, i.e. the two intervals \( A \) and \( B \) have the same length \( l \), and the length of interval between them is set to be \( 1 \).
Any other configuration can be transformed into a symmetric one by a conformal transformation which does not change the value of \( I(A:B|E) \).

We first analyze the case that the \( m \) intervals of \( E \) locate only between \( A \) and \( B \).
As the MPT diagrams in Figure \ref{MPT1} suggest, in such symmetric configurations, the intervals of \( E \) must also be symmetric with respect to \( A \) and \( B \) when \( I(A:B|E) \) reaches its upper bound, since the phase transition configurations are symmetric.

For instance, if \( m=1 \), let the length of \( E \) be \( e_1 \) and the length of the interval between \( A \) and \( E \) be \( d_1 \) (the length between \( B \) and \( E \) is also \( d_1 \) due to the symmetric configuration and the symmetric phase transition configuration), the phase transition condition \( A/E \) (and \( B/E \)) becomes 
\begin{align}
    le_1=d_1(l+d_1+e_1),\\
    e_1+2d_1=1,
\end{align}
and can be solved to be 
\begin{align}\label{TP}
\left\{\begin{matrix}
d_1=\frac{1}{2}(1+3l-\sqrt{1+2l+9l^2})\\
e_1=-3l+\sqrt{1+2l+9l^2}.
\end{matrix}\right.
\end{align}
According to the multi-entanglement phase transition rule, \( I(A:B|E)\) takes its maximum at the transition point \eqref{TP} such that the upper bound for \( m=1 \) is 
\begin{equation}
    2\log\frac{l^2 e_1}{d_1^2(2l+1)}=2\log\frac{1 + 2 l + 3 l^2 + (1 + l) \sqrt{1 + 2 l + 9 l^2}}{2(2l+1)}.
\end{equation}
The fraction inside \( \log \) happens to be the positive root of the following quadratic polynomial equation
\begin{equation}
    (2 l + 1) x^2 - (3 l^2 + 2 l + 1) x - l^2 = 0.
\end{equation}
Note that the cross ratio of \( AB \) is 
\begin{equation}
    CR=\frac{l^2}{2l+1}.
\end{equation}
Thus the above equation can then be rearranged to be 
\begin{equation}\label{poly1}
    x^2 - (3 CR + 1) x - CR = 0,
\end{equation}
which is consistent with the previous result that the upper bound depends only on the cross ratio of \( AB \).

Therefore, the maximum value of \( I(A:B|E)\) is a solution of the polynomial equation (\ref{poly1}) for $m=1$. For general $m>1$ and for the case where $E$ lives in both gap regions between and outside $A$ and $B$, the upper bound \( I(A:B|E)\) is more complicated to get. However, we expect for these cases, the maximum value of \( I(A:B|E)\) is still a zero point of a certain polynomial for each $n$. In the following, we will use an induction method to get these generating polynomials for more general cases.

\subsection{Generating polynomials for $m>1$ in the case of $E$ living in one gap region. }

\begin{figure}[H]
	\centering
	\includegraphics[scale=0.8]{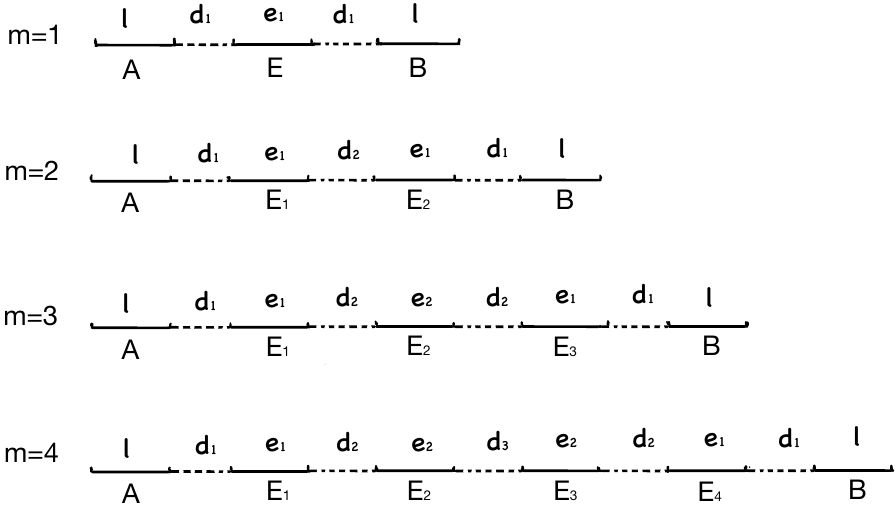}
	\caption{An illustration for the definition of the lengths \(e_1, e_2, ...\) and \(d_1, d_2, ...\) of the intervals in the case of \(E\) living in one gap region. According to the phase transition condition, the intervals are arranged to be reflection symmetric.}\label{interval1}
\end{figure}

\noindent For general integer \( m \geq 1 \), we could make a list of the phase transition configuration conditions and the upper bound of \( \exp\left( \frac{I(A:B|E)}{2} \right) \) , see Table \ref{table1}. Here $l$ is the length of region $A$ and $B$. $e_i$ represents the length of the $i$-th interval of $E$ counted from left. $d_i$ is the length of the $i$-th interval in the gap region between $A$ and $E_1$, $E_1$ and $E_2$, ... as shown in Figure \ref{interval1}.
\begin{table}[H]
	\resizebox{\textwidth}{!}{%
		\begin{tabular}{|c|c|l|}
			 \hline
			\multicolumn{1}{|c|}{\(m\)} & \multicolumn{1}{l|}{Upper bound of \(\exp\left( \frac{I(A:B|E)}{2} \right)\)} & Phase transition condition                                                                                                                                                                                                                                                                                                        \\ \hline
			1                          & \( \frac{l^2 e_1}{(2l+1)d_1^2} \)                                   & \(\frac{l e_1}{d_1(l+d_1+e_1)}=1\), \(2d_1+e_1=1\)                                                                                                                                                                                                                                                                                \\ \hline
			2                          & \( \frac{l^2 e_1^2}{(2l+1)d_1^2 d_2} \)                             & \(\frac{l e_1}{d_1(l+d_1+e_1)}=1\), \(\frac{(l+d_1+e_1) e_1}{d_2(l+d_1+e_1+d_2+e_1)}=1\), \(2d_1+d_2+2e_1=1\)                                                                                                                                                                                                                     \\ \hline
			3                          & \( \frac{l^2 e_1^2 e_2}{(2l+1)d_1^2 d_2^2}\)                        & \begin{tabular}[c]{@{}l@{}}\(\frac{l e_1}{d_1(l+d_1+e_1)}=1\), \(\frac{(l+d_1+e_1) e_2}{d_2(l+d_1+e_1+d_2+e_2)}=1\), \\ \(\frac{(l+d_1+e_1+d_2+e_2) e_1}{d_2(l+d_1+e_1+d_2+e_2+d_2+e_1)}=1\), \(2d_1+2d_2+2e_1+e_2=1\)\end{tabular}                                                                                               \\ \hline
			4                          & \( \frac{l^2 e_1^2 e_2^2}{(2l+1)d_1^2 d_2^2 d_3} \)                 & \begin{tabular}[c]{@{}l@{}}\(\frac{l e_1}{d_1(l+d_1+e_1)}=1\), \(\frac{(l+d_1+e_1) e_2}{d_2(l+d_1+e_1+d_2+e_2)}=1\), \\ \(\frac{(l+d_1+e_1+d_2+e_2) e_2}{d_3(l+d_1+e_1+d_2+e_2+d_3+e_2)}=1\), \(\frac{(l+d_1+e_1+d_2+e_2+d_3+e_2) e_1}{d_2(l+d_1+e_1+d_2+e_2+d_3+e_2+d_2+e_1)}=1\), \\ \(2d_1+2d_2+d_3+2e_1+2e_2=1\)\end{tabular} \\ \hline
			\(\cdots\)                 & \(\cdots\)                                                          & \(\cdots\)                                                                                                                                                                                                                                                                                                                       
		\end{tabular}%
	}
	\caption{Phase transition configuration conditions of the maximum CMI configuration and the corresponding maximum value of CMI for $m=1,2,3,4$ in the case of $E$ living in the gap region between $A$ and $B$.}\label{table1}
\end{table}
As can be seen from Table \ref{table1}, with \( m \) intervals of \( E \) between \( A \) and \( B \), the phase transition condition is a system of \( m+1 \) quadratic equations with \( m+1 \) variables of \( e_i \) which are the lengths of regions \( E_i \), and the lengths of all the gaps \( d_i \).
Therefore the solution of each of these variables should be a positive root of a polynomial equation and the upper bounds of \( I(A:B|E) \), which are merely multiplications and divisions of \( e_i \) and \( d_j \), should also be some positive roots of some different polynomial equations.
If an upper bound is indeed a positive root of a polynomial equation, we call the corresponding polynomial as ``generating"  the upper bound.

However, for general \( m>1 \), it is not that easy to find the corresponding polynomial equation directly from the analytic expression of {the CMI at} the maximum value. The explicit maximum value is lengthy for \(m>1\).
Instead, we take another strategy that takes advantage of \(\mathsf{Mathematica}\) and find this generating polynomial for general $l$ utilizing the patterns in the polynomial for integer values of $l$. For each integer \( m \), we evaluate the exponential of the maximum values for various integer values of \( l \), starting from \( l=1 \).
\(\mathsf{Mathematica}\) gives the maximum value of CMI for each \( l \) in terms of a minimal polynomial that generates the maximum value.
It turns out that the largest exponent of the polynomial is \( m+1 \), if there are \( m \) intervals of \( E \).
Then we assume the general polynomial for integer \( l \) is 
\begin{equation}
    \bar{P}_m(l)=\sum_{j=0}^{m+1}c_j(l)x^j,
\end{equation}
where $c_j(l)$ is the $l$ dependent coefficient of $x^j$ in this polynomial. We then try to induce from the polynomials obtained from these discrete \( l \) a polynomial that is adapted to arbitrary values of \( l \), i.e. find general formula for the sequence \( ( c_j(1), c_j(2), c_j(3), \) \( c_j(4),... ) \) for each \( j=0,1,...,m+1\).
In order to obtain all the \( m+1 \) sequences in \( \bar{P}_m(l) \), we only need to evaluate the polynomials with the first several different integer values of \( l \), e.g \( l=1,2,3,4 \).
We find that the difference between the two terms \( b_j(l)=c_j(l+1)-c_j(l) \), as a sequence of \( l \), are always arithmetic progressions for each of \( j=0,1,...,m+1\) and all the integer \( m \), which implies that \( c_j(l) \)s are quadratic in \( l \).
We emphasize here that although the general formulae for \( c_j(l) \) are obtained by an enumeration and induction of integer values of \( l \), we have checked that they also apply to any non-integer \( l \).

To make the above calculation and induction procedure explicit, we take the case of \( m=7 \) as an example.
The polynomials given by \(\mathsf{Mathematica}\) that generate the upper bounds for \( l=1,2,3,4 \) are 
\begin{equation}
\begin{aligned}
    \bar{P}_7(1)=&3x^8-36x^7+56x^6-196x^5+70x^4-140x^3+0x^2-12x-1&\\
    \bar{P}_7(2)=&5x^8-95x^7+77x^6-539x^5+35x^4-413x^3-49x^2-41x-4&\\
    \bar{P}_7(3)=&7x^8-184x^7+84x^6-1064x^5-70x^4-840x^3-140x^2-88x-9&\\
    \bar{P}_7(4)=&9x^8-303x^7+77x^6-1771x^5-245x^4-1421x^3-273x^2-153x-16&
\end{aligned}
\end{equation}
One can find the general formulae for the sequence \(  c_j(l) \), \( l=1,2,3,...\) for each of \( j=0,1,2...,8\) by noting that \( b_j(l)=c_j(l+1)-c_j(l) \), \( l=1,2,3,...\) are arithmetic progressions. 
The general polynomial as a function of \( l \) is 
\begin{align}
    \bar{P}_7(l)=&(2l+1)x^8-(15l^2+14l+7)x^7-7(l^2-6l-3)x^6-7(13l^2+10l+5)x^5\notag\\
    &-35(l^2-2l-1)x^4-7(11l^2+6l+3)x^3-7(3l^2-2l-1)x^2-(9l^2+2l+1)x-l^2.
\end{align}
After divided by \( 2l+1 \), this polynomial for $m=7$ becomes
\begin{align}\label{P7}
    &x^8-(15CR+7)x^7-(7CR-21)x^6-(91CR+35)x^5-(35CR-35)x^4\notag\\
    &-(77CR+21)x^3-(21CR-7)x^2-(9CR+1)x-CR,
\end{align}
which depends only on the cross ratio \( CR=\frac{l^2}{2l+1} \).

With such kind of calculation, we obtain the polynomial
\begin{equation}
    P_m(CR):=\frac{\bar{P}_m(l)}{2l+1}
\end{equation}
that generate the upper bound of \( I(A:B|E) \) for \( m=0,1,2,3,... \). 
All we have to do next is to find the dependence of coefficients \( c_j \) on \( m \) such that a single polynomial works for all the values of \( CR \) and \( m \) would be obtained.
We first note that \( P_m(CR) \) can be separated into two parts: one with the terms that are proportional to \( CR \) and the other that does not depend on $CR$.
Thus we write
\begin{equation}
	P_m(CR)=f_m(x)+CR\;g_m(x).
\end{equation}
In the following equation, we list \(  P_m(CR) \) {from $m=0$ to $m=7$}. {The first line in \eqref{Pm} is the polynomial of \( m=0 \) which generates the mutual information \(  \exp\left(\frac{I(A:B)}2\right) \), i.e. the three-partite information with region \( E \) absent.}
\begin{equation}\label{Pm}
\begin{aligned}
    &x-CR\\
    &x^2-x-CR (3x+1)\\
    &x^3-2x^2+x-CR(5x^2+2x+1)\\
    &x^4-3x^3+3x^2-x-CR(7x^3+3x^2+5x+1)\\
    &x^5-4x^4+6x^3-4x^2+x-CR(9x^4+4x^3+14x^2+4x+1)\\
    &x^6-5x^5+10x^4-10x^3+5x^2-x-CR(11x^5+5x^4+30x^3+10x^2+7x+1)\\
    &x^7-6x^6+15x^5-20x^4+15x^3-6x^2+x-CR(13x^6+6x^5+55x^4+20x^3+27x^2\\&+6x+1)\\
    &x^8-7x^7+21x^6-35x^5+35x^4-21x^3+7x^2-x-CR(15x^7+7x^6+91x^5+35x^4\\&+77x^3+21x^2+9x+1).
\end{aligned}
\end{equation}
Note that the last line in \eqref{Pm} is just an rearrangement of \eqref{P7}.

It is apparent that the coefficients in \( f_m(x) \) form a variant version of the Pascal's Triangle in which the {elements} in the odd-numbered columns of the original Pascal's Triangle are replaced with their opposite numbers, while the ones in the even-numbered columns remain unchanged.
Therefore, we deduce 
\begin{equation}\label{fn}
	f_m(x)= x(x-1)^m
\end{equation}
according to the binomial theorem.

As for \( g_m(x) \), the coefficients read 
\begin{equation}\label{Coeffgn}
	\begin{matrix}
		1&  &  &  &  &  &  & \\
		3&  1&  &  &  &  &  & \\
		5&  2&  1&  &  &  &  & \\
		7&  3&  5&  1&  &  &  & \\
		9&  4&  14&  4&  1&  &  & \\
		11&  5&  30&  10&  7&  1&  & \\
		13&  6&  55&  20&  27&  6&  1& \\
		15&  7&  91&  35&  77&  21&  9& 1
	\end{matrix}
\end{equation}
and we denote the coefficient at row \(m\) and column \(q\) as \(g_m^q\) with \(m=0,1,2,...\), \(q=1,2,3,...,m+1\).
We find that there is also a {pattern} behind these numbers.
All the even-numbered columns are the same as that of Pascal's Triangle, while any number (except the 1s in the first line of each column) in an odd-numbered column is the sum of two numbers in the next column: the one lies in the same row and the one in the row below, i.e.\footnote{Since \(2j+1\) may be larger than \(m\) when \(j=[\frac{m}{2}]\), we define \(C_{m}^{m+1}=0\).}
\begin{equation}\label{godd}
	 g_m^{2j+1}=g_m^{2j+2}+g_{m+1}^{2j+2}=C_{m}^{2j+1}+C_{m+1}^{2j+1}, \;\;\;\; j=0,1,2,...,[\frac{m}{2}],
\end{equation}
where \([\frac{m}{2}]\) denotes the round down of \(\frac{m}{2} \).
Using this {pattern}, we find 
\begin{equation}\label{gn}
	\begin{aligned}
		g_m(x)&=\sum_{j=0}^{[\frac{m}{2}]}g_m^{2j+1}x^{m-2j}+\sum_{j=0}^{[\frac{m-1}{2}]}g_{m}^{2j+2}x^{m-2j-1}\\
		&=\sum_{j=0}^{[\frac{m}{2}]}(C_{m}^{2j+1}+C_{m+1}^{2j+1})x^{m-2j}+\sum_{j=0}^{[\frac{m-1}{2}]}C_{m}^{2j+1}x^{m-2j-1}\\
		&=\sum_{j=0}^{[\frac{m}{2}]}(C_{m}^{2j+1}+C_{m+1}^{2j+1}-C_{m}^{2j})x^{m-2j}+\left(\sum_{j=0}^{[\frac{m}{2}]}C_{m}^{2j}x^{m-2j}+\sum_{j=0}^{[\frac{m-1}{2}]}C_{m}^{2j+1}x^{m-2j-1}\right)\\
		&=\sum_{j=0}^{[\frac{m}{2}]}2C_{m}^{2j+1}x^{m-2j}+(x+1)^m\\
		&=(x+1)^{m+1}-x(x-1)^m
	\end{aligned}	
\end{equation}

Combining \eqref{fn} and \eqref{gn}, we arrive at our final expression for \( P_m(CR) \) for $m$ intervals of $E$ living in the gap region between $A$ and $B$. 
\begin{equation}
	P_m(CR)=(1+CR)x(x-1)^m-CR (x+1)^{m+1}.
\end{equation}   
It is checked that for general $m$, the zero point of this polynomial exactly coincides with the value of $\exp \left(\frac {I(A:B|E)}2\right)$ which has been obtained using the iterative method.
%%%%%%%%%%%%%%%%%%%%%%%%%%%%%%%%    
%%%%%%%%%%%%%%%%%%%%%%%%%%%%%%%%
 
\subsection{Generating polynomials for the case of $E$ living in both gap regions: with $m$ odd. }
  
\begin{figure}[H]
	\centering
	\includegraphics[scale=0.8]{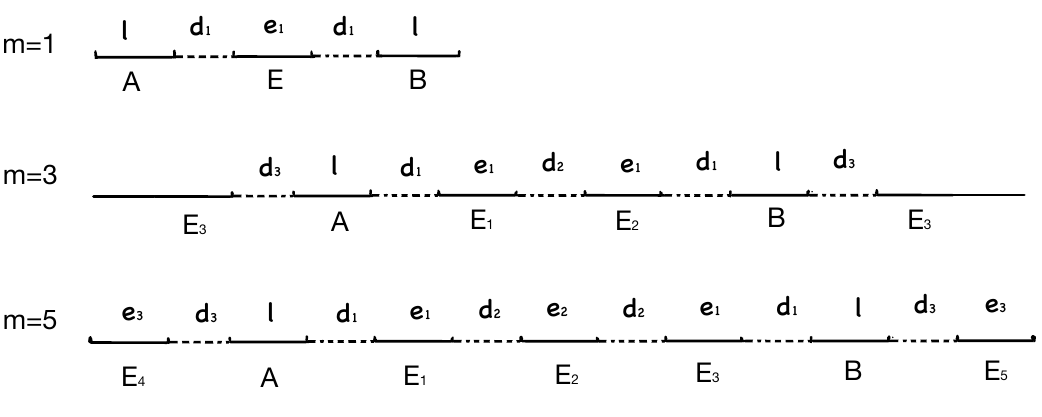}
	\caption{An illustration for the definition of the lengths \(e_1, e_2, ...\) and \(d_1, d_2, ...\) of the intervals in the case of odd \(m\). In this case, the phase transition condition demands that the maximal configuration should also have the reflection symmetry. Thus if \(m=3\), the subregion \(E_3\), which is outside \(A\) and \(B\), consists two semi-infinite intervals, while if \(m=5\), the two subregions outside \(A\) and \(B\) (i.e. \(E_4\) and \(E_5\)) are finite intervals with equal length. }\label{interval2}
\end{figure}

\noindent As shown in Figure \ref{I_n2} in the main text, one can find that, different from the one-gap cases, the maximum value of \( \exp\left(\frac{I(A:B|E)}2\right) \) seems like a perfect quadratic function of \( m \). 
The same as the one gap case, through the same method of induction, we find that the maximum value of \( I=\exp\left(\frac{I(A:B|E)}2\right) \) is generated by a polynomial.
Again, we use the freedom of conformal transformation to make the lengths of \(A\) and \(B\) to be \(l\) and the gap between them to be \(1\). We denote $e_i$ to be the lengths of the $i$-th interval in $E$ and $d_i$ to be the lengths for intervals in the gap regions, as shown in Figure \ref{interval2}.

In the case of odd \(m\), the phase transition condition, as depicted in the right MPT diagram of Figure \ref{MPT45} for an example of \( m=5 \), has {the reflection} symmetry.
Thus, the lengths \(e_j\) and \(d_j\) should also have {the reflection} symmetry.
Then we could calculate the upper bound \( I \) for each odd number \(m\) from the phase transition condition dictated by the corresponding MPT diagrams.
The values of \( I \) and the phase transition conditions for the first several odd number \( m \) are listed in the following table.
\begin{table}[H]
	\resizebox{\textwidth}{!}{%
		\begin{tabular}{|c|c|l|}
		\hline
			\(m\)      & Upper bound of \( \exp\left(\frac{I(A:B|E)}2\right)\)                                                               & Phase transition condition                                                                                                                                                                                   \\ \hline
			1                          & \( \frac{l^2 e_1}{(2l+1)d_1^2} \)                                                    & \(\frac{l e_1}{d_1(l+d_1+e_1)}=1\), \(2d_1+e_1=1\)                                                                                                                                                                                  \\ \hline
			3                          & \( \frac{(2d_3+2l+2d_1+2e_1+d_2)l^2 e_1^2}{d_1^2 d_2 d_3^2} \)                       & \begin{tabular}[c]{@{}l@{}}\(\frac{l e_1}{d_1(l+d_1+e_1)}=1\), \(\frac{e_1(d_2+e_1+d_1+l+d_3)}{d_2(d_1+l+d_3)}=1\),\\ \(\frac{(l+d_1+e_1)(2d_3+2l+2d_1+2e_1+d_2)}{d_3(d_2+e_1+d_1+l+d_3)}\), \(2d_1+d_2+2e_1=1\)   \end{tabular}                     
			\\ \hline
			5                          & \( \frac{l^2 e_1^2 e_2 e_3^2}{d_1^2 d_2^2 d_3^2(2e_3+2d_3+2l+2d_1+2e_1+2d_2+e_2)} \) & \begin{tabular}[c]{@{}l@{}l@{}}\(\frac{l e_1}{d_1(l+d_1+e_1)}=1\), \(\frac{e_3 (l+d_1+e_1)}{d_2(e_3+d_3+l+d_1+e_1)}=1\), \(\frac{e_2 (e_3+d_3+l+d_1+e_1)}{d_2(e_3+d_3+l+d_1+e_1+d_2+e_2)}=1\),\\
			 \(\frac{e_3(e_3+d_3+l+d_1+e_1+d_2+e_2)}{(d_2+e_1+d_1+l+d_3)(e_3 + d_3 + l + d_1 + e_1 + d_2 + e_2 + d_2 + e_1 + d_1 + l + d_3 + e_3)}=1\),\\ 
			\(\frac{e_1 (d_2 + e_1 + d_1 + l + d_3)}{d_2 (d_1 + l + d_3)}=1\) \(2d_1+2d_2+2e_1+e_2=1\)\end{tabular} \\ \hline
			\(\cdots\)                 & \(\cdots\)                                                                           & \(\cdots\)                                                                                                                                                                                                                         
		\end{tabular}%
	}
	\caption{Phase transition configuration conditions of the maximum CMI configuration and the corresponding maximum value of CMI for $m=1,3,5$ in the case of $E$ living in both the gap regions between and outside $A$ and $B$.}\label{table2}
\end{table}

Our strategy is the same as the one gap case. We first evaluate the upper bounds with integers \(l=1,2,3,4\) by \(\mathsf{Mathematica}\) and our program would give us the minimal polynomials \(\bar{P}_m(1), \bar{P}_m(2), \bar{P}_m(3), \bar{P}_m(4)\) that generate these upper bounds.
Then, one is able to find the pattern of the polynomials \(\bar{P}_m(l)\) that are adapted to arbitrary values of \(l\) for each \(m\).
With the normalization \(P_m(CR):=\frac{\bar{P}_m(l)}{2l+1}\), the polynomials should be linear in \(CR\): \(P_m(CR)=f_m+CR\;g_m\).
At last, a single polynomial can be deduced by a comparison of the coefficients in \(f_m\) and \(g_m\) with the numbers in the Pascal's Triangle respectively.

The polynomials \(P_m(CR)\) for \(m=1,3,5\) are found to be
\begin{equation}
	\begin{aligned}
		&x^2-x-CR (3x+1)\\
		&x^4-3x^3+3x^2-x-CR(3x+1)(5x^2+10x+1)\\
		&x^6-5x^5+10x^4-10x^3+5x^2-x-CR(5x^2+10x+1)(7x^3+35x^2+21x+1)\\
		&\cdots	
	\end{aligned}
\end{equation}
By comparing with the Pascal's Triangle, \(f_m(x)\) is identified as \(x(x-1)^m\) with \(m\) an odd integer, while \(g_m(x)\) can be factorized into two polynomials, the coefficients of both polynomial also appear in the Pascal's Triangle.
More specifically, the coefficients of the first and second polynomial are respectively the numbers in the odd-numbered columns of row \(m\) and \(m+2\) of the Pascal's Triangle.
Therefore, we have
\begin{equation}
	\begin{aligned}
		g_m(x)&=\left(\sum_{j=0}^{(m-1)/2}C_m^{2j}x^{2j}\right)\left(\sum_{j=0}^{(m+1)/2}C_{m+2}^{2j}x^{2j}\right)\\
		&=\frac{1}{4}\left(\left(\sqrt{x}+1\right)^m-\left(\sqrt{x}-1\right)^m\right)\left(\left(\sqrt{x}+1\right)^{m+2}-\left(\sqrt{x}-1\right)^{m+2}\right)\\
		&=\frac{1}{4}\left(\left(\sqrt{x}+1\right)^{m+1}-\left(\sqrt{x}-1\right)^{m+1}\right)^2-(x-1)^m,
	\end{aligned}
\end{equation}
and 
\begin{equation}\label{Pn2o}
		P_m(CR)=x(x-1)^m-\frac{CR}{4}\left(\left(\sqrt{x}+1\right)^{m+1}-\left(\sqrt{x}-1\right)^{m+1}\right)^2+CR(x-1)^m,
\end{equation}
where \( m=1,3,5,7,\cdots \) .

\subsection{Generating polynomials for the case of $E$ living in both gap regions: with $m$ even. }

\begin{figure}[H]
	\centering
	\includegraphics[scale=0.8]{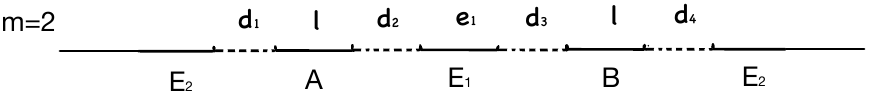}
	\caption{An illustration for the definition of the lengths \(e_1, e_2, ...\) and \(d_1, d_2, ...\) of the intervals in the case of \(m=2\). In this case, the phase transition condition does not have the reflection symmetry. Therefore, all the gaps should be different. The subregion \(E_2\) should be two semi-infinite intervals, otherwise the phase transition condition does not have a real positive solution for all the intervals. }\label{interval3}
\end{figure}

\noindent In the case of even \( m \), the {reflection} symmetry of the phase transition configuration condition, as depicted in the left MPT diagram of Figure \ref{MPT45} for an example of \( m=4 \), is broken.
%Taking advantage of this symmetry, the lengths of \(E_3\) and \(E_1\) are equal and can be set to be \(e_1\) as well as the lengths of \(E_2\) and \(E_4\) are set to be \(e_2\).
%Similarly, the gaps between \( A \)\(E_1\) and \( B \)\(E_3\) are set to be \(d_1\), the gaps between \( A \)\(E_4\) and \( B \)\(E_2\) are set to be \(d_2\), while the remaining gap between \( E_2\)\(E_3\) is \(d_3\).
As a consequence, all the lengths of intervals and gaps are independent, which makes the computation more involved. 
Even so, the upper bound of \( I=\exp\left(\frac{I(A:B|E)}{2}\right) \) for each even number \(m\) and the phase transition condition could still be represented by the lengths of intervals \(e_1, e_2, \cdots\), gaps \(d_1, d_2, \cdots\), and \(l\).
For example, at $m=2$ the upper bound of \( I \) is
\begin{equation}
	\text{max}\left\{I\right\}=\frac{l^2 e_1 (d_1 + d_2 + d_3 + d_4 + 2 l + e_1)}{d_1 d_2 d_3 d_4},
\end{equation}
and the phase transition conditions are
\begin{equation}
	\begin{aligned}
		l e_1&=d_2(l+d_2+e_1),\\
		(l+d_2+e_1)(d_1 + d_2 + d_3 + d_4 + 2 l + e_1)&=d_1(d_3+l+d_4),\\
		l(d_1 + d_2 + d_3 + d_4 + 2 l + e_1)&=d_4(d_1+l+d_2+e_1+d_3),\\
		e_1(d_1+l+d_2+e_1+d_3)&=d_3(d_1+l+d_2),\\
		d_2+e_1+d_3&=1,   
	\end{aligned}
\end{equation}
where the definitions for the lengths of the subregions and gaps are shown in Figure \ref{interval3}.

Through the same method used in the above cases, we obtain the polynomials \(P_m(CR)\) that generate the upper bounds for each even \( m \), which are
\begin{equation}
	\begin{aligned}
		&x^3-2x^2+x-CR (3x+1)^2\\
		&x^5-4x^4+6x^3-4x^2+x-CR(5x^2+10x+1)^2\\
		&x^7-6x^6+15x^5-20x^4+15x^3-6x^2+x-CR(7x^3+35x^2+21x+1)^2\\
		&\cdots	
	\end{aligned}
\end{equation}
Note that the polynomials are all linear in \(CR\), such that we can write \(P_m(CR)=f_m+CR\;g_m\).
The same as the odd case, \(f_m(x)=x(x-1)^m\) except that \(m\) is even now.
Moreover, \(g_m\) is found to be the square of a polynomial whose coefficients appear exactly as the numbers that {locate} at the odd-numbered columns of the \( (m+1) \)th row in the Pascal's Triangle.

Thus, we have
\begin{equation}
	\begin{aligned}
		g_m(x)&=\left(\sum_{j=0}^{m/2}C_{m+1}^{2j}x^{2j}\right)^2\\
		&=\frac{1}{4}\left[\left(\sqrt{x}+1\right)^{m+1}-\left(\sqrt{x}-1\right)^{m+1}\right]^2,
	\end{aligned}
\end{equation}
and 
\begin{equation}\label{Pn2e}
	P_m(CR)=x(x-1)^m-\frac{CR}{4}\left(\left(\sqrt{x}+1\right)^{m+1}-\left(\sqrt{x}-1\right)^{m+1}\right)^2,
\end{equation}
where \( m=2,4,6,8,\cdots \) .

The two polynomials \eqref{Pn2o} and \eqref{Pn2e} that govern the upper bound of \(I\) respectively in the odd and even case can be written together as
\begin{equation}\label{algebraic2inappendix}
		x(x - 1)^m - \frac{CR}{4} \left[ (\sqrt{x} + 1)^{m + 1} - (\sqrt{x} - 1)^{m + 1} \right]^2 + CR(x - 1)^m \delta,
\end{equation}
where \( \delta = 0 \) for even \( m \) and \( \delta = 1 \) for odd \( m \).

\bibliography{reference}

\providecommand{\href}[2]{#2}\begingroup\raggedright\begin{thebibliography}{10}

\bibitem{Maldacena:1997re}
J.~M. Maldacena, \emph{{The Large N limit of superconformal field theories and supergravity}}, \href{http://dx.doi.org/10.4310/ATMP.1998.v2.n2.a1}{\emph{Adv. Theor. Math. Phys.} {\bf 2} (1998) 231--252}, [\href{http://arxiv.org/abs/hep-th/9711200}{{\tt hep-th/9711200}}].

\bibitem{Ryu_2006}
S.~Ryu and T.~Takayanagi, \emph{Holographic derivation of entanglement entropy from the anti{\textendash}de sitter space/conformal field theory correspondence}, \href{http://dx.doi.org/10.1103/physrevlett.96.181602}{\emph{Physical Review Letters} {\bf 96} (may, 2006) }.

\bibitem{Maldacena_2013}
J.~Maldacena and L.~Susskind, \emph{Cool horizons for entangled black holes}, \href{http://dx.doi.org/10.1002/prop.201300020}{\emph{Fortschritte der Physik} {\bf 61} (aug, 2013) 781--811}.

\bibitem{VanRaamsdonk:2010pw}
M.~Van~Raamsdonk, \emph{{Building up spacetime with quantum entanglement}}, \href{http://dx.doi.org/10.1142/S0218271810018529}{\emph{Gen. Rel. Grav.} {\bf 42} (2010) 2323--2329}, [\href{http://arxiv.org/abs/1005.3035}{{\tt 1005.3035}}].

\bibitem{Akers:2019gcv}
C.~Akers and P.~Rath, \emph{{Entanglement Wedge Cross Sections Require Tripartite Entanglement}}, \href{http://dx.doi.org/10.1007/JHEP04(2020)208}{\emph{JHEP} {\bf 04} (2020) 208}, [\href{http://arxiv.org/abs/1911.07852}{{\tt 1911.07852}}].

\bibitem{Ju:2024hba}
X.-X. Ju, W.-B. Pan, Y.-W. Sun and Y.~Zhao, \emph{{Holographic multipartite entanglement from the upper bound of $n$-partite information}},  \href{http://arxiv.org/abs/2411.07790}{{\tt 2411.07790}}.

\bibitem{Ju:2023dzo}
X.-X. Ju, B.-H. Liu, W.-B. Pan, Y.-W. Sun and Y.-T. Wang, \emph{{Squashed Entanglement from Generalized Rindler Wedge}},  \href{http://arxiv.org/abs/2310.09799}{{\tt 2310.09799}}.

\bibitem{Basak:2024uwc}
J.~K. Basak, V.~Malvimat and J.~Yoon, \emph{{A New Genuine Multipartite Entanglement Measure: from Qubits to Multiboundary Wormholes}},  \href{http://arxiv.org/abs/2411.11961}{{\tt 2411.11961}}.

\bibitem{Bao:2017nhh}
N.~Bao and I.~F. Halpern, \emph{{Holographic Inequalities and Entanglement of Purification}}, \href{http://dx.doi.org/10.1007/JHEP03(2018)006}{\emph{JHEP} {\bf 03} (2018) 006}, [\href{http://arxiv.org/abs/1710.07643}{{\tt 1710.07643}}].

\bibitem{Ju:2024xcn}
X.-X. Ju, T.-Z. Lai, B.-H. Liu, W.-B. Pan and Y.-W. Sun, \emph{{Entanglement structures from modified IR geometry}}, \href{http://dx.doi.org/10.1007/JHEP07(2024)181}{\emph{JHEP} {\bf 07} (2024) 181}, [\href{http://arxiv.org/abs/2404.02737}{{\tt 2404.02737}}].

\bibitem{Hubeny_2014}
V.~E. Hubeny, \emph{Covariant residual entropy}, \href{http://dx.doi.org/10.1007/jhep09(2014)156}{\emph{Journal of High Energy Physics} {\bf 2014} (sep, 2014) }.

\bibitem{Czech_2015}
B.~Czech, P.~Hayden, N.~Lashkari and B.~Swingle, \emph{The information theoretic interpretation of the length of a curve}, \href{http://dx.doi.org/10.1007/jhep06(2015)157}{\emph{Journal of High Energy Physics} {\bf 2015} (jun, 2015) }.

\bibitem{Czech:2014wka}
B.~Czech, X.~Dong and J.~Sully, \emph{{Holographic Reconstruction of General Bulk Surfaces}}, \href{http://dx.doi.org/10.1007/JHEP11(2014)015}{\emph{JHEP} {\bf 11} (2014) 015}, [\href{http://arxiv.org/abs/1406.4889}{{\tt 1406.4889}}].

\bibitem{Myers:2014jia}
R.~C. Myers, J.~Rao and S.~Sugishita, \emph{{Holographic Holes in Higher Dimensions}}, \href{http://dx.doi.org/10.1007/JHEP06(2014)044}{\emph{JHEP} {\bf 06} (2014) 044}, [\href{http://arxiv.org/abs/1403.3416}{{\tt 1403.3416}}].

\bibitem{Headrick:2014eia}
M.~Headrick, R.~C. Myers and J.~Wien, \emph{{Holographic Holes and Differential Entropy}}, \href{http://dx.doi.org/10.1007/JHEP10(2014)149}{\emph{JHEP} {\bf 10} (2014) 149}, [\href{http://arxiv.org/abs/1408.4770}{{\tt 1408.4770}}].

\bibitem{Balasubramanian:2018uus}
V.~Balasubramanian and C.~Rabideau, \emph{{The dual of non-extremal area: differential entropy in higher dimensions}}, \href{http://dx.doi.org/10.1007/JHEP09(2020)051}{\emph{JHEP} {\bf 09} (2020) 051}, [\href{http://arxiv.org/abs/1812.06985}{{\tt 1812.06985}}].

\bibitem{Balasubramanian:2013lsa}
V.~Balasubramanian, B.~D. Chowdhury, B.~Czech, J.~de~Boer and M.~P. Heller, \emph{{Bulk curves from boundary data in holography}}, \href{http://dx.doi.org/10.1103/PhysRevD.89.086004}{\emph{Phys. Rev. D} {\bf 89} (2014) 086004}, [\href{http://arxiv.org/abs/1310.4204}{{\tt 1310.4204}}].

\bibitem{Czech:2014ppa}
B.~Czech and L.~Lamprou, \emph{{Holographic definition of points and distances}}, \href{http://dx.doi.org/10.1103/PhysRevD.90.106005}{\emph{Phys. Rev. D} {\bf 90} (2014) 106005}, [\href{http://arxiv.org/abs/1409.4473}{{\tt 1409.4473}}].

\bibitem{Balasubramanian:2013rqa}
V.~Balasubramanian, B.~Czech, B.~D. Chowdhury and J.~de~Boer, \emph{{The entropy of a hole in spacetime}}, \href{http://dx.doi.org/10.1007/JHEP10(2013)220}{\emph{JHEP} {\bf 10} (2013) 220}, [\href{http://arxiv.org/abs/1305.0856}{{\tt 1305.0856}}].

\bibitem{Ju:2023bjl}
X.-X. Ju, W.-B. Pan, Y.-W. Sun and Y.-T. Wang, \emph{{Generalized Rindler Wedge and Holographic Observer Concordance}},  \href{http://arxiv.org/abs/2302.03340}{{\tt 2302.03340}}.

\bibitem{Vidal:2014aal}
G.~Vidal and Y.~Chen, \emph{{Entanglement contour}}, \href{http://dx.doi.org/10.1088/1742-5468/2014/10/P10011}{\emph{J. Stat. Mech.} {\bf 2014} (2014) P10011}, [\href{http://arxiv.org/abs/1406.1471}{{\tt 1406.1471}}].

\bibitem{Wen:2018whg}
Q.~Wen, \emph{{Fine structure in holographic entanglement and entanglement contour}}, \href{http://dx.doi.org/10.1103/PhysRevD.98.106004}{\emph{Phys. Rev. D} {\bf 98} (2018) 106004}, [\href{http://arxiv.org/abs/1803.05552}{{\tt 1803.05552}}].

\bibitem{Wen:2019iyq}
Q.~Wen, \emph{{Formulas for Partial Entanglement Entropy}}, \href{http://dx.doi.org/10.1103/PhysRevResearch.2.023170}{\emph{Phys. Rev. Res.} {\bf 2} (2020) 023170}, [\href{http://arxiv.org/abs/1910.10978}{{\tt 1910.10978}}].

\bibitem{Nozaki:2013wia}
M.~Nozaki, T.~Numasawa and T.~Takayanagi, \emph{{Holographic Local Quenches and Entanglement Density}}, \href{http://dx.doi.org/10.1007/JHEP05(2013)080}{\emph{JHEP} {\bf 05} (2013) 080}, [\href{http://arxiv.org/abs/1302.5703}{{\tt 1302.5703}}].

\bibitem{Bhattacharya:2014vja}
J.~Bhattacharya, V.~E. Hubeny, M.~Rangamani and T.~Takayanagi, \emph{{Entanglement density and gravitational thermodynamics}}, \href{http://dx.doi.org/10.1103/PhysRevD.91.106009}{\emph{Phys. Rev. D} {\bf 91} (2015) 106009}, [\href{http://arxiv.org/abs/1412.5472}{{\tt 1412.5472}}].

\bibitem{Shimaji:2018czt}
T.~Shimaji, T.~Takayanagi and Z.~Wei, \emph{{Holographic Quantum Circuits from Splitting/Joining Local Quenches}}, \href{http://dx.doi.org/10.1007/JHEP03(2019)165}{\emph{JHEP} {\bf 03} (2019) 165}, [\href{http://arxiv.org/abs/1812.01176}{{\tt 1812.01176}}].

\bibitem{Alishahiha:2014jxa}
M.~Alishahiha, M.~R. Mohammadi~Mozaffar and M.~R. Tanhayi, \emph{{On the Time Evolution of Holographic n-partite Information}}, \href{http://dx.doi.org/10.1007/JHEP09(2015)165}{\emph{JHEP} {\bf 09} (2015) 165}, [\href{http://arxiv.org/abs/1406.7677}{{\tt 1406.7677}}].

\bibitem{LoMonaco:2023xws}
G.~Lo~Monaco, L.~Innocenti, D.~Cilluffo, D.~A. Chisholm, S.~Lorenzo and G.~M. Palma, \emph{{Quantum scrambling via accessible tripartite information}}, \href{http://dx.doi.org/10.1088/2058-9565/accd92}{\emph{Quantum Sci. Technol.} {\bf 8} (2023) 035006}, [\href{http://arxiv.org/abs/2305.19334}{{\tt 2305.19334}}].

\bibitem{Hayden_2013}
P.~Hayden, M.~Headrick and A.~Maloney, \emph{Holographic mutual information is monogamous}, \href{http://dx.doi.org/10.1103/physrevd.87.046003}{\emph{Physical Review D} {\bf 87} (feb, 2013) }.

\bibitem{bengtsson2016brief}
I.~Bengtsson and K.~Zyczkowski, \emph{A brief introduction to multipartite entanglement}, {\emph{arXiv preprint arXiv:1612.07747} (2016) }.

\bibitem{Dutta:2019gen}
S.~Dutta and T.~Faulkner, \emph{{A canonical purification for the entanglement wedge cross-section}}, \href{http://dx.doi.org/10.1007/JHEP03(2021)178}{\emph{JHEP} {\bf 03} (2021) 178}, [\href{http://arxiv.org/abs/1905.00577}{{\tt 1905.00577}}].

\bibitem{Calabrese:2012ew}
P.~Calabrese, J.~Cardy and E.~Tonni, \emph{{Entanglement negativity in quantum field theory}}, \href{http://dx.doi.org/10.1103/PhysRevLett.109.130502}{\emph{Phys. Rev. Lett.} {\bf 109} (2012) 130502}, [\href{http://arxiv.org/abs/1206.3092}{{\tt 1206.3092}}].

\bibitem{Kusuki:2019zsp}
Y.~Kusuki, J.~Kudler-Flam and S.~Ryu, \emph{{Derivation of holographic negativity in AdS$_3$/CFT$_2$}}, \href{http://dx.doi.org/10.1103/PhysRevLett.123.131603}{\emph{Phys. Rev. Lett.} {\bf 123} (2019) 131603}, [\href{http://arxiv.org/abs/1907.07824}{{\tt 1907.07824}}].

\bibitem{Gadde:2022cqi}
A.~Gadde, V.~Krishna and T.~Sharma, \emph{{New multipartite entanglement measure and its holographic dual}}, \href{http://dx.doi.org/10.1103/PhysRevD.106.126001}{\emph{Phys. Rev. D} {\bf 106} (2022) 126001}, [\href{http://arxiv.org/abs/2206.09723}{{\tt 2206.09723}}].

\bibitem{Penington:2022dhr}
G.~Penington, M.~Walter and F.~Witteveen, \emph{{Fun with replicas: tripartitions in tensor networks and gravity}}, \href{http://dx.doi.org/10.1007/JHEP05(2023)008}{\emph{JHEP} {\bf 05} (2023) 008}, [\href{http://arxiv.org/abs/2211.16045}{{\tt 2211.16045}}].

\bibitem{Yuan:2024yfg}
M.-K. Yuan, M.~Li and Y.~Zhou, \emph{{Reflected multi-entropy and its holographic dual}},  \href{http://arxiv.org/abs/2410.08546}{{\tt 2410.08546}}.

\bibitem{Bao:2015bfa}
N.~Bao, S.~Nezami, H.~Ooguri, B.~Stoica, J.~Sully and M.~Walter, \emph{{The Holographic Entropy Cone}}, \href{http://dx.doi.org/10.1007/JHEP09(2015)130}{\emph{JHEP} {\bf 09} (2015) 130}, [\href{http://arxiv.org/abs/1505.07839}{{\tt 1505.07839}}].

\bibitem{Hubeny:2018trv}
V.~E. Hubeny, M.~Rangamani and M.~Rota, \emph{{Holographic entropy relations}}, \href{http://dx.doi.org/10.1002/prop.201800067}{\emph{Fortsch. Phys.} {\bf 66} (2018) 1800067}, [\href{http://arxiv.org/abs/1808.07871}{{\tt 1808.07871}}].

\bibitem{Hubeny:2018ijt}
V.~E. Hubeny, M.~Rangamani and M.~Rota, \emph{{The holographic entropy arrangement}}, \href{http://dx.doi.org/10.1002/prop.201900011}{\emph{Fortsch. Phys.} {\bf 67} (2019) 1900011}, [\href{http://arxiv.org/abs/1812.08133}{{\tt 1812.08133}}].

\bibitem{He:2019ttu}
T.~He, M.~Headrick and V.~E. Hubeny, \emph{{Holographic Entropy Relations Repackaged}}, \href{http://dx.doi.org/10.1007/JHEP10(2019)118}{\emph{JHEP} {\bf 10} (2019) 118}, [\href{http://arxiv.org/abs/1905.06985}{{\tt 1905.06985}}].

\bibitem{HernandezCuenca:2019wgh}
S.~Hern\'andez~Cuenca, \emph{{Holographic entropy cone for five regions}}, \href{http://dx.doi.org/10.1103/PhysRevD.100.026004}{\emph{Phys. Rev. D} {\bf 100} (2019) 026004}, [\href{http://arxiv.org/abs/1903.09148}{{\tt 1903.09148}}].

\bibitem{He:2020xuo}
T.~He, V.~E. Hubeny and M.~Rangamani, \emph{{Superbalance of Holographic Entropy Inequalities}}, \href{http://dx.doi.org/10.1007/JHEP07(2020)245}{\emph{JHEP} {\bf 07} (2020) 245}, [\href{http://arxiv.org/abs/2002.04558}{{\tt 2002.04558}}].

\bibitem{Avis:2021xnz}
D.~Avis and S.~Hern\'andez-Cuenca, \emph{{On the foundations and extremal structure of the holographic entropy cone}}, \href{http://dx.doi.org/10.1016/j.dam.2022.11.016}{\emph{Discrete Appl. Math.} {\bf 328} (2023) 16--39}, [\href{http://arxiv.org/abs/2102.07535}{{\tt 2102.07535}}].

\bibitem{Fadel:2021urx}
M.~Fadel and S.~Hern\'andez-Cuenca, \emph{{Symmetrized holographic entropy cone}}, \href{http://dx.doi.org/10.1103/PhysRevD.105.086008}{\emph{Phys. Rev. D} {\bf 105} (2022) 086008}, [\href{http://arxiv.org/abs/2112.03862}{{\tt 2112.03862}}].

\bibitem{Bao:2024azn}
N.~Bao, K.~Furuya and J.~Naskar, \emph{{Towards a complete classification of holographic entropy inequalities}},  \href{http://arxiv.org/abs/2409.17317}{{\tt 2409.17317}}.

\bibitem{Brown:1986nw}
J.~D. Brown and M.~Henneaux, \emph{{Central Charges in the Canonical Realization of Asymptotic Symmetries: An Example from Three-Dimensional Gravity}}, \href{http://dx.doi.org/10.1007/BF01211590}{\emph{Commun. Math. Phys.} {\bf 104} (1986) 207--226}.

\bibitem{Li_2018}
K.~Li and A.~Winter, \emph{Squashed entanglement, {\textdollar}{\textdollar}{\textbackslash}mathbf $\lbrace$k$\rbrace${\textdollar}{\textdollar} k -extendibility, quantum markov chains, and recovery maps}, \href{http://dx.doi.org/10.1007/s10701-018-0143-6}{\emph{Foundations of Physics} {\bf 48} (feb, 2018) 910--924}.

\bibitem{Li_2014}
K.~Li and A.~Winter, \emph{Relative entropy and squashed entanglement}, \href{http://dx.doi.org/10.1007/s00220-013-1871-2}{\emph{Communications in Mathematical Physics} {\bf 326} (jan, 2014) 63--80}.

\bibitem{Wilde_2016}
M.~M. Wilde, \emph{Squashed entanglement and approximate private states}, \href{http://dx.doi.org/10.1007/s11128-016-1432-7}{\emph{Quantum Information Processing} {\bf 15} (sep, 2016) 4563--4580}.

\bibitem{Avis_2008}
D.~Avis, P.~Hayden and I.~Savov, \emph{Distributed compression and multiparty squashed entanglement}, \href{http://dx.doi.org/10.1088/1751-8113/41/11/115301}{\emph{Journal of Physics A: Mathematical and Theoretical} {\bf 41} (mar, 2008) 115301}.

\bibitem{Yang_2009}
D.~Yang, K.~Horodecki, M.~Horodecki, P.~Horodecki, J.~Oppenheim and W.~Song, \emph{Squashed entanglement for multipartite states and entanglement measures based on the mixed convex roof}, \href{http://dx.doi.org/10.1109/tit.2009.2021373}{\emph{{IEEE} Transactions on Information Theory} {\bf 55} (jul, 2009) 3375--3387}.

\bibitem{Christandl_2004}
M.~Christandl and A.~Winter, \emph{{\textquotedblleft}squashed entanglement{\textquotedblright}: An additive entanglement measure}, \href{http://dx.doi.org/10.1063/1.1643788}{\emph{Journal of Mathematical Physics} {\bf 45} (mar, 2004) 829--840}.

\bibitem{Brand_o_2011}
F.~G. S.~L. Brand{\~{a}}o, M.~Christandl and J.~Yard, \emph{Faithful squashed entanglement}, \href{http://dx.doi.org/10.1007/s00220-011-1302-1}{\emph{Communications in Mathematical Physics} {\bf 306} (aug, 2011) 805--830}.

\bibitem{Bhattacharjee:2024ceb}
A.~Bhattacharjee and J.~Naskar, \emph{{Revisiting holographic codes with fractal-like boundary erasures}},  \href{http://arxiv.org/abs/2411.02825}{{\tt 2411.02825}}.

\bibitem{Takayanagi:2017knl}
T.~Takayanagi and K.~Umemoto, \emph{{Entanglement of purification through holographic duality}}, \href{http://dx.doi.org/10.1038/s41567-018-0075-2}{\emph{Nature Phys.} {\bf 14} (2018) 573--577}, [\href{http://arxiv.org/abs/1708.09393}{{\tt 1708.09393}}].

\bibitem{Mori:2024gwe}
T.~Mori and B.~Yoshida, \emph{{Does connected wedge imply distillable entanglement?}},  \href{http://arxiv.org/abs/2411.03426}{{\tt 2411.03426}}.

\bibitem{Jain:2022csf}
P.~Jain, N.~Jokela, M.~Jarvinen and S.~Mahapatra, \emph{{Bounding entanglement wedge cross sections}}, \href{http://dx.doi.org/10.1007/JHEP03(2023)102}{\emph{JHEP} {\bf 03} (2023) 102}, [\href{http://arxiv.org/abs/2211.07671}{{\tt 2211.07671}}].

\bibitem{Ju:2025}
X.-X. Ju, W.-B. Pan, Y.-W. Sun and Y.~Zhao, \emph{{In progress}}, .

\bibitem{Erdmenger:2017gdk}
J.~Erdmenger, D.~Fernandez, M.~Flory, E.~Megias, A.-K. Straub and P.~Witkowski, \emph{{Time evolution of entanglement for holographic steady state formation}}, \href{http://dx.doi.org/10.1007/JHEP10(2017)034}{\emph{JHEP} {\bf 10} (2017) 034}, [\href{http://arxiv.org/abs/1705.04696}{{\tt 1705.04696}}].

\end{thebibliography}\endgroup
\bibliographystyle{JHEP}

\end{document}